\documentclass[10pt, twocolumn]{aastex631}
\usepackage{amsmath}
\usepackage{amsthm}
\usepackage{mathtools}
\usepackage{graphicx}
\usepackage{enumitem}
\usepackage[T1]{fontenc}
\usepackage{bm}

\newcommand{\be}{\begin{equation}}
\newcommand{\ee}{\end{equation}}
\newcommand{\ba}{\begin{eqnarray}}
\newcommand{\ea}{\end{eqnarray}}

\newcommand{\nn}{\mbox{} \nonumber \\ \mbox{}}
\newcommand{\rmax}{r_{\rm max}}
\newcommand{\rmin}{r_{\rm min}}
\newcommand{\epszero}{\hat\varepsilon_0}
\newcommand{\epsone}{\hat{\varepsilon}_1}
\newcommand{\epstwo}{\hat{\varepsilon}_2}

\newcommand{\rmsBeta}{\delta{\bm\beta}}
\newcommand{\ki}{\hat{k}_0}
\newcommand{\kf}{\hat{k}_1}

\newcommand{\norm}{(\ki \times \epszero)}
\newcommand{\maxdenschange}{\delta n_\pm^{\rm max}}
\newcommand{\thetaext}{\theta_{\rm ext}}
\newcommand{\phiext}{\phi_{\rm ext}}
\newcommand{\Eorb}{\mathcal{E}}
\newcommand{\Lang}{\mathcal{L}_z}
\newcommand{\rISCO}{r_{\rm ISCO}}
\newcommand{\rCorona}{r_{\rm corona}}
\newcommand{\nequil}{n_{\rm eq}}
\newcommand{\ZAMO}{{\cal Z}}
\newcommand{\disk}{{\cal D}}
\newcommand{\particle}{{\cal P}}

\newcommand{\gt}{\tilde{g}_{tt}}
\newcommand{\gr}{g_{rr}}
\newcommand{\gtheta}{g_{\theta \theta}}
\newcommand{\gphi}{g_{\phi \phi}}

\newcommand{\tetBLZ}{e_{{\rm BL}\rightarrow \ZAMO}}

\newcommand{\simbase}{\texttt{a90}}
\newcommand{\simdisk}{\texttt{a50\_plungeII}}
\newcommand{\simoutflow}{\texttt{a90\_flow}}
\newcommand{\simlowflow}{\texttt{a50\_flow}}
\newcommand{\simlowspin}{\texttt{a50\_plunge}}

\newcommand{\simnodisk}{\texttt{a90\_nodisk}}

\begin{document}

\title{Pair-Rich Corona of an Accreting Kerr Black Hole}

\author{Jonathan Zhang}
\affiliation{Department of Physics, McLennan Physical Labs, Toronto, ON M5S 1A7, Canada}
\email{jon.zhang@mail.utoronto.ca}

\author{Christopher Thompson}
\affiliation{Canadian Institute for Theoretical Astrophysics, 60 St. George St., Toronto, ON M5S 3H8, Canada}

\begin{abstract}
    We build a self-consistent model of a warm scattering corona near an accreting black hole in
    Kerr geometry, in the regime of slow ($\sim 0.01$ Eddington) mass accretion.  An iterative
    Monte Carlo procedure is developed that incorporates self-consistently the effects of
    Compton scattering and electron-positron pair creation, as well as 
    general relativistic lensing and frame dragging effects.  Soft thermal
    photons are seeded in the inner disk and the velocity dispersion of the electrons
    and positrons adjusted to yield a fixed seed luminosity amplification through Compton
    scattering.  A simple kinematic prescription is also added for bulk outflow.  
    Pair creation by photon collisions
    raises significantly the density of scattering charges in and around the innermost ion disk, which
    is assumed to be geometrically thick and rarefied compared with the disk outside 10
    gravitational radii.  The self-consistent pair cloud is concentrated closer to the BH.
    The spectrum and polarization of the escaping X-rays are recorded
    as a function of the observer's orientation.  The temperature and Compton parameter
    measured from the output spectra using the compPS package are consistent with fits to binary BH
    data in the hardest spectral state; the polarization degree rises to $4-10\%$ through 
    the 2-8 keV band with allowance for $e^\pm$ upflow from the BH equator.  
\end{abstract}

\section{Introduction}\label{sec:intro}

A stellar-mass black hole (BH) accreting from a second, less evolved star forms a bright X-ray binary (XRB). Galactic BHXRBs have been observed to transition between soft and hard spectral states \citep{Belloni2010}.
The hardest detected X-ray spectra show a prominent power law component that is consistent with repeated Compton scattering of soft, quasi-thermal X-rays by sub-relativistic electrons  \citep{Shapiro1976,Sunyaev1980}. 
The measured intensity of hard X-rays and gamma-rays may be high enough to form a substantial density of electrons and positrons if the underlying dissipation is concentrated around the innermost stable orbit \citep{Guilbert1983, Done1989, Fabian2015}.

This paper investigates the self-regulating formation of a $e^\pm$ corona around an accreting BH.
A Monte Carlo (MC) procedure is developed that combines the effects of relativistic ray propagation and Klein-Nishina (KN) scattering with $e^\pm$ heating, photon collisions, and pair annihilation.  The model produces an inhomogeneous corona that is simultaneously in energy and $e^\pm$ creation/annihilation equilibrium;  it can probe how the observer-dependent spectrum and polarization depend on BH spin, heating profile, and photon seeding mechanism.

The enrichment of coronal plasma in pairs has frequently been studied in the approximation of local kinetic equilibrium \citep{Lightman1982,Svensson1982,Poutanen1996,Chan2026}.   
The reduction in mass per scattering charge facilitates the formation of an outflow of coronal plasma, which has
interesting implications for spectra and polarimetry \citep{Beloborodov1999, Malzac2001, Poutanen2023}. Zoom-in kinetic simulations demonstrate that energy injection via reconnection and turbulence can self-consistently produce the required density of pairs and the observed hard spectra \citep{Groselj2024,Nattila2024}.  However, global simulations of photon transport have not previously combined the effects of relativistic gravity, particle heating, and pair creation.

The scattering electrons move through some combination of kinetic and hydromagnetic motions, whose coupling to the photons is enhanced by pair creation \citep{Thompson1994,Socrates2004,Beloborodov2017}.
Fluid simulations that incorporate radiative transport and energy exchange with plasma often do not retain spectral information, instead evolving the total photon energy \citep{Farris2008, White2023}. MC codes that implement Compton scattering for ray tracing \citep{Dolence2009, Schnittman2013, Kawashima2023} and energy exchange with the plasma \citep{Ryan2015} have been developed, but do not track pair production.
Non-linear MC codes have also been used to model photon-pair interactions, including cooling, Comptonization and pair creation \citep{Stern1995, Dove1997, Dove1997_2}; while these self-consistently set the optical depth and temperature of the pair corona, the geometry itself is fixed and the calculation performed in flat spacetime. 
General relativistic magnetohydrodynamic (MHD) and kinetic simulations capture the distribution of ion-electron plasma 
(under specific assumptions about the initial magnetic field and source plasma), 
but only local treatments of $e^\pm$ injection have so far
been implemented, without directly representing photon collisions \citep{Mehlhaff2026}.  

In this paper, we consider a disk structure with a zone of relatively low ion density and large scale height outside the innermost stable circular orbit (ISCO).  The MC model starts from a simplified ion-electron distribution, adds stochastic motions to the electrons dependent on a fixed ratio of Compton energy loss to soft photon input, and follows the general relativistic trajectories of X-ray photons including redistribution by scattering.  By an iterative procedure, we re-compute the density of $e^\pm$ from the ensemble distribution of hard X-rays and gamma-rays via the process $\gamma + \gamma \leftrightarrow e^+ + e^-$. 
The spatial distribution of $e^\pm$ expands considerably beyond the seed profile of orbiting ions, extending well inside the ISCO and above the seed ion cloud.  
  
We test how the X-ray spectrum and measured polarization \citep{Krawczynski2022, Podgorny2024, Ewing2025} vary with the orientation of the observer, and how they are connected with the structure of
the corona and the presence of a cold, outer disk.  This procedure accounts self-consistently for
the effects of gravitational redshift, gravitational lensing, and orbital velocity shear.
In the MC model, dissipation follows the expanding $e^\pm$ cloud, with the expansion
being limited by the efficacy of photon collisions.
To probe the possible role of magnetic dissipation in the plunging zone, we allow the coronal heating 
either to be concentrated outside the ISCO, or alternatively to follow the $e^\pm$ plasma into the innermost plunging flow.  But otherwise there is no explicit representation of magnetic stresses.

The MC model provides insight into other open questions, such as soft photon seeding mechanisms, and tension in the spin measurement of BHXRBs.
In principle, seed photons may be emitted by cyclo-synchrotron emission of energized, non-thermal electrons \citep{Wardzinski2000, Poutanen2009, Veledina2013}, or electron-ion bremsstrahlung from cold, dense clumps of the accretion disk \citep{Celotti1992, Poutanen2018, Liska2022}.  We test the impact of these distinct emission processes in an approximate manner by tying the emission of soft seed X-rays either to 
the background ion flow, or to the self-consistent $e^\pm$ profile.

Measurements of BHXRBs favor high spin, with some approaching the theoretical maximum \citep{Gou2014, Qin2019, Draghis2025}. This is in tension with constraints from gravitational wave (GW) detections of merging binary BHs, which prefer zero or low spin BHs \citep{LVK2025}. Selection effects may ease this tension; high spin BHXRBs may be brighter and thus preferentially detected \citep{Qin2019, Sen2021}, while GW measurements may allow for a limited subpopulation of high spin BHs \citep{Fishbach2022}. BHXRB spin measurements also rely on accretion disk modeling; spin is measured through X-ray reflection or continuum fitting of the spectra, which both assume the disk truncates at the ISCO \citep{Reynolds2021}. Modifications to the disk model, such as a warm Comptonizing region above the disk, can result in much lower spin fits of $a \leq 0.1$ \citep{Zdziarski2024}. Even in thin accreting disks, magnetic stress may support the disk further inside the ISCO \citep{Noble2010, Sadowski2016}, which can lead to underestimates of the ISCO and a corresponding overestimate of the spin.

The plan of the paper is as follows. Section \ref{sec:outline} highlights a number of key results of our model. 
The next three sections and the Appendices describe basic elements of the MC procedure:  the adopted orbital velocity profile and 
input disk structure (Section \ref{sec:disk});  
the emission, propagation and scattering of photons (Section \ref{sec:photon});
and the creation of pairs and iteration of the coronal heating rate and $e^\pm$ density
(Section \ref{sec:iteration}).
Various astrophysical implications of the model are investigated in Section \ref{sec:obs_results},
including the dependence of the spectropolarimetric data on BH spin, the imprint of free electron
scattering off an outer, cold disk, and the influence of the X-ray seeding mechanism.
Section \ref{sec:conclusion} summarizes the main results of the work and its application to 
observations of hard-state BHXRBs.
Throughout we use units in which the speed of light $c$, Newton's constant $G$ and reduced Planck's constant $\hbar$ are unity. 

\section{Outline}
\label{sec:outline}
This section summarizes some key outcomes of our model.  These connect non-local 
$e^\pm$ pair creation around a stellar-mass BH with the spectrally
hard and low-luminosity states of galactic X-ray binaries.  We describe
the $e^\pm$ distribution around the BH, and the observer-dependent spectrum and polarization of escaping hard X-rays.  The observational 
effects of relativistic gravity and photon scattering off the outer, cold disk are quantified. 

The computation of the X-ray spectrum and $e^\pm$ distribution starts with a simplified 
background ion-electron flow in Kerr spacetime (Table \ref{tab:parameters}, Section \ref{sec:disk}).
We take this flow to be radiatively inefficient inside some radius $\rCorona$ outside the ISCO,
in the sense that thermal emission
processes (electron-ion bremsstrahlung or cyclo-synchrotron emission) release less than
a fraction $1/A \sim 0.1$ of the gravitational accretion energy.  The amplification factor
$A$ relating this seed thermal emission to the total radiated power (dominated by $50-100$ keV
photons) is empirically determined \citep{Burke2017};  we comment below on the implications for the soft photon
source.  Compton heating is agnostic to the source of stochastic $e^\pm$ motions (particle heating
versus bulk hydromagnetic motions).

There are several theoretical advantages to considering a $e^\pm$-dominated corona.
First, pair creation is non-local and can extend inside the ISCO and above the ion-dominated flow.
Second, the injection of pairs strengthens the coupling between the X-ray radiation field and the 
dynamic magnetic field, even where kinetic heating may be relatively weak.
Third, as our model calculations demonstrate, the scattering optical depth of the pairs
and the X-ray spectrum are self-regulating over a wide range of accretion luminosities and
for both moderate and high BH spins.

\subsection{Model Components}

\begin{table}
    \centering \small
    \begin{tabular}{ll|l}
         Parameter & Variable & Values  \\
         \hline \hline
         Black Hole Parameters && \\
         \hline \hline
         Normalized BH Spin& $a/M$ & $\{0.5, 0.9\}$\\
         Corresponding ISCO radius & $\rISCO/M$ & $\{4.23, 2.32\}$\\
         BH Mass& $M$ & $10 \, M_\odot$ \\
         Inner disk aspect ratio& $\theta_{d, {\rm in}}$ & $0.5$ \\
         Inner disk ion column depth & $\tau_{{\rm T, in}}$ & 1 \\
         Outer radius of hot corona & $\rCorona$ & $4\,\rISCO$ \\
         Outer disk aspect ratio & $\theta_{d, {\rm out}}$ & $0.1\,\theta_{d, {\rm in}}$ \\ 
         Outer disk column optical depth & $\tau_{{\rm T, out}}$ & $100\,\tau_{{\rm T},i}$\\
         Ion surface density radial index & $\alpha$ & -1 \\
         Seed blackbody temperature & $k_{\rm B}T_{\rm bb}$ & $0.5 \; {\rm keV}$ \\
         Seed luminosity & $L_{\rm seed}$ & $10^{-3} L_{\rm Edd}$ \\
         Non-thermal Amplification & $A$ & 10 \\ 
         \hline \hline 
         Interior Grid Parameters  && \\
         \hline \hline
         Outer grid radius & $\rmax$ & $32\,M$ \\
         \# radial grid cells & $N_r$ & 32 \\
         Grid polar angle cutoff & $\theta_{\rm jet}$ & 0.5 \\
         \# polar grid cells & $N_\theta$ & 64 \\
         \# azimuthal grid cells & $N_\phi$ & 32\\
         \# polar angular cells & $N_{\theta_k}$ & 8 \\
         Max \# azimuthal angular cells & $N_{\phi_k, {\rm max}}$ & 8 \\
         \# of energy bins & $N_\omega$ & 64 \\
         \hline \hline
         Exterior Grid Parameters && \\
         \hline \hline
         \# external polar cells & $N_{\thetaext}$ & 32 \\ 
         \# external azimuthal cells & $N_{\phiext}$ & 32 \\
         \# external energy bins & $N_{\omega_{\rm ext}}$ & 128 \\
         \hline \hline 
         Monte Carlo Run Parameters && \\
         \hline \hline
         Max time fraction & $\epsilon_t^{\rm max}$ & 1.0 \\
         Initial limit on density change & $\delta n_{\pm}^{{\rm max}, 0}$& 100\% \\ 
         Initial photon $\#$ per $\rmsBeta$ trial & $N_{\delta\beta}$ & $2 \times 10^5$ \\
         Initial photon $\#$ per density update & $N_{\rm density}$ & $10^7$
    \end{tabular}
    \caption{MC model parameters}
    \label{tab:parameters}
\end{table}

\begin{table}
    \centering \small
    \begin{tabular}{l|llll}
         Model Name & $a/M$ & $\delta \beta_{\disk}$ &${\rm min} (r_{\rmsBeta > 0})$ & Other \\
         \hline \hline
         $\simbase$ & 0.9 & 0.75 & $\rISCO$ & N/A \\
         $\simnodisk$ & 0.9 & 0.77 & $\rISCO$ & No outer disk \\
         $\simoutflow$ & 0.9 & 0.55 &$\rISCO$ & Outflow \\
         $\simlowspin$ & 0.5 & 0.79 &$\rmin$ & N/A\\
         $\simdisk$ & 0.5 & 0.81 &$\rmin$ & Disk seeding\\
         $\simlowflow$ & 0.5 & 0.58 &$\rmin$ & Outflow
    \end{tabular}
    \caption{List of MC models with given shorthand name, and varied parameters: BH spin $a/M$, converged r.m.s. velocity $\delta \beta_\disk$, minimum radius where $\rmsBeta > 0$, and other special conditions. Model $\simnodisk$ uses the same set-up as $\simbase$, but with no cold, outer disk. All models with spin $a/M = 0.5$ are assumed to have dissipation in the plunging zone. Model $\simdisk$ has emission tied to the seed
    electron-ion density, at a rate $\propto n_d^2$. Models labeled $\texttt{flow}$ include a poloidal outflow 
    in the velocity of each scattering $e^\pm$, following Equation (\ref{eq:outflow}).}
    \label{tab:sims}
\end{table}

\begin{enumerate}[wide, labelwidth=!, labelindent=0pt]

\item {\it Background spacetime.}  We consider BHs with both moderate and rapid rotation, corresponding
to Kerr spin parameter $a/M = 0.5$ and $0.9$ (models labeled ``{\tt a50}'' and ``{\tt a90}''; 
see Table \ref{tab:sims}).  The BH mass is $M = 10\,M_\odot$.
The spatial grid is described in Table \ref{tab:parameters} and Appendix \ref{app:grid}.

\item {\it Accretion rate and ion column.}   We consider a low default accretion luminosity, 
$L = 0.01\,L_{\rm edd}$, where the Eddington luminosity $L_{\rm edd} = 4\pi GM c/\kappa_{\rm T}$ is 
expressed in terms of the Thomson electron scattering opacity 
$\kappa_{\rm T}$.  At this accretion rate, a strong amplification and buffering 
of the $e^\pm$ density around the ISCO is evident in the model.
The ion column at the inner boundary of the coronal disk
is taken to be $\Sigma_{\rm in} \sim 1\,\kappa_{\rm T}^{-1}$;  this corresponds to an inflow speed about 
$10\%$ of the orbital speed and a baseline scattering depth $\tau_{\rm T,in} = 
\kappa_{\rm T}\Sigma_{i,\rm in} \sim 1$.  The contribution of the ion-electron flow to 
X-ray scattering is neglected in the plunging zone inside the ISCO, where the created $e^\pm$ 
dominate the population of scattering charges.

\item {\it Volumetric dissipation.}  Compton heating is applied in the simplest manner possible:
a dissipation zone is established over a certain range of radius, and then all $e^\pm$ in this zone
are given a velocity $\rmsBeta_\disk$ with random orientation and fixed magnitude in the frame of 
the background flow.  The effects of photoelectric absorption and 
line emission are ignored.  The models presented here include dissipation close to the BH:
either due to centrifgual support in the presence of high BH spin, or due to the presence
of a dynamic magnetic field.
In the $a/M = 0.5$ models (e.g. $\simlowspin$), the heating extends over both the plunging zone ($r < \rISCO$) and
the coronal disk ($\rISCO < r < \rCorona$).
In the $a/M = 0.9$ models (e.g. $\simbase$), dissipation is restricted 
to the coronal disk.  
For any given distribution of $e^\pm$, we iterate on $|\rmsBeta_\disk|$ to achieve the desired
amplification $A = 10$.
It is sometimes necessary to limit the Compton $y$-parameter in the most optically
thick zones close to the BH horizon, so as to avoid runaway pair creation;
but here the photons are almost all advected inward (Section \ref{sec:rms_beta}).

\item {\it Mean orbital flow.}  We impose a self-consistent relativistic orbital velocity field (Section
\ref{sec:disk}).   This flow is circular outside the ISCO and falls ballistically toward the BH 
at smaller radius. The flow profile away from the BH equator is obtained by 
restricting the flow to surfaces of constant $\theta$, implying the presence of an appropriate 
latitudinal matter force.  The coronal disk is geometrically thick
(angular scale height $\theta_{d,i} \sim 0.5$ rad).  The outer, cold disk, when present, is a factor $0.1$ 
thinner than the coronal disk (scale height $\theta_{d,o} = \theta_{d,i}/10$).
To test the influence of outflow on X-ray polarization, in two models we add in an upflow velocity
when computing photon-electron scattering (Section \ref{sec:outflow}).

\item  {\it Photon seeding.}
Seed thermal X-ray photons are injected into the
corona with a 0.5 keV rest-frame temperature and
a spatial distribution proportional to the square of the disk-frame density (Section \ref{sec:photon}).  
This is either taken to be $n_d^2$ (here $n_d$ is the seed density of neutralizing electrons)
or more commonly $n_\pm^2$ (where $n_\pm = n_{e^+} + n_{e^-} = n_d + 2n_{e^+}$ is the total
density of positrons and electrons).  These two prescriptions are associated qualitatively with 
photon seeding by electron-ion bremsstrahlung, or alternatively by cyclo-synchrotron emission.
(The latter scaling implies a positive correlation between $n_\pm$ and both a sub-population of relativistic pairs and the magnetic field energy density; Section \ref{sec:seeding}.)

\item {\it Photon propagation and electron scattering.}
The photons are propagated along geodesics and allowed either to escape to infinity or be
absorbed by the BH (Section \ref{sec:photon}).
Electron scattering is treated fully relativistically with the KN cross section
(Section \ref{sec:scatter}).   In this way, a self-consistent photon spectrum is
established as a function of observer angle.  The photon distribution function is stored
in each spatial cell, as a function of photon energy and direction (Table \ref{tab:parameters}
and Appendix \ref{app:grid}).

\item  {\it Electron-positron pair creation.}   Once the photon spectrum is established,
for a given spatial distribution of charges and fixed amplification $A$, the $e^\pm$
density field is recomputed.  The calculation of $n_\pm$ takes into account photon collisions, 
$\gamma + \gamma \rightarrow e^+ + e^-$, annihilation of pairs back into gamma rays, and 
(inside the ISCO) the inward advection of pairs (Section \ref{sec:iteration}).  
These steps are repeated until the $e^\pm$ density converges to less than $10\%$ maximum variance.
Axisymmetry is imposed on the photon and $e^\pm$ density fields.

\item {\it Hard X-ray polarization.}  X-ray polarization is evolved self-consistently
under KN scattering (Section \ref{sec:polarization} and Appendix \ref{app:scattering}).  
The soft seed X-ray emission is assumed to be unpolarized. 
Stokes parameters are stored at null infinity as a function of photon energy and observer
orientation.
\end{enumerate}

\begin{figure}
    \centering
    \includegraphics[width=\linewidth]{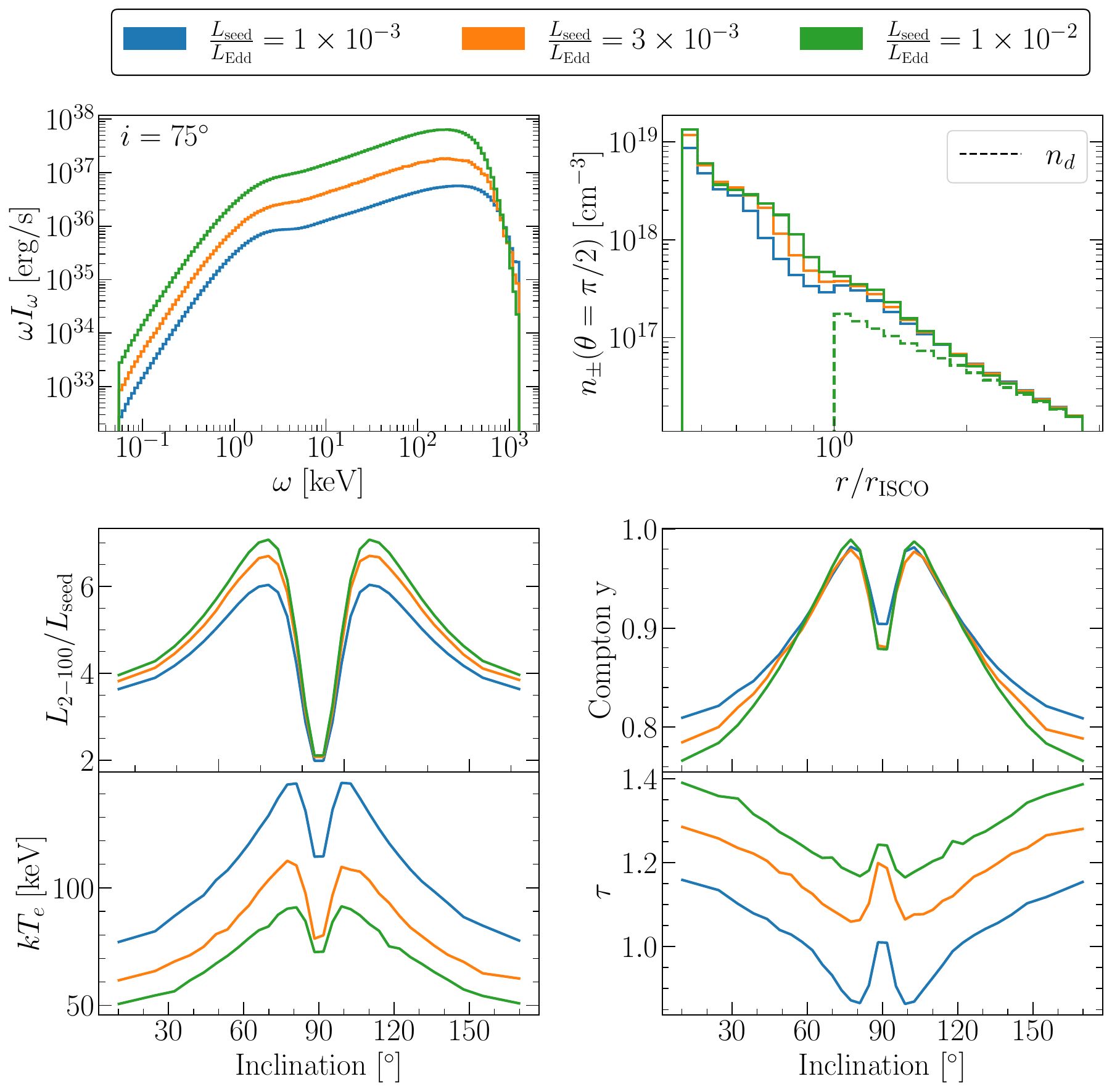}
    \caption{Changes in spectra, mid-plane density $n_\pm$, and \texttt{Xspec} fits as a function of seed luminosity. Orange and green curves show increasing seed luminosity compared to default model $\simlowspin$ (blue).
    The equilibrium density shows moderate increase in response to increasing seed luminosity. Top left: spectra for observers at observer-disk inclination $i = 75^\circ$. Top right: radial $n_\pm$ profile at the mid-plane. Spectral data and \texttt{Xspec} fits are also shown: the $2-100$ keV luminosity, normalized by seed luminosity (middle left), fitted Compton $y$ (middle right), fitted coronal temperature (bottom left), and fitted optical depth (bottom right, Equation \ref{eq:optical_fit}).}
    \label{fig:accretion_rates}
\end{figure}

\subsection{Coronal Variation with Luminosity}\label{sec:coronalum}

The MC model demonstrates that a pair-dominated corona can self-consistently reproduce the observed hard state spectra of BHXRBs over a range of luminosities.  The density of pairs is tied to the exponential Boltzmann tail of
the photon energy distribution through the reaction $\gamma + \gamma \leftrightarrow e^+ + e^-$,
and also to the density of seed photons through the Compton parameter $y$.  In a static corona, $y$ is
defined in terms of the mean number of scatterings $N_{\rm scatt}$ and the mean square $e^\pm$ momentum
by $y \equiv {4\over 3} \langle N_{\rm scatt}\rangle \langle\gamma_e^2\beta_e^2\rangle$.
The $1-100$ keV spectral slope is in the range observed for $y = O(1)$;  this is also the value where
energy exchange between $e^\pm$ and photons is balanced \citep{Haardt1993,Thompson1994}.  

Because the photon collision rate is sensitive to the high-energy spectral cutoff, the scattering depth and the 
cutoff energy are expected to vary only slowly with luminosity.  Figure \ref{fig:accretion_rates} demonstrates
this effect in model $\simlowspin$ as we vary the luminosity $L = A\,L_{\rm seed}$ over the range 
$(0.01-0.1)\,L_{\rm edd}$, corresponding to a seed keV luminosity $L_{\rm seed} = (0.001-0.01)L_{\rm edd}$. 
At most distances from the BH,
the $e^\pm$ density varies by a factor 2 or less, and the X-ray flux maintains an almost constant
profile with respect to observer orientation.  

Figure \ref{fig:accretion_rates} also shows the fit to the electron temperature $T_e$ and coronal
scattering depth $\tau$ that are obtained from the output spectrum using
the {\tt Xspec} package {\tt compPS} and assuming a slab corona.
The temperature decreases moderately toward higher $L$ in order to compensate the growing photon density;
$\tau$ correspondingly increases, even while $y$ remains almost constant.

\begin{figure}[ht]
    \centering
    \includegraphics[width=\linewidth]{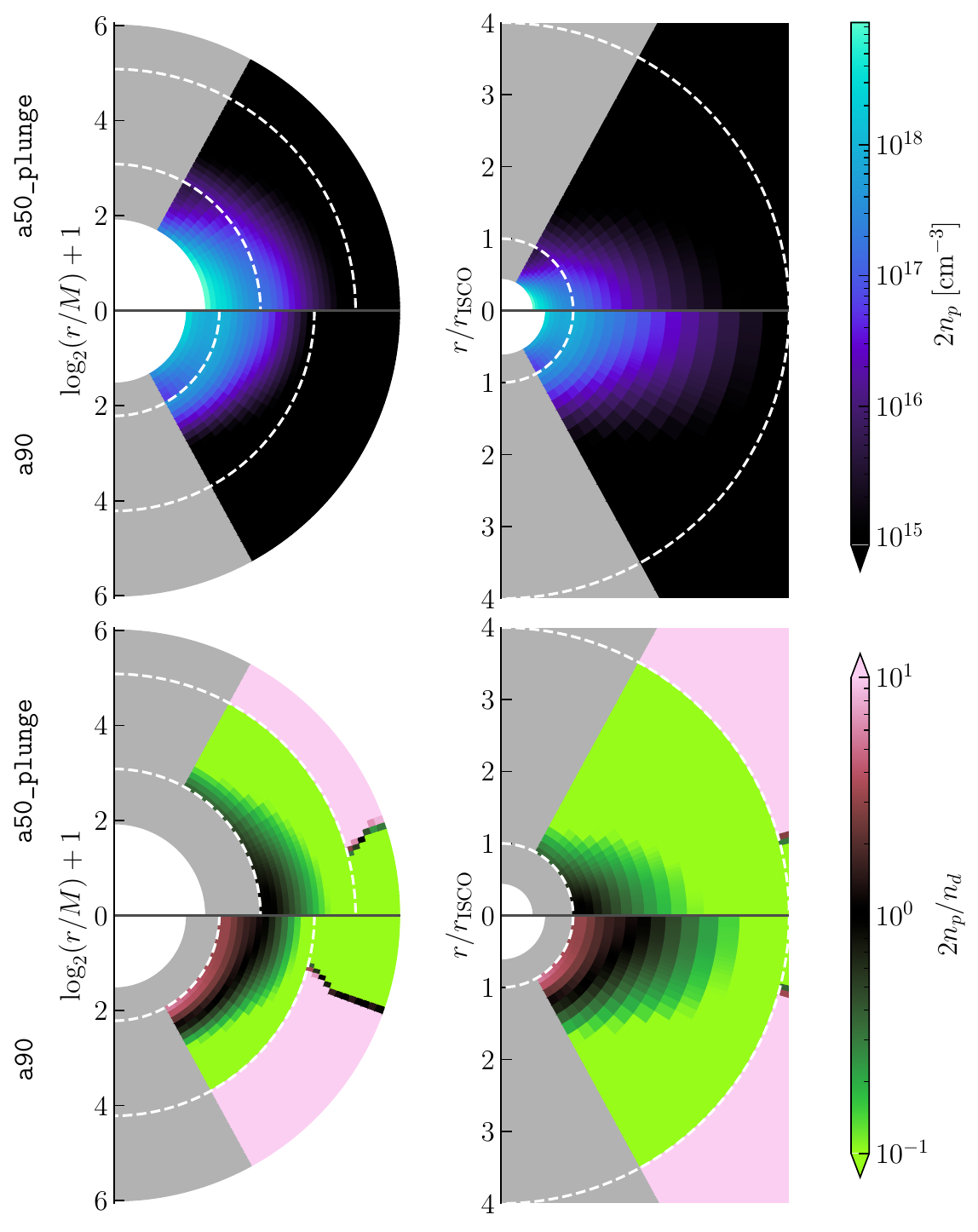}
    \caption{ Top panels: poloidal profile of coronal pair density.  Bottom panels: ratio of
    converged total pair density $n_p$ to input electron-ion density $n_d$.   The upper half of each
    panel shows results for model $\simlowspin$ and the lower half for model $\simbase$.
    Left panels show the full MC grid with radial coordinate $\log_2(r/M) + 1$;  right
    panels show an expanded view between the ISCO (inner dashed curve) and outer boundary of
    the coronal disk (radius $\rCorona$; outer dashed curve).
    The input ion density is assumed to be negligible compared with pairs inside the ISCO.}
    \label{fig:pair_dist}
\end{figure}
\begin{figure}
    \centering
    \includegraphics[width=\linewidth]{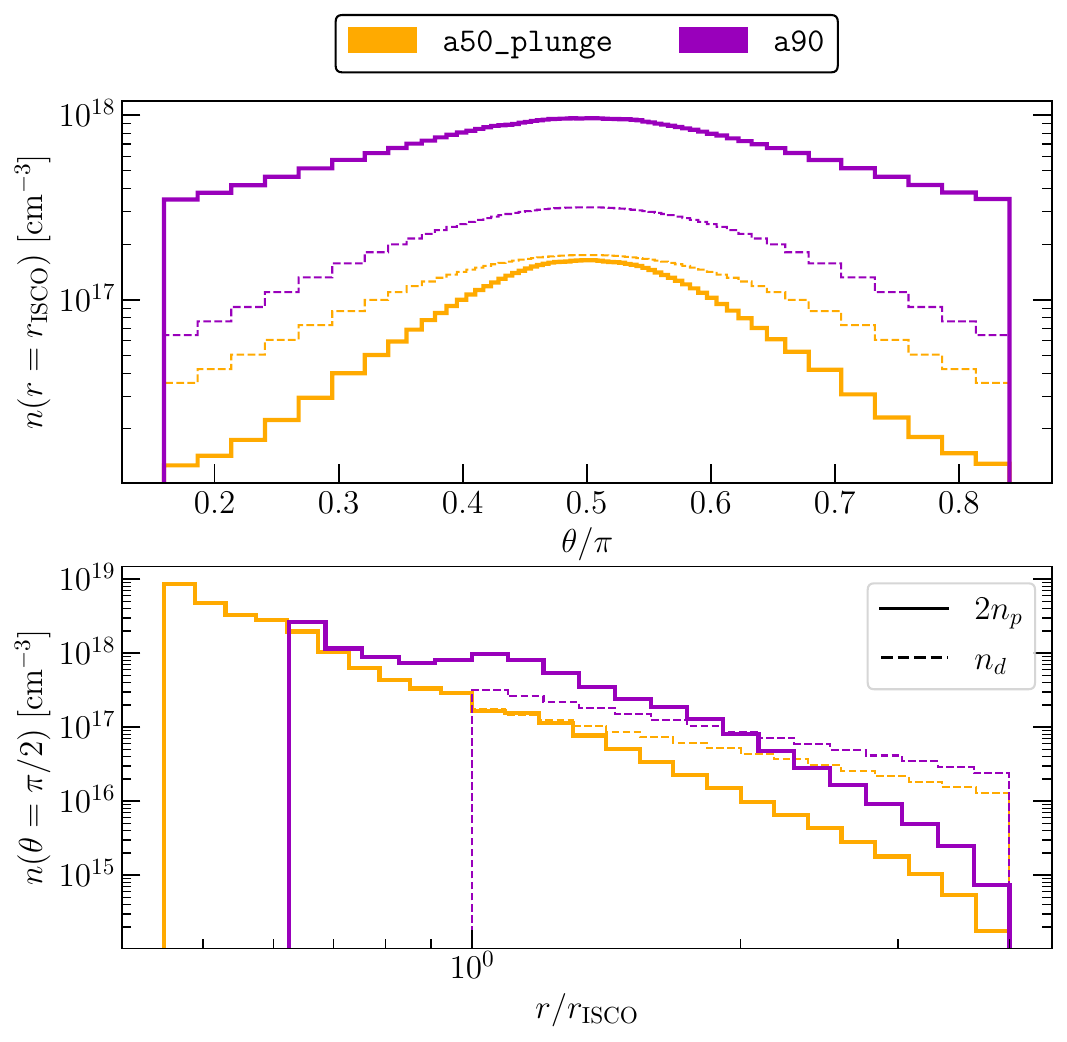}
    \caption{Top panel:  latitudinal profile of the total pair density $2n_p$ and the input electron-ion density $n_d$ as measured at the ISCO.   Bottom panel:  radial profile of the two density fields
    as measured at the mid-plane.  Purple (orange) curves show models with high (moderate) BH spin.}
    \label{fig:dens_spin}
\end{figure}

\subsection{Pair Spatial Distribution}

The models $\simlowspin$ and $\simbase$ have different BH spins and heating profiles.
In the moderate-spin model $\simlowspin$, heating extends inside the ISCO; the pair cloud is therefore
concentrated closer to the BH and is found to be more inwardly peaked (Figure \ref{fig:pair_dist}).
Outside the ISCO, the produced pairs in model $\simbase$ dominate over the input disk density out to 
radius $\sim 2\,\rISCO$.  
Figure \ref{fig:dens_spin} compares the radial density distributions of pairs and ions at the BH equator,
and the angular distributions at the ISCO.  In all models, the contribution of the electron-ion
flow to coronal scattering is neglected in the plunging zone. 

\begin{figure}[t]
    \centering    \includegraphics[width=\linewidth]{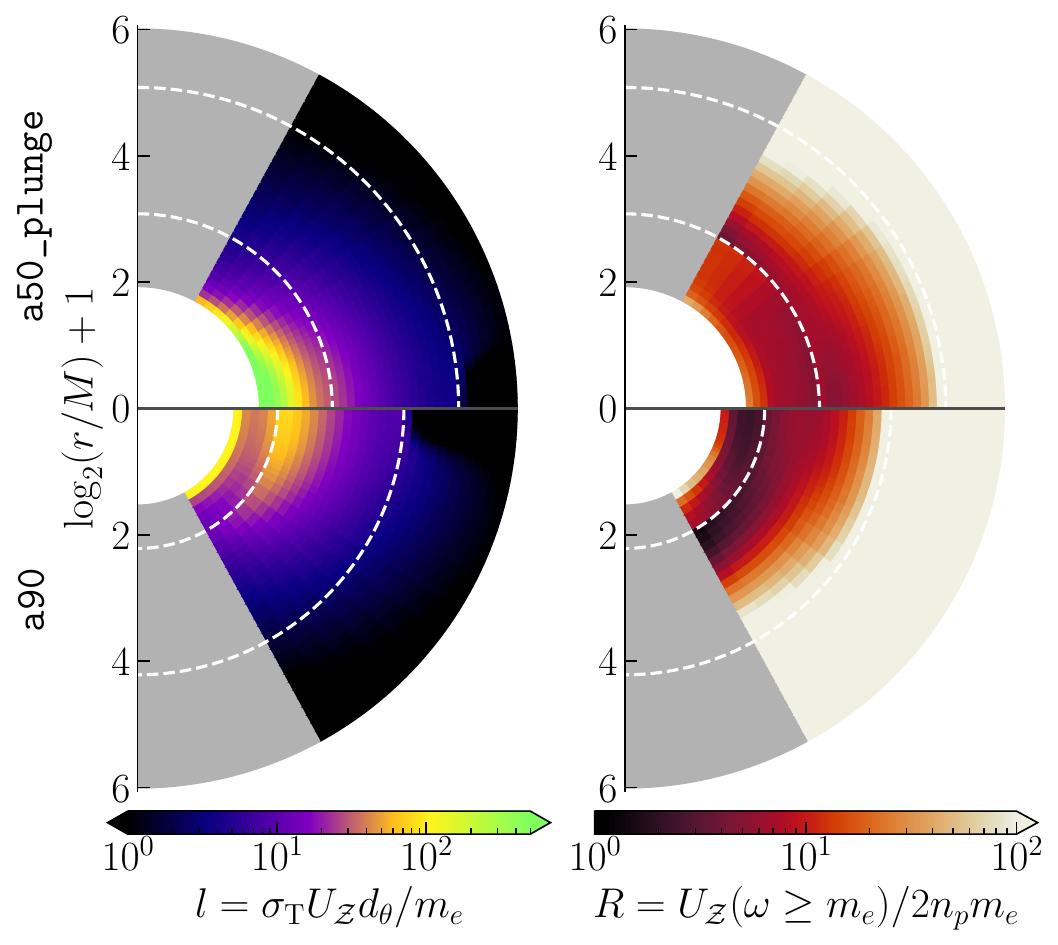}
    \caption{Radiation compactness $l$ and ratio $R$ of photon energy density to pair rest-energy density, for 
    models $\simlowspin$ (top half of panels) and $\simbase$ (bottom half).  Dashed white curves mark the
    ISCO and outer boundary of the coronal disk.  Left column:  bolometric compactness 
    (Equation (\ref{eq:lum_compactness})). Right column: $R$ evaluated for hard photon energy density.
    Model $\simlowspin$ includes heating and photon emission inside the ISCO;  model $\simbase$ does not.
    The compactness and photon/pair energy ratio both show strong inhomogeneity and anisotropy.} 
    \label{fig:compactness}
\end{figure}

Efficient pair creation is connected with a local compactness $l$ in high-energy photons 
exceeding unity.   Figure \ref{fig:compactness} shows the quantity
\begin{equation}
    \label{eq:lum_compactness}
    l = \frac{\sigma_{\rm T} U_{\ZAMO} d_\theta}{m_e}; \quad d_\theta = \sqrt{\gtheta} 
    \times (1 \; {\rm rad}).
\end{equation}
Here $U_\ZAMO$ is the total photon energy density in the frame of a zero angular momentum observer
(ZAMO; Section \ref{sec:disk}), $\sigma_{\rm T}$ the Thomson cross section, and $m_e$ 
the electron mass.  
In model $\simlowspin$, $l$ traces the high plasma density in the plunging zone. 
Model $\simbase$ does not allow for $e^\pm$ heating in the plunging zone,
and so the compactness is relatively smaller inside the ISCO. 
We also plot the ratio of energy density in hard photons ($\omega \geq m_e$) to that in pair rest energy,
$R = U_\ZAMO/2n_p m_e$.   This ratio is buffered around the ISCO as photons equilibrate with pairs; 
farther from the BH, escaping hard photons dominate over the relatively sparse pairs. 

\begin{figure}[ht]
    \centering    \includegraphics[width=\linewidth]{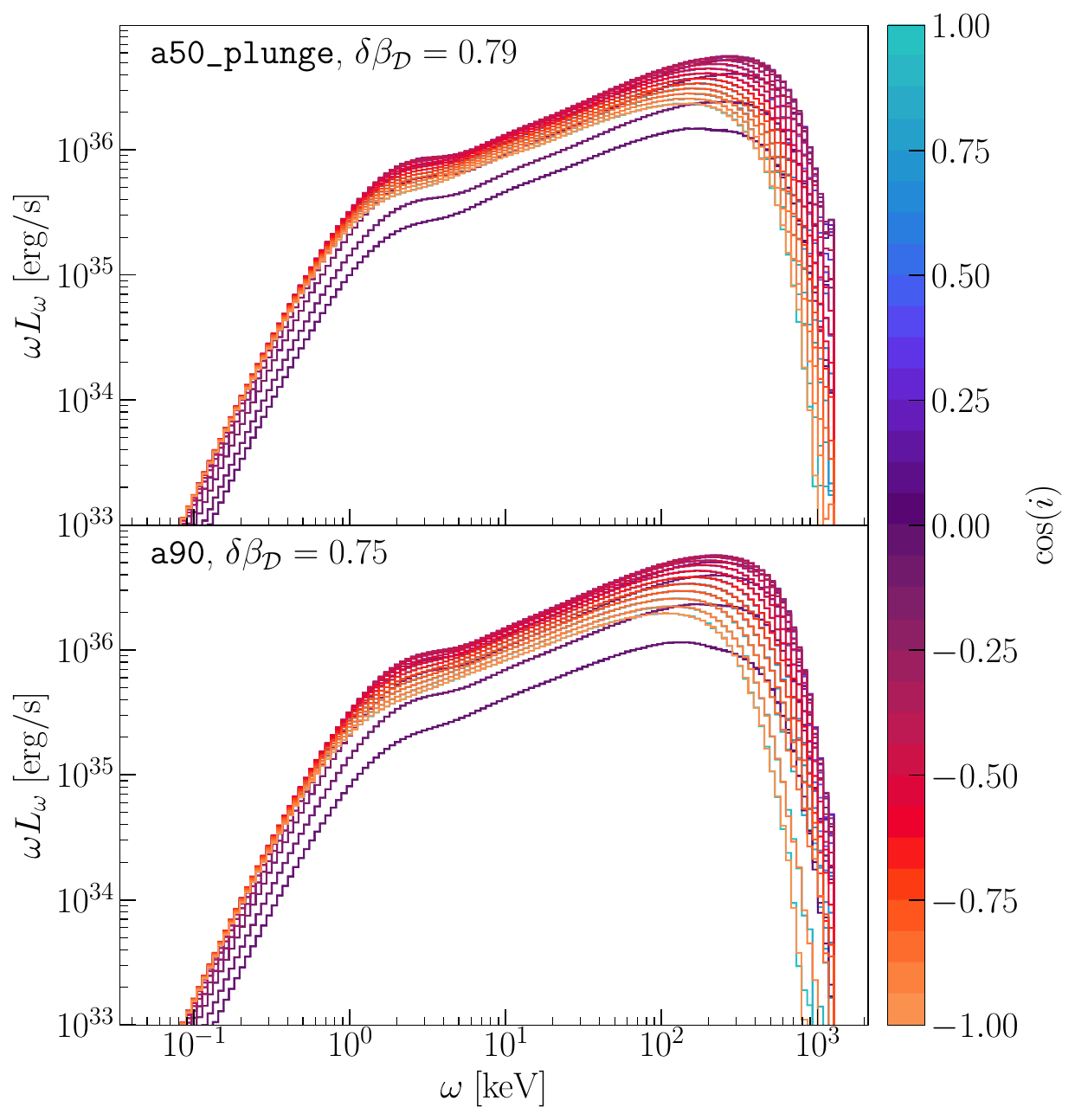}
    \caption{Output spectra for models $\simlowspin$ (top) and $\simbase$ (bottom) in observer-disk inclination bins uniformly spread in $\cos(i)$.  Polar observers are shown in blue (north) and orange (south), while equatorial observers are shown in purple. The photon flux is suppressed around the equator due to occlusion by the cold, outer disk.}
    \label{fig:spec_data}
\end{figure}
\begin{figure*}[t]
    \centering
    \includegraphics[width=\linewidth]{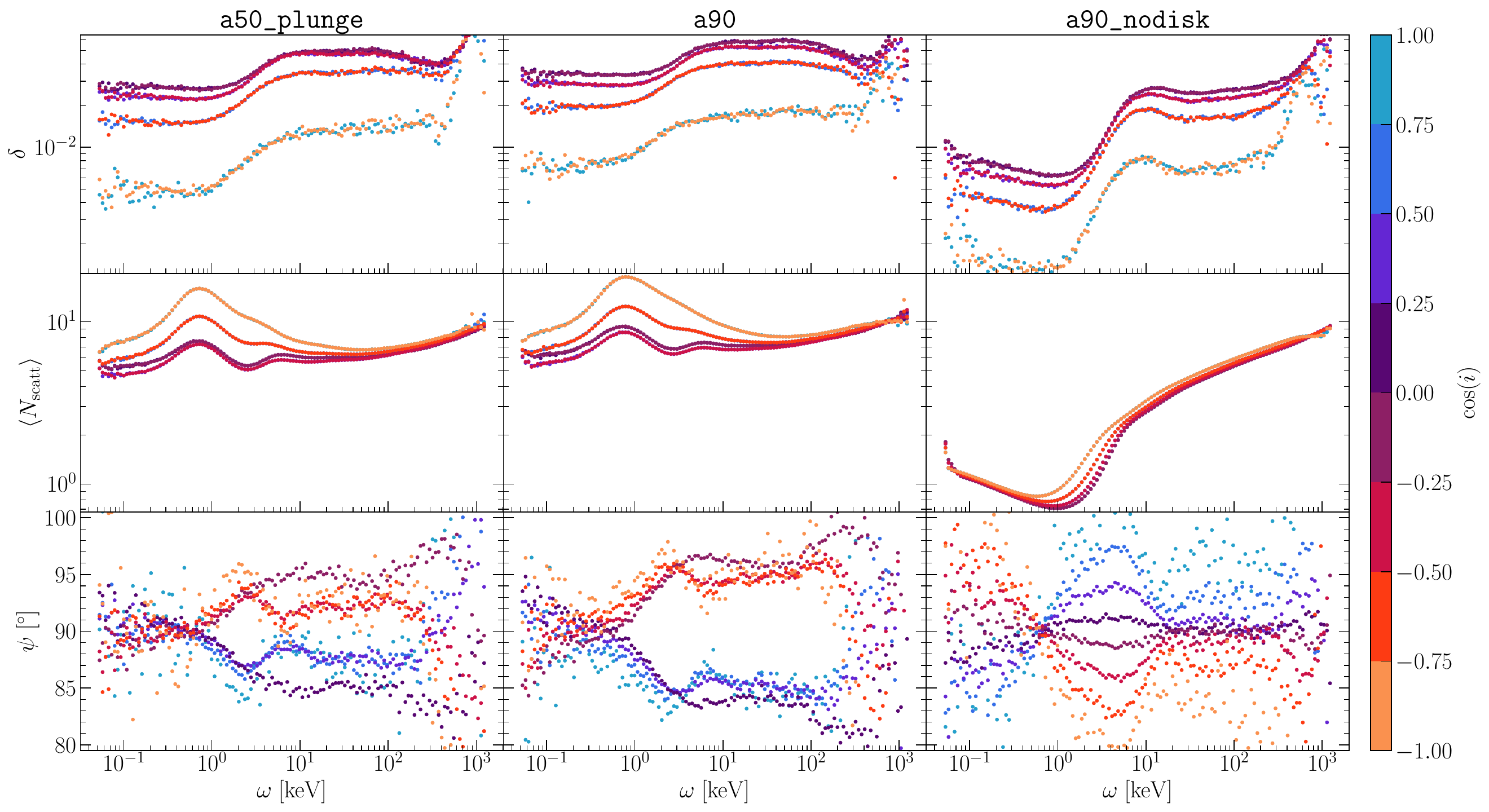}
    \caption{Polarization degree (top row), average number of scatterings (middle row), and polarization angle (bottom row) as a function of detected photon energy, for eight different observer-disk inclinations. Data is presented for three models: $\simlowspin$ (left), $\simbase$ (middle) and 
    $\simnodisk$ (right). Polarization angles are shown with respect to basis $\{\hat{\phi}, -\hat{\theta}\}.$}
    \label{fig:pol_comp}
\end{figure*}
\subsection{Observer-Dependent Spectra and Polarization}

The spectrum of escaping X-rays is plotted as a function of observer-disk 
inclination\footnote{Inclination angle $i = 0$ corresponds to an observer positioned along the spin axis
of the BH or -- from the perspective of the observer -- to an alignment between the BH equatorial plane 
and the plane of the sky.  For equatorial inclination $i = \pi/2$, these two planes are orthogonal.} angle $i$ 
in Figure \ref{fig:spec_data} for models $\simbase$ and $\simlowspin$.  The converged value of
$|\rmsBeta_\disk|$ is similar in both models, resulting in similar spectra.
All spectra show a rising power law (in spectral luminosity $\omega L_\omega$) and a sharp cutoff above
a $\sim100$ keV peak, both features being representative of the hard spectral state.  The spectral peak is softer 
and the received luminosity lower for observers at equatorial inclinations, due to partial occultation by
the cold, outer disk.  The same effect is observed at polar inclinations,
for different reasons.  First, photons tend to be somewhat beamed along
the azimuthal direction of the orbiting plasma; and second, photons escaping toward the poles
are typically scattered backward in the disk frame and experience a net Doppler redshift.
\begin{figure}
    \centering
    \includegraphics[width=\linewidth]{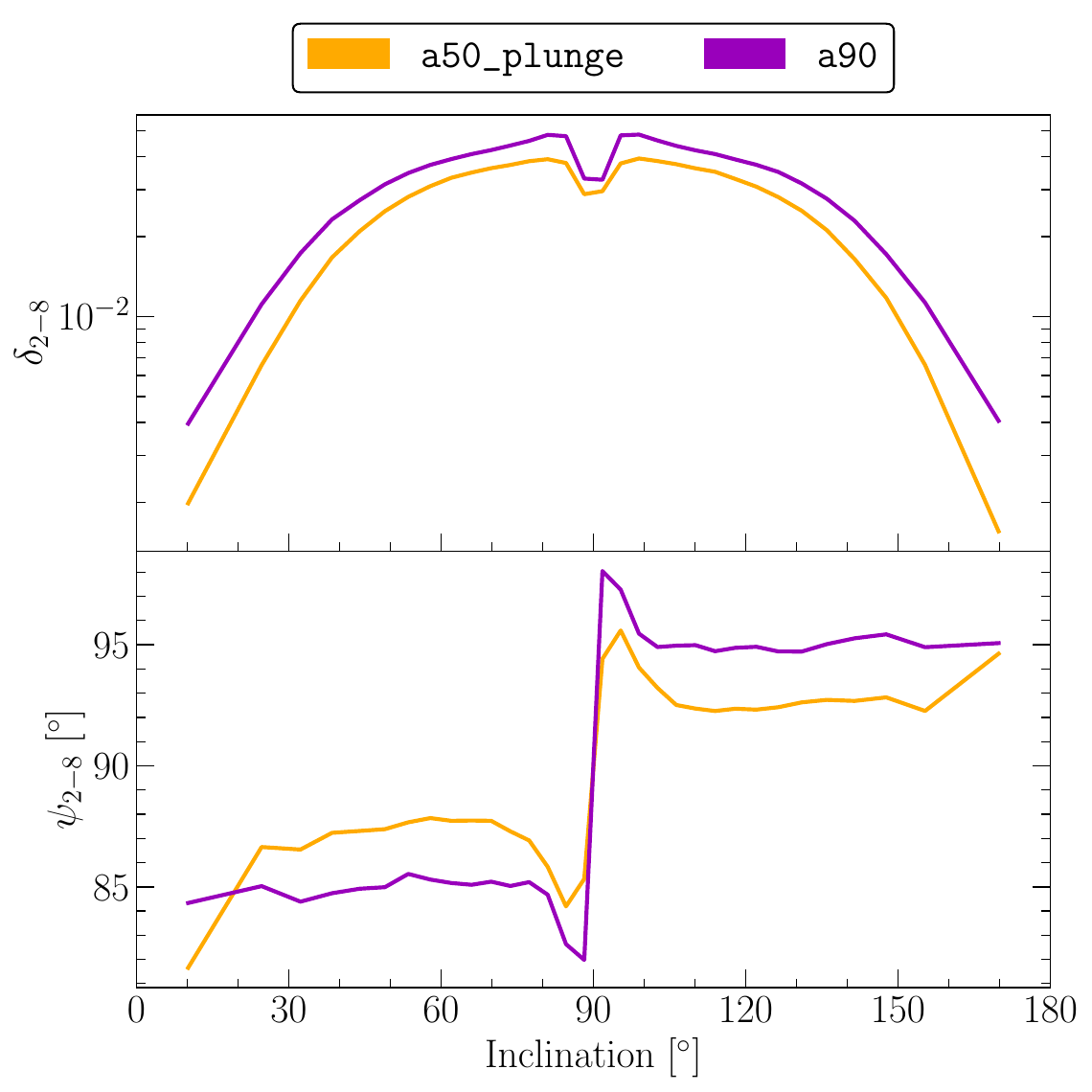}
    \caption{Band-averaged $2-8$ keV polarization degree (top) and angle (bottom) for models $\simlowspin$ and $\simbase$.}
    \label{fig:pol_2_8}
\end{figure}

The seed keV X-rays are taken to be unpolarized at emission.  Then, as Compton scattering does not generate
circular polarization, the polarization of the escaping X-rays is defined by the linear polarization 
degree $\delta$ and angle $\psi$ (or, equivalently, the Stokes parameters $Q$ and $U$; Section \ref{sec:polarization}).  The linear polarization direction at null infinity is recorded as
$\pm\cos\psi \; \hat{\phi} \mp \sin\psi \; \hat{\theta}$. Figure \ref{fig:pol_comp} shows 
the polarization data, versus photon energy, for models $\simbase$ and $\simlowspin$ and 
observer-disk inclination varying from $0^\circ$ to $180^\circ$. 

The polarization degree increases in the $2-8$ keV band, as unpolarized seed photons gain both polarization and energy via scattering. A cleaner measure of this effect is obtained by removing the outer, cold disk (model $\simnodisk$; right panels of Figure \ref{fig:pol_comp}). Here
soft X-rays are no longer trapped and scatter a large number of times in the outer disk, and one sees 
a sharper increase in polarization with photon energy.
As with simpler flat-space coronal slab models (\citealt{Poutanen1996,Poutanen2023} and references therein),
the polarization degree is greater at equatorial inclinations ($\cos i \rightarrow 0$), because scattering from
a median depth in the corona is more concentrated toward $90^\circ$ scattering angle. The polarization degree saturates above 10 keV, 
maximizing at $5\%$ for observers near the equator.

The polarization angle is generally aligned with the spin axis ($90^\circ$), with the exception of 
equatorial observers (Figure \ref{fig:pol_2_8}).  The higher-spin model $\simbase$ has a slightly higher polarization degree and a misalignment of the polarization 
with the spin axis.
As discussed further in Section \ref{sec:disk_effects}, this misalignment is due to parallel transport from the dominant scattering region to the observer; the cold, outer disk occludes mid-plane emission, and so photons are sourced mainly off the mid-plane.  
The cold, outer disk is also a source of highly polarized reflected photons, which are absent in model $\simnodisk$, which therefore shows weaker polarization.

\subsection{Polarization Imprint of Outflow}
\label{sec:outflow}

Hard-state BHXRBs have been observed with relatively strong linear polarization, in excess of $\delta = 9\%$ \citep{Ewing2025}. An outflowing corona could generate a similar degree of polarization even in the
absence of a secondary and more highly polarized emission component \citep{Poutanen2023}. 

We incorporate a bulk disk outflow into models $\simoutflow$ and $\simlowflow$ using the following procedure.
This starts with the converged density profile from models $\simbase$ and $\simlowspin$, 
respectively.  At each photon emission or scattering point, a bulk latitudinal velocity is
added with the following angular profile,
\begin{align}
    \label{eq:outflow}
    {\bm\beta}_{O, \disk} = \pm\delta\beta_\disk \, {\rm max}\left(1, \frac{|\pi/2 - \theta|}{\theta_{\rm d, in}} \right) \hat \theta.
\end{align}
This velocity increases linearly away from the BH equator up to an angle $\Delta\theta = \pm\theta_{\rm d,in}$,
above which it saturates.
The random velocity of the $e^\pm$ in each scattering event is modified by calculating $\rmsBeta_\disk$ in
the usual way and combining it with ${\bm\beta}_{O,\disk}$ using the standard relativistic velocity summation.
The saturation value of $\beta_{O,\disk}$ is taken to equal
$|\rmsBeta_\disk|$. 

\begin{figure}
    \centering
    \includegraphics[width=\linewidth]{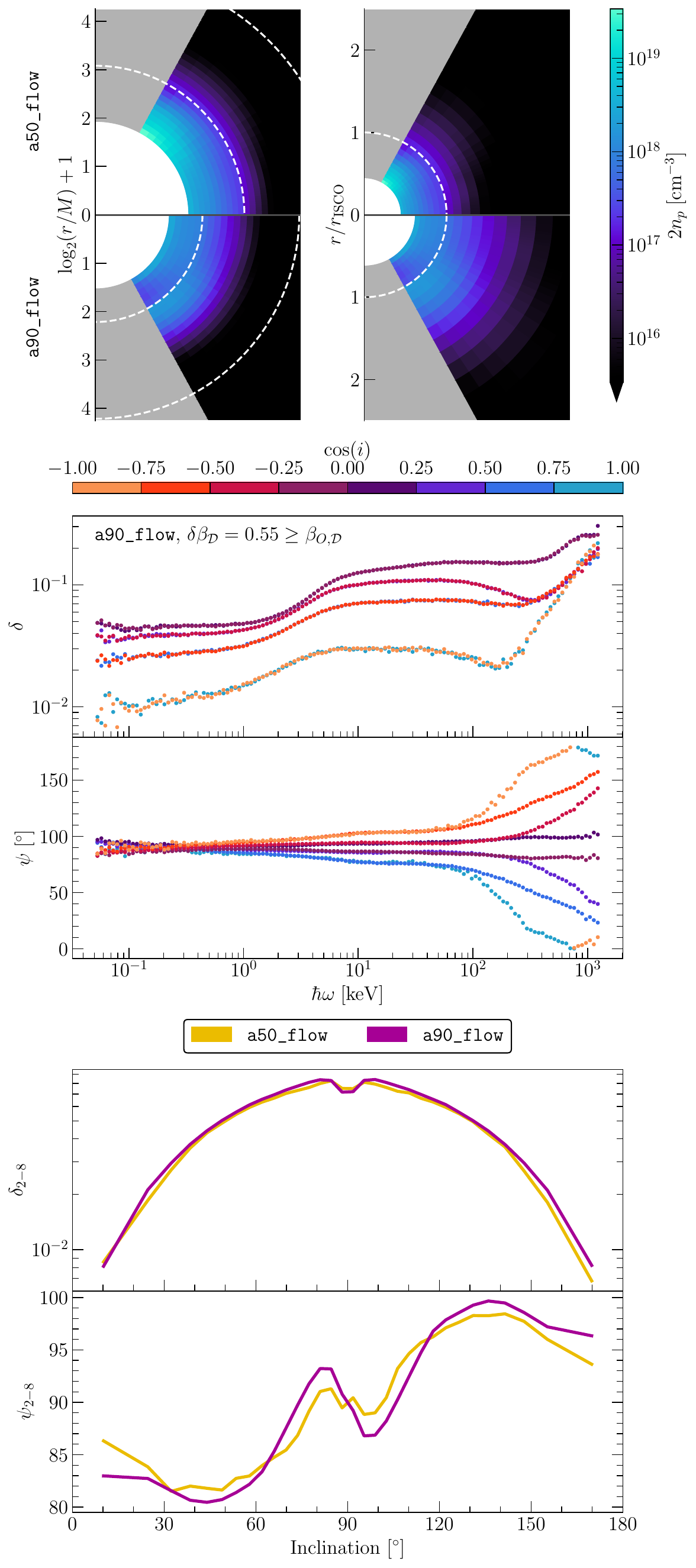}
    \caption{Effect of plasma outflow on $e^\pm$ density distribution (top panel), 
    frequency-dependent polarization (middle panel) and band-averaged $2-8$ keV polarization
    (bottom panels).      Input
    electron-ion distribution, soft-photon seeding prescription, and heating profile are the same
    as in models $\simbase$ and $\simlowspin$, but with an outflow velocity (\ref{eq:outflow}) added
    to the randomly selected flow-frame velocity $\rmsBeta_\disk$ of each scattering charge.}
    \label{fig:outflow}
\end{figure}

The addition of the outflow velocity therefore reduces the amplitude of the stochastic $e^\pm$ motion
that will upscatter enough photons to pair-producing energies.  The plasma density profile is altered 
by the outflow, as photons are preferentially scattered away from the mid-plane (Figure \ref{fig:outflow}).  
Off-center density lobes form in both hemispheres.  We enforce a reflection symmetry in the photon distribution
across the mid-plane;  otherwise a strong asymmetry in the $e^\pm$ distribution appears to develop spontaneously.

Figure \ref{fig:outflow} also shows the polarization data for both outflow models.
Similarly to the case without outflow, the $2-8$ keV polarization signal is insensitive to the BH spin. 
However, the polarization degree is significantly boosted even by a modest outflow velocity
$(\beta_{O, \disk} \leq 0.5-0.6)$; the spike in polarization angle observed near the equator is
also significantly reduced.
As the $e^\pm$ outflow biases the photon flux away from the mid-plane, the cold disk is less effective at occluding the hard, scattered X-rays from equatorial observers. 
A small deviation in polarization angle (with respect to spin alignment) also appears near the poles.

\section{Bulk Flow and Stochastic Motions}\label{sec:disk}

Seed electrons that initiate soft photon emission and Compton scattering are carried inward to
the BH by an ion-electron flow.  The equilibrium pair distribution near the BH turns out to be
weakly sensitive to details of this flow, so we adopt a simple analytic profile for
the seed electrons.   The accretion flow rotates in the same direction as the BH 
spacetime.  The Kerr metric describing a hole of mass $M$ and spin angular momentum $aM$ is
\begin{align}
\label{eq:metric}
ds^2 &= \gt dt^2 + \gphi[d(\phi-\omega t)]^2 + \gr dr^2 + \gtheta d\theta^2 \nn
&= -\frac{\Sigma \Delta}{A} dt^2 + \frac{A \sin^2 \theta}{\Sigma} [d(\phi - \omega t)]^2 + \frac{\Sigma}{\Delta} dr^2 + \Sigma d\theta^2
\end{align}
in Boyer-Lindquist (BL) coordinates.  Here,
\be\label{eq:omZ}
    \omega(r,\theta) = -\frac{g_{t \phi}}{\gphi} = \frac{2Mar}{A}
\ee
is the angular frequency of a ZAMO, and
\begin{align} 
\gt &= g_{tt} - \omega^2 \gphi; \nn 
\Sigma(r, \theta) &= r^2 + a^2 \cos^2 \theta; \nn 
\Delta(r) &= r^2 - 2Mr + a^2; \nn
A(r, \theta) &= (r^2 + a^2)^2 - a^2 \Delta \sin^2 \theta.
\end{align}

Centrifugal support near the BH equator is then stable outside the standard ISCO
for material test particles \citep{Bardeen1972}, 
\ba\label{eq:risco}
    {\rISCO\over M} &=& {3 + Z_2 - [(3-Z_1)(3+Z_1 + 2Z_2)]^{1/2}}; \nn 
    Z_1 &\equiv& 1 + (1-a^2)^{1/3}[(1+a)^{1/3} + (1-a)^{1/3}]; \nn
    Z_2 &\equiv& (3a^2 + Z_1^2)^{1/2}.
\ea
The ion density profile at $r > \rISCO$ is chosen to approximate an isothermal 
distribution in hydrostatic balance in the $\theta$-direction,
\begin{equation}\label{eq:rho}
    n_d(\theta,r) = n_{d,0}(r) \exp \left(- \frac{\cos^2\theta}{2\sin^2 \theta_d} \right).
\end{equation}
This angular profile is independent of radius within each segment of the disk,
corresponding to a disk aspect ratio $h/r \sim \theta_d$.
The equatorial density $n_{d, 0}$ can be related to the non-radial electron column, 
$N_e(r) = \tau_{\rm T}/\sigma_{\rm T}$ and corresponding Thomson scattering depth 
$\tau_{\rm T}(r)$.  Integrating Equation (\ref{eq:rho}) near the equator gives
\be
    N_e(r) = \int_0^\pi n_{d, 0}(r,\theta) \sqrt{g_{\theta\theta}} d\theta
    \simeq \sqrt{2\pi} \theta_d r n_{d, 0}(r).
\ee

In order to take into account the possible presence of a jet emerging from the
ergo-region of the BH, the corona plasma and ion disk are excluded from the polar zones
$\theta < \theta_{\rm jet}$ and $\theta > \pi - \theta_{\rm jet}$.   We make
no attempt to represent the plasma state in the jet zones (e.g. \citealt{Parfrey2019,Yuan2025} and
references therein) and their influence
on the X-ray output of the BH is ignored.  The effect of a transrelativistic upflow within
the coronal disk on the X-ray spectrum is examined in Section \ref{sec:outflow}.

The seed ion-electron column has a simple radial profile,
\be\label{eq:disk_density}
    N_e (r) = \frac{\tau_{\rm T}} {\sigma_{\rm T}} \left(\frac{r}{\rISCO}\right)^\alpha.
\ee
The index $\alpha = -1$ is taken to be the same in the inner coronal disk 
($\rISCO < r < \rCorona$) and outer, cold disk ($r > \rCorona$), with the outer disk 
a factor $10$ thinner.  The ion column is taken to be $(\theta_{d,{\rm in}}/\theta_{d,{\rm out}})^2 
\sim 100$ times larger in the outer disk, as is expected from the simplest model of angular 
momentum transport with a uniform ratio of viscous stress to pressure. 

In the configuration considered here, there is a single disk structure in the zone $\rISCO < r < \rCorona$,
without a division into thin disk and corona.  If the electron-ion plasma in this zone were to be the main scattering medium
over at least a factor $\sim 100$ in luminosity, while preserving a Compton parameter $y = O(1)$
and an effective electron temperature $50-200$ keV, then fine tuning of the disk column would be implied.  
Pair plasma can alternatively be the main source of scattering charges, as we show, especially
inside the ISCO (see Figures \ref{fig:accretion_rates} and \ref{fig:dens_spin}).
The seed electron-ion density remains subdominant at the ISCO if the accretion rate 
is not too high ($0.01-0.1$ Eddington) and angular momentum transport is very efficient,
corresponding to $\kappa_{\rm T}\Sigma \lesssim 1-2$.  In an $\alpha$-disk model with angular velocity
$\Omega$ approximated as Keplerian, the accretion rate is 
\ba
    \dot{M} &\sim& 3\pi \alpha h^2 \Omega \Sigma \nn
    &=& \frac{1}{\epsilon_{\rm rad} c^2} \left( \frac{L_{\infty}}{L_{\rm Edd}} \right) \frac{4 \pi G M c}
    {\kappa_{\rm T}},
\ea
with corresponding seed scattering depth,
\ba
    \tau_{\rm T}(\rISCO) &=& \kappa_{\rm T}\Sigma(\rISCO) \nn
    &\sim& 0.5\, \frac{L_{\infty}/L_{\rm Edd}}{0.01} \frac{0.1}{\alpha} \frac{0.1}{\epsilon_{\rm rad}} \left({h\over r}\right)^{-2}
    \left({\rISCO\over 6\,r_g}\right)^{-1/2}.\nn
\ea 
Limiting the disk column to this value at higher accretion rates requires efficient angular momentum transport within the coronal disk, which may be possible if it is strongly magnetized (e.g. \citealt{Squire2025} and references therein).  

The radial flow outside the ISCO is assumed to be significantly slower than the ballistic
inflow at $r < \rISCO$.  By contrast, the density of created pairs that is sourced non-locally
by photon collisions is found to peak inside the ISCO.
We therefore neglect the contribution of the ion-electron flow to the scattering particle density inside the ISCO;  photon scattering develops self-consistently inside the ISCO via pair creation.

To summarize, the background ion-electron mass profile in the inner coronal disk at $\rISCO < r < \rCorona$ is fixed 
by three parameters: 
the column depth $\tau_{\rm T,ISCO}$ at the ISCO, the angular width $\theta_d$, and the surface density index $\alpha$.

\subsection{Disk Rotation Outside the ISCO}

We require a self-consistent model for orbital motions away from the rotational equator 
of the BH.   This defines a `disk frame' $\disk$ both in the zone of quasi-circular motion
outside the ISCO and the interior zone where matter plunges toward the BH.  The rotation profile is
obtained by (i) neglecting the radial pressure gradient and (ii) restricting the inflow to surfaces of
constant $\theta$. 

Outside of the ISCO, the mean angular rotation rate of the disk on the equator is \citep{Bardeen1972},
\be 
    \label{eq:eq_circ}
    \omega_d(r,\theta = \pi/2) = {d\phi_d\over dt}\biggr|_{\theta = \pi/2} = \frac{(Mr)^{1/2}}{r^2 + a(Mr)^{1/2}}.
\ee 
Off the equator, we assume the presence of a pressure gradient $\nabla_\theta P$ that enforces
$d\theta_d/d\tau \rightarrow 0$, where $\tau$ is the proper time of the disk matter and
$dx^\mu_d/d\tau$ is the tangent vector of the disk flow.   Then the radial component of the geodesic equation,
\be
\frac{d^2r_d}{d\tau^2} = -\Gamma^{r}_{\alpha \beta} \frac{dx^\alpha_d}{d\tau} \frac{dx^\beta_d}{d\tau} = 0,
\ee
gives a quadratic equation for $\omega_d = d\phi_d/dt$,
\ba
&&\frac{\Delta p}{\Sigma} \left(\frac{dt}{d\tau} \right)^2 + \frac{\Delta \sin^2 \theta}{\Sigma} \left(r + p a^2 \sin^2 \theta \right) \left( \frac{d\phi_d}{d\tau}\right)^2 \nn 
 && \quad - 2 \left( \frac{a \Delta p \sin^2 \theta} {\Sigma} \right) \frac{d\phi_d}{d\tau} \frac{dt}{d\tau} = 0,
\ea
where $p = [M(\Sigma - 2Mr)]/\Sigma^2$.  We obtain the angular velocity profile 
(within the zone $\rISCO < r < \rCorona$)
\be\label{eq:omc}
  \omega_d(r,\theta) = \frac{a \sin^2 \theta - \sqrt{-r \sin^2 \theta / p}}
  {\frac{r}{p} \sin^2 \theta + a^2 \sin^4 \theta};
\ee
this reduces to Equation (\ref{eq:eq_circ}) on the equator.   This profile will describe
both the motions of seed electron-ion plasma and created pairs.

\subsection{Infall Inside the ISCO}
Inside the ISCO, all components of the plasma fall inward on timelike geodesics, constrained as before to 
lie on surfaces of constant $\theta$.  One may solve
for $u^\mu_d = dx^\mu_d/d\tau$ by setting the specific orbital energy ${\cal E} = 
-g_{t\mu}u^\mu_d$ and specific angular momentum ${\cal L}_z = g_{\phi\mu}u^\mu_d$ equal to their values 
${\cal E}_{\rm ISCO}$, ${\cal L}_{z,\rm ISCO}$ at the inner boundary of the centrifgually supported disk.
In between the ISCO and the BH horizon at radius $r_H = M + \sqrt{M^2 - a^2}$, we have
\begin{align}\label{eq:plunge1}
    u^t_d &= -g^{tt} \Eorb_{\rm ISCO} + g^{t \phi} {\cal L}_{z,\rm ISCO}; \nn
    u^\phi_d &= -g^{\phi t} \Eorb_{\rm ISCO} + g^{\phi \phi} {\cal L}_{z,\rm ISCO}.
\end{align}
Imposing our assumption that $u^\theta_d = 0$, we find
\be\label{eq:plunge2}
u^r_d = -\frac{1}{\gr^{1/2}} \Bigl[1 + \gt (u^t_d)^2 + \gphi (u^\phi_d - \omega u^t_d)^2 \Bigr]^{1/2}.
\ee
The constants ${\cal E}_{\rm ISCO}$ and ${\cal L}_{z,\rm ISCO}$ are obtained in the approximation that
the boundary of the plunging zone sits at constant radius $r_{\rm ISCO}$ (Equation (\ref{eq:risco})), 
independent of polar angle.  Then
\ba\label{eq:ISCOints}
    \Eorb_{\rm ISCO}(\theta) &= \left[-\gt u^t_d + \omega \mathcal{L}_{z,\rm ISCO}\right]_{\rISCO,\theta}, \nn
    {\cal L}_{z,\rm ISCO}(\theta) &= \left[\gphi \left(u^\phi_d - \omega u^t_d \right)\right]_{\rISCO,\theta},
\ea    
where at $r \geq \rISCO$ we can approximate $u^r_d \rightarrow 0$ and
\be\label{eq:uplunge}
    u^t_d = \left[\frac{-1}{\gt + \gphi (\omega_d - \omega)^2 } \right]^{1/2}; \quad
    u^{\phi}_d = \omega_d u^t_d.
\ee

\begin{figure}
    \centering
    \includegraphics[width=\linewidth]{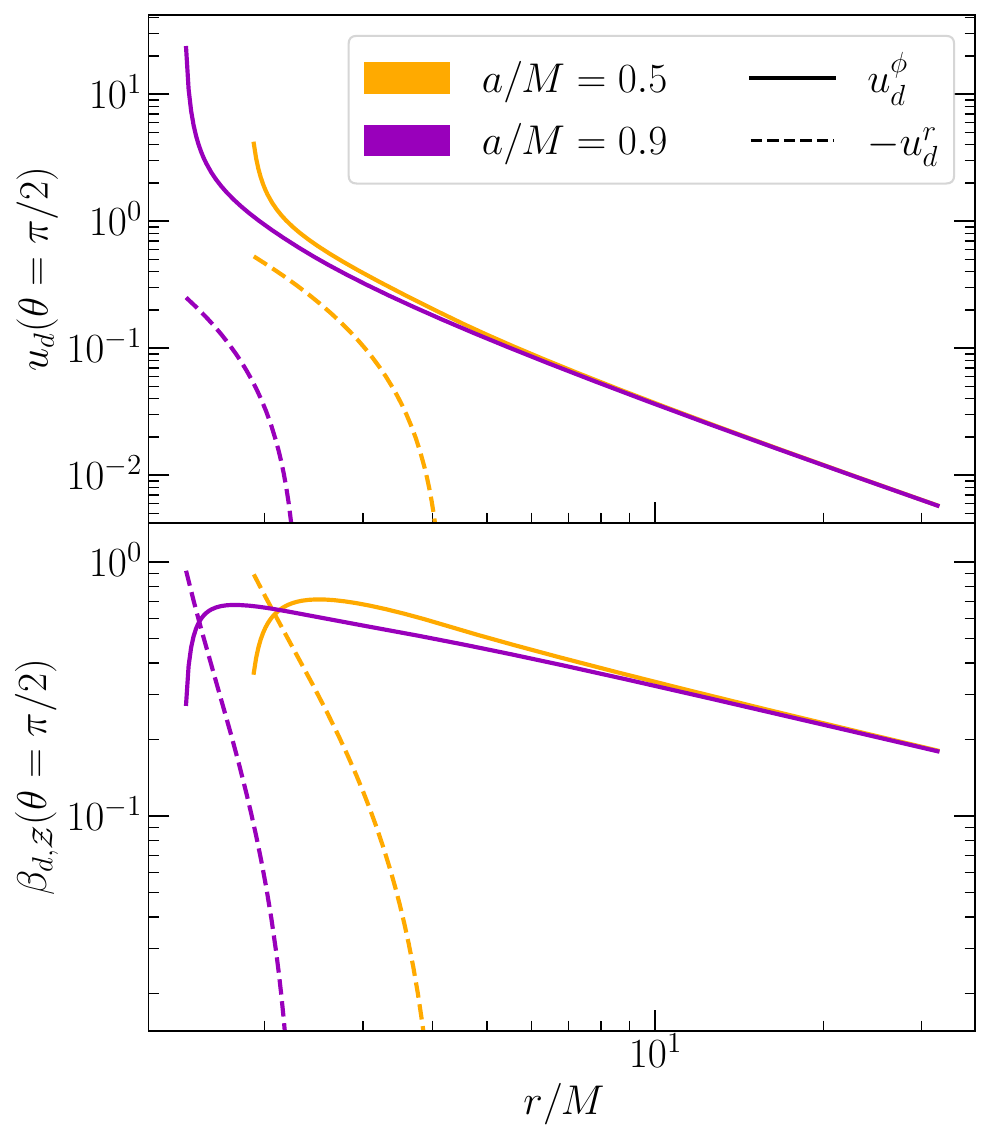}
    \caption{Radial profile of BL-frame four-velocity and ZAMO-frame three-velocity components within the mid-plane.
    Purple (orange) curves show flows around BHs with spin $a/M = 0.9$ ($a/M = 0.5$).
    Outside the ISCO, the flow is taken to be nearly circular;  inside, the infalling flow is fixed by the
    constants of motion $\{ \Eorb, \Lang \}$.}
    \label{fig:fluid_velocity}
\end{figure}

On the equator, Equations (\ref{eq:plunge1}) and (\ref{eq:plunge2}) reduce to the 
solution for the disk flow in the plunging zone \citep{Mummery2022}:
\begin{align}
    u_d^t &= \frac{1}{\Delta} \left[\left(r^2 + a^2 +   2M\frac{a^2}{r}\right) \Eorb_{\rm ISCO} - 
    \frac{2 M a {\cal L}_{z,\rm ISCO}}{r}\right]; \nn
    u_d^r &= -\sqrt{\frac{2M}{3\rISCO}} \left(\frac{\rISCO}{r} - 1 \right)^{3/2} ; \nn 
    u_d^{\theta} &= 0; \nn
    u_d^{\phi} &= \frac{1}{\Delta}\left[\frac{2Ma\Eorb_{\rm ISCO}}{r} + 
    \left(1 - \frac{2M}{r}\right){\cal L}_{z,\rm ISCO}\right].
\end{align}
Here, $\Eorb_{\rm ISCO}$ and ${\cal L}_{z,\rm ISCO}$ have simplified to
\begin{align}
    \Eorb_{\rm ISCO} &= \left(1 - \frac{2M}{3\rISCO}\right)^{1/2}; \nn
    {\cal L}_{z,\rm ISCO} & = 2\sqrt{3}M\left(1 - \frac{2a}{3 \sqrt{M\rISCO}}\right).
\end{align}
Under the assumption of motion with constant $\theta$, fluid geodesics are constructed that smoothly transition from outside the ISCO to the plunging zone. The mid-plane fluid velocity components are shown in Figure \ref{fig:fluid_velocity}.

\subsection{Stochastic Motion of Scattering Particles}
\label{sec:rms_beta}

Scattering $e^\pm$ in heated zones carry an additional random velocity $\rmsBeta_\disk$ 
in addition to the fixed bulk motion described by Equations (\ref{eq:omc}) and 
(\ref{eq:plunge1})-(\ref{eq:ISCOints}).
The quantity $\rmsBeta_\disk$, defined in the disk frame, represents some combination
of turbulent and kinetic $e^\pm$ motions;  no attempt is made to separate these contributions. As a simplified treatment of dissipation, we assume a constant magnitude and isotropic distribution
for $\rmsBeta_\disk$, which is adjusted iteratively to maintain a set amplification $A$ of 
escaping seed photon energy (Equation (\ref{eq:amplif})). The magnitude $|\rmsBeta_\disk|$ 
vanishes at $r > \rCorona$;  we also consider separate cases where (i) $|\rmsBeta_\disk|$ 
is the same in the plunging zone and the coronal disk (the $a/M = 0.5$ models); and 
(ii) $|\rmsBeta_\disk| = 0$ at $r < \rISCO$ (the $a/M = 0.9$ models).

The choice of uniform $|\rmsBeta_\disk|$ must lead to inconsistent behavior in 
very optically thick plasma, 
where trapped photons scatter many times and a stable balance between pair creation and photon escape
cannot then be maintained.  We find that this effect is noticeable only well inside the ISCO, where photons tend to
be advected toward the BH.  Convergence in the $e^\pm$ density field is maintained by
imposing a maximum effective Compton parameter,
\begin{align}
    y_{\rm eff} &\equiv \rmsBeta_\disk^2 \,{\rm max}(\tau_\pm, \tau_\pm^2) \;\leq\; y_{\rm max};\nn
    \tau_\pm &= n_{\pm, \disk} d_\theta \sigma_{\rm T}.
\end{align}
Here, the characteristic optical depth $\tau_\pm$ is obtained by projecting the local $n_\pm$ density 
across a distance $d_\theta = \sqrt{g_{\theta\theta}}$ (1 radian in latitude). 
Choosing $y_{\rm max} = 10$ only affects the pair distribution in the densest regions inside the ISCO,
where the magnitude of $\rmsBeta_\disk$ is limited to
\begin{equation}\label{eq:suppress}
    |\rmsBeta_\disk| \rightarrow {\rm min}\left(|\rmsBeta_\disk|, 
    \sqrt{y_{\rm max} / {\rm max}(\tau_\pm, \tau_\pm^2)} \right).
\end{equation}
This procedure has a minimal effect on the spectra and polarization of escaping photons, which
are heated in zones where the cutoff does not apply; this we have checked by varying the value of 
$y_{\rm max}$.

\subsection{Frame Transformations}

In order to compute scattering of photons off disk material, it is useful to know 
the disk four-velocity in the ZAMO frame, which is locally flat.  This is given by 
the transformation $\gamma_{d,\,\ZAMO}(1,\bm{\beta}_{d,\ZAMO}) = (\tetBLZ)_{\;\nu}^\mu u^\mu_d$, using
\begin{equation}\label{eq:Lambda}
    (\tetBLZ)_{\;\nu}^\mu = \left( 
    \begin{array}{cccc}
    \sqrt{-\gt} & 0 & 0 & 0 \\
    0 & \sqrt{\gr} & 0 & 0\\
    0 & 0 & \sqrt{\gtheta} & 0 \\
    -\omega \sqrt{\gphi} & 0 & 0 & \sqrt{\gphi}
    \end{array}
    \right).
\end{equation}
The time lapsed in the ZAMO frame is $dt_\ZAMO = \sqrt{-\gt}\,dt$, giving
\be
   \gamma_{d,\ZAMO} = {dt_\ZAMO\over d\tau} = \sqrt{-\gt}{dt\over d\tau}.
\ee
Hence the inflow velocity as seen by a ZAMO is
\begin{align}
    \label{eq:inner_velocity}
    \beta_{d, \ZAMO}^r &= {1\over\gamma_{d,\ZAMO}}\sqrt{\gr} {dr\over d\tau}; \nn 
    \beta_{d, \ZAMO}^{\theta} &= 0; \nn
    \beta_{d, \ZAMO}^{\phi} &= {1\over\gamma_{d,\ZAMO}} \sqrt{\gphi}
    \left({d\phi\over d\tau} - \omega {dt\over d\tau}\right)
    = {\sqrt{\gphi}\over\sqrt{-\gt}}\left({d\phi\over dt} - \omega\right),
\end{align}
with total speed $\beta_{d, \ZAMO} = \sqrt{1 - \gamma_{d, \ZAMO}^{-2}}$.
Outside the ISCO, where the radial velocity is small and the disk rotates with 
angular velocity $\omega_d$, as given by Equation (\ref{eq:omc}), this reduces to 
\be
    \label{eq:disk_velocity}
    \bm{\beta}_{d,\ZAMO} = 
    \frac{\sqrt{\gphi}}{\sqrt{-\gt}} (\omega_d - \omega)\,\hat\phi. 
\ee

The particle densities transform between the disk and ZAMO frames according to
\begin{align}\label{eq:dentrans}
    n_{d, \disk} &= n_{d, \ZAMO} / \gamma_{d, \ZAMO};\nonumber\\
    n_{\pm, \disk} &= n_{e^+, \disk} + n_{e^-, \disk} = n_{\pm, \ZAMO} / \gamma_{d, \ZAMO}.
\end{align}

\section{Photon Seeding and Propagation} \label{sec:photon}

We now describe the MC algorithm controlling the emission of seed keV photons around the BH,
their propagation in the curved spacetime, and their scattering off ambient $e^\pm$. 
Photons are emitted in the frame of the disk; general relativistic effects influence the
effective ZAMO-frame
emission rate and are accounted for by a weighting scheme (Section \ref{sec:emission}). 
Disk-frame wavevectors are isotropically distributed on emission and then converted to BL-frame vectors (Section \ref{sec:emission_wave});  the volumetric emission rate is scaled to the spatial
distribution of emitting particles (Section \ref{sec:emission_space}). While initially unpolarized, 
photons develop linear polarization by scattering (Section \ref{sec:polarization}). 
Photon wavevectors and polarization experience parallel transport in the strong gravitational field,
with the scattering rate sensitive to bulk and random plasma motions (Section \ref{sec:propagation}). 
Scattering particles inside the corona are given a random velocity to model heating via dissipation;
the treatment of photon-$e^\pm$ scattering is fully relativistic (Section \ref{sec:scatter}, Appendix \ref{app:scattering}).  The photon occupancy in grid cells is recorded and used to compute the local pair
creation rate;  escaping photons are recorded to build observer-dependent spectra and polarization data (Section \ref{sec:record}).

\subsection{Monte Carlo Treatment of Photon Emission}
\label{sec:emission}

The emission probability is influenced both by the plasma density distribution and also by
differences in the rates at which photons emitted in different directions would pass through
the neighboring spacetime.
For example, in the simplest case, the emission may be taken to be isotropic about any
emission point, as measured by an observer comoving with the disk.  Photons may be 
emitted repeatedly in some direction at some fraction of the wave frequency
$k^t_\disk$, as measured in the disk rest frame $\disk$.  Shifting to the BL coordinate system,
these photons would then be emitted at the same fraction of the wave frequency 
${\cal E}_{\rm em}$.  

We record the occupancy of photons in the local ZAMO frame, which orbits
with azimuthal frequency $\omega(r,\theta)$ (Equation (\ref{eq:omZ})).  
At a fixed point in such a differentially rotating frame, 
the electromagnetic phase is $\phi_{\rm em} = 
-({\cal E}_{\rm em} - \omega {\cal L}_{z,\rm em})t + g_{rr}k_r r + g_{\theta\theta}k_\theta \theta$.  
A photon emitted at position ${\bm r}_{\rm em}$, with energy
${\cal E}_{\rm em}$ and angular momentum ${\cal L}_{z,\rm em}$, is therefore given a weight
\be\label{eq:weight}
w = {{\cal E}_{\rm em} - \omega({\bm r}_{\rm em}){\cal L}_{z,\rm em} \over k^t_\disk}
= -{\gt({\bm r}_{\rm em})\over k^t_\disk}{dt\over d\lambda}\biggr|_{\rm em}.
\ee
The weight factor (\ref{eq:weight}) calibrates the relative rate (per unit global time $t$) 
at which successive photons with
the same emission location and direction would pass by neighboring observers and be
detected at infinity.  An important feature is that $w$ is positive definite, in contrast
with the energy integral ${\cal E}_{\rm em}$, which may be negative inside the BH ergosphere. 
In addition, $w$ vanishes toward the BH horizon, thereby downweighting seed photons that
are emitted deep in the gravitational potential.

The quantity $w$ accounts for the difference between ``emitted' and ``received'' power that is characteristic
of radiation from relativistic charges \citep{RL1986}.   This effect can be illustrated by
considering emission and detection probabilities in the ZAMO frame.  The net rate of emission
from a small patch of the disk, as measured in this frame, is suppressed by a 
factor $d\tau/dt_\ZAMO = 1/\gamma_{d,\ZAMO}$, whereas in the direction of emission the
detection rate would be enhanced by a factor $k^t_\ZAMO/k^t_\disk \sim \gamma_{d,\ZAMO}$.  
As expected, the received power is enhanced by a factor $\sim \gamma_{d,\ZAMO}^2$.

As a proxy for the luminosity $L_{\rm seed}$ of keV photons, we track the total weighted seed energy of escaping photons:
\begin{align}
E_{\rm seed} &= \sum_i w_i \mathcal{E}_{{\rm em},i}.
\end{align}
The weighted total energy at infinity after scattering, $E_\infty = \sum_i w_i \mathcal{E}_{\infty,i}$, will be constrained to be a fixed multiple $A$ of the seed energy,
\be\label{eq:amplif}
    E_\infty = A\,E_{\rm seed}.
\ee    
This is achieved by iterating on the stochastic $e^\pm$ velocity 
$|\rmsBeta_\disk|$ (Section \ref{sec:iteration}).

\subsection{Emission Wavevector}\label{sec:emission_wave}

In the disk frame $\disk$, each emitted photon has a wavevector ${\bm k}_\disk$ and a
frequency $k^t_\disk = |{\bm k}_\disk|$ that is drawn randomly from a blackbody distribution
of temperature $T_{\rm bb}$.  For the models described in this paper, the emission is isotropic
in the disk frame. 

The ray tangent vector in the BL frame is determined by successive boosts 
to the ZAMO frame and then to the BL frame. 
For emission outside the ISCO, where the bulk disk motion is azimuthal, we have
\begin{align}
    \label{eq:frame_transform}
    \frac{dt}{d\lambda} &= \frac{\gamma_{d, \ZAMO}(k_\disk^t + \beta_{d, \ZAMO} k_\disk^\phi)}{\sqrt{-\gt}}; \nn
    \frac{d\phi}{d\lambda} &= \frac{\gamma_{d, \ZAMO}}{\sqrt{\gphi}} \left[(\beta_{d, \ZAMO} + \beta_{{\rm BL}, \ZAMO}) k^t_\disk + (1 +\beta_{d, \ZAMO} \beta_{{\rm BL}, \ZAMO}) k^\phi_\disk \right]; \nn
    \frac{dr}{d\lambda} &= \frac{k_\disk^r}{\sqrt{\gr}}; \quad\; \frac{d\theta}{d\lambda} = \frac{k^\theta_\disk}{\sqrt{\gtheta}} ; \quad \beta_{{\rm BL}, \ZAMO} \equiv  \frac{\sqrt{\gphi}}{\sqrt{-\gt}} \omega.
\end{align}
The corresponding orbital energy and angular momentum integrals are
\begin{align}\label{eq:integrals}
    \mathcal{E} &= -\gt \frac{dt}{d\lambda} + \omega \mathcal{L}_z; \nn 
    \mathcal{L}_z &= \gphi \left(\frac{d\phi}{d\lambda} - \omega \frac{dt}{d\lambda} \right).
\end{align}

\subsection{Spatial Distribution of Emission}\label{sec:emission_space}

The soft-photon emissivity $dU_{\rm soft}/dt$ (energy per unit volume 
and time) is chosen to scale with plasma density in such a way that
the volume-integrated output is weighted
to smaller radius:  we take $dU_{\rm soft}/dt \propto n_{d, \disk}^2$ or $n_{\pm,\disk}^2$
in the disk frame, where $n_\pm = n_{e^-} + n_{e^+} = n_d + 2n_{e^+}$ is the total $e^\pm$ density.
Starting with no pairs and an ion surface density profile
$\Sigma \propto r^{-1}$ ($\alpha = 1$) and scale height $h \propto r$, we expect a seed luminosity
$dL_{\rm soft}/d\ln r \propto (\Sigma/h)^2 \cdot r^2 h \propto r^{-1}$.  
The same density scaling is expected for electron-ion bremsstrahlung emission;  a non-linear
scaling is also expected for cyclo-synchrotron emission by pairs as long as $n_\pm$ correlates
positively with magnetic energy density.
We note that the profile of the created pair density
field is even more strongly peaked at small radius (Section \ref{sec:outline});  
this would allow the photon emission to
concentrate in the inner disk even for a power-law scaling $dU_{\rm soft}/dt \propto n_\pm^\delta$ with
$\delta$ well below 2. 

Seed emission is allowed wherever the plasma is heated; models with $a/M = 0.5$ 
then allow emission inside the ISCO.  The density peaks close to the horizon, where any emitted
photons tend to be advected inward;
to avoid these trapped photons dominating the MC, emission in the innermost two radial 
layers of cells is suppressed.
 
To determine the emission location and photon wavector, we adopt the following procedure.  The emission rates from different cells are ordered by the local disk-frame value.
For example, electron-ion plasma in a spatial cell of size $\Delta V_\disk
= \gamma_{d,\ZAMO}\Delta V$ would emit at a rate $\Gamma_{\rm em} \sim n_{d,\disk}^2 \Delta V_\disk$.  
The emitting region is spanned by 
$N_{e, r} = 128$ separate radial shells, uniformly spaced in $\log(r)$ (Appendix \ref{app:grid}).
The quantity $n_{d,\disk}^2$ or $n_{\pm,\disk}^2$ 
is first summed over volume in each radial shell;  this sum is
then splined over radius to obtain a cumulative radial emission probability.
The $\theta$-profile within each radial shell is also splined over
$N_{e,\theta} = 256$ angular cells (uniformly distributed in the range
$\theta_{\rm jet} < \theta < \pi - \theta_{\rm jet}$).  
Once the emission radius is determined by drawing a random variable,
the $\theta$-splines of the two closest radial shells are linearly combined to determine the emission colatitude.
Given that the background disk is axisymmetric, we can also randomly draw
the emission azimuth $\phi \in [0,2\pi)$ from a uniform distribution.
Once the emission location
and initial wavevector are determined, the trial photon is then assigned a weight $w$ 
(Equation (\ref{eq:weight})), where ${\cal E}_{\rm em}$ and ${\cal L}_{z,\rm em}$ are
computed from Equations (\ref{eq:frame_transform}) and (\ref{eq:integrals}).

\subsection{Polarization}\label{sec:polarization}

X-ray photons are taken to be unpolarized at emission.  KN scattering does not generate circular polarization from a linearly polarized state;
we can therefore express the polarization of each photon by the linear Stokes parameters $\{Q, U\}$, 
which are defined with respect to a local polarization basis.  At emission in the disk frame, the basis 
states are set to 
\be
    \hat\varepsilon_{1, \disk} = \frac{\hat k_\disk \times \hat \theta}{|\hat k_\disk \times \hat \theta|}; \quad\quad
    \hat\varepsilon_{2, \disk} = \hat k_\disk \times \hat \varepsilon_{1, \disk}.
\ee
The polarization degree $\delta$ and linear polarization angle $\psi$ are defined in terms of $\{Q,U\}$ as
\ba
\delta &=& \sqrt{Q^2 + U^2}; \nn
\psi &=& 0.5 \arctan{(U/Q)},
\ea
where we have set intensity $I \rightarrow 1$.

The Stokes parameters are conserved during photon propagation, whereas the polarization basis states
evolve by parallel transport.   At any point along a ray trajectory, 
each basis state can be recovered from the conserved, complex-valued Walker-Penrose constant \citep{Penrose1970,Connors1977}:
\ba
\kappa_{\rm WP} &=& (r - ia \cos \theta)\Big( \mathcal{A}^{tr} + a \sin^2 \theta \mathcal{A}^{r \phi} \nn
&& -i \sin \theta \left[\left(r^2 + a^2\right) \mathcal{A}^{\phi \theta} - a\mathcal{A}^{t\theta}\right]\Big) 
; \nn 
\mathcal{A}^{\mu \nu} &\equiv& \frac{dx^\mu}{d\lambda} \varepsilon^\nu - \frac{dx^\nu}{d\lambda} \varepsilon^\mu.
\ea 
This constant, which being complex comprises two conserved variables, is initialized in the BL frame
after transforming the basis states from the disk frame.  We have checked that imposing constant $\kappa_{\rm WP}$
reproduces the evolution of basis states by parallel transport.

\subsection{Propagation}
\label{sec:propagation}

Each photon, once emitted, is propagated in BL coordinates until it scatters,
reaches the horizon, or escapes to infinity.  The $t-$ and $\phi-$components of the tangent vector 
are fixed by the ray integrals ${\cal E}$ and ${\cal L}_z$.  The $r-$ and $\theta-$components 
are evolved by parallel transport,
\be 
\frac{d^2 x^\mu}{d\lambda^2} = -\Gamma^\mu_{\alpha \beta} \frac{dx^\alpha}{d\lambda} \frac{dx^\beta}{d\lambda}.
\ee 
A complete solution by quadrature can also be obtained for the ray trajectory by computing the Carter constant \citep{Carter1968}, but the stochastic effects of scattering are more readily represented by the
present method, especially near a turning point of the geodesic.

The scattering probability is evaluated by integrating the corresponding optical depth, $\tau$,
along the ray.  The effect of the mean rotation and plunging motion on the scattering process
is readily accommodated in the ZAMO frame, where
\begin{equation}
    \label{eq:scatter_depth}
    \Gamma_\ZAMO = \frac{d\tau}{dt_\ZAMO} = n_{\pm, \ZAMO} (1 - {\bm\beta}_{d,\ZAMO} \cdot {\hat{k}_\ZAMO}) \sigma(\epsilon).
\end{equation}
Here ${\bm\beta}_{d,\ZAMO}$ is the bulk ion-electron flow velocity as defined by 
angular velocity (\ref{eq:omc}) outside the ISCO, and by Equations 
(\ref{eq:plunge1})-(\ref{eq:ISCOints}) in the plunging zone interior to $\rISCO$.
The total scattering cross section $\sigma(\epsilon)$ is a function of $\epsilon = k^t_\disk / m_e$. 
Transforming to the BL frame gives
\begin{equation}
    \frac{d\tau}{d\lambda} = \Gamma_\ZAMO \frac{dt_\ZAMO}{dt}\frac{dt}{d\lambda} = 
    (-\gt)^{1/2} \Gamma_\ZAMO \frac{dt}{d\lambda}.
\end{equation}

The scattering rate depends on the velocity distribution of electrons and positrons, as described in Section \ref{sec:rms_beta}.
The scattering rate differs from the result for cold $e^\pm$ according to \citep{Coppi1990}:
\begin{align}
    \sigma(\epsilon) &= \int_{-1}^{+1} \frac{d\mu}{2} (1 - \delta\beta_\disk \mu) 
    \sigma_{\rm KN}\bigl[\delta\gamma_\disk(1-\delta\beta_\disk \mu)\epsilon\bigr] \nn
    &= \frac{1}{8 \delta\gamma_\disk^2 \delta\beta_\disk \epsilon^2}
  \int_{2\delta\gamma_\disk(1 - \delta\beta_\disk)\epsilon}^{2\delta\gamma_\disk(1 + \delta\beta_\disk)\epsilon} x \sigma_{\rm KN}(x/2) dx,
\end{align}
where $x = 2 \delta\gamma_\disk (1-\delta\beta_\disk \mu) \epsilon$, $\delta\beta_\disk\mu = 
\rmsBeta_\disk\cdot\hat k_\disk$, $\delta\gamma_\disk \equiv (1-\delta\beta_\disk^2)^{-1/2}$, and $\sigma_{\rm KN}$ is the KN cross-section.
Scattering occurs over an interval $\Delta\lambda$
when a randomly drawn number $X_{\rm scatt} \in U[0,1]$ falls in the range 
$X_{\rm scatt} < P(\Delta\lambda) = 1 - {\rm exp}[-\int_0^{\Delta \lambda} (d\tau/d\lambda) d\lambda]$, where $P(\Delta\lambda)$ is the probability.

Attenuation by $\gamma-\gamma$ collisions is ignored compared with
repeated KN scattering when calculating the propagation of photons up to energy $2.5 m_ec^2$
in the exponential tail of the spectrum.  We show this is a reasonable approximation, in the sense
that the absorption rate is a small fraction of the scattering rate;  photons are replenished
by scattering more rapidly than they can be lost to a collision with another photon, or
injected by $e^-e^-$ annihilation (Appendix \ref{app:photon_consumption}).  This approximation
would break down in the presence of a power-law tail to the $e^\pm$ energy distribution.

The quantities to be integrated over a ray trajectory are stored in a dimensionless vector 
\begin{equation}
    \mathbf{X} = \left({r\over M}, \theta, \phi, {t\over M}, \frac{dr}{d\lambda}, M\frac{d\theta}{d\lambda}, \tau \right),
\end{equation}
alongside the conserved quantities $\{{\cal E}, {\cal L}_z, Q, U, \kappa_{\rm WP}\}$. 
The vector ${\bm X}$ is integrated 
using a Runge-Kutta-Fehlberg (or RKF45) method with adaptive step size.
The error $\delta{\bm X}$ is constrained to be ${\rm max}(\delta \mathbf{X}) \leq 10^{-12} (1 + |\mathbf{X}|)$.
The error in the geodesic solution can be tested by computing the change in the conserved Carter constant;
this remains below $10^{-13}$ for a large fraction of escaping rays,
increasing to $\sim 10^{-11}$ for a small fraction of rays experiencing a turning point near the poles.

Photons reaching the horizon radius $r_H$ are removed and their energy is lost.  Photons escaping the computational grid at radius $\rmax$ are allowed to propagate without scattering out to a radius $32\,\rmax = 1024\,M$, where their position, frequency and polarization are recorded (Section \ref{sec:record}).

\subsection{Photon Update After Scattering} \label{sec:scatter}

Photon data are updated following KN scattering by a standard procedure 
(Appendix \ref{sec:KN}). First, the random velocity $\rmsBeta_\disk$ for the scattering electron or positron is drawn (Section \ref{sec:rms_beta}). The total ZAMO-frame velocity ${\bm\beta}_{d,\ZAMO}$ of the scattering
particle is then obtained by relativistic summation with the velocity of the mean flow. 
A representative polarization vector $\varepsilon^\mu$ is next selected, as outlined in Appendix 
\ref{app:scattering}. The ray tangent vector and selected polarization are transformed to the rest frame of the scattering particle, as $k_\particle^\mu$ and $\varepsilon_\particle^\mu$.
After scattering, the updated $k_\particle^\mu$ is then converted to the BL frame ray tangent vector for further propagation, along with the
updated Stokes parameters and Walker-Penrose constant. 

\subsection{Test of Scattering and Polarization Transport}

\begin{figure}[ht]
    \centering
    \includegraphics[width=0.9\linewidth]{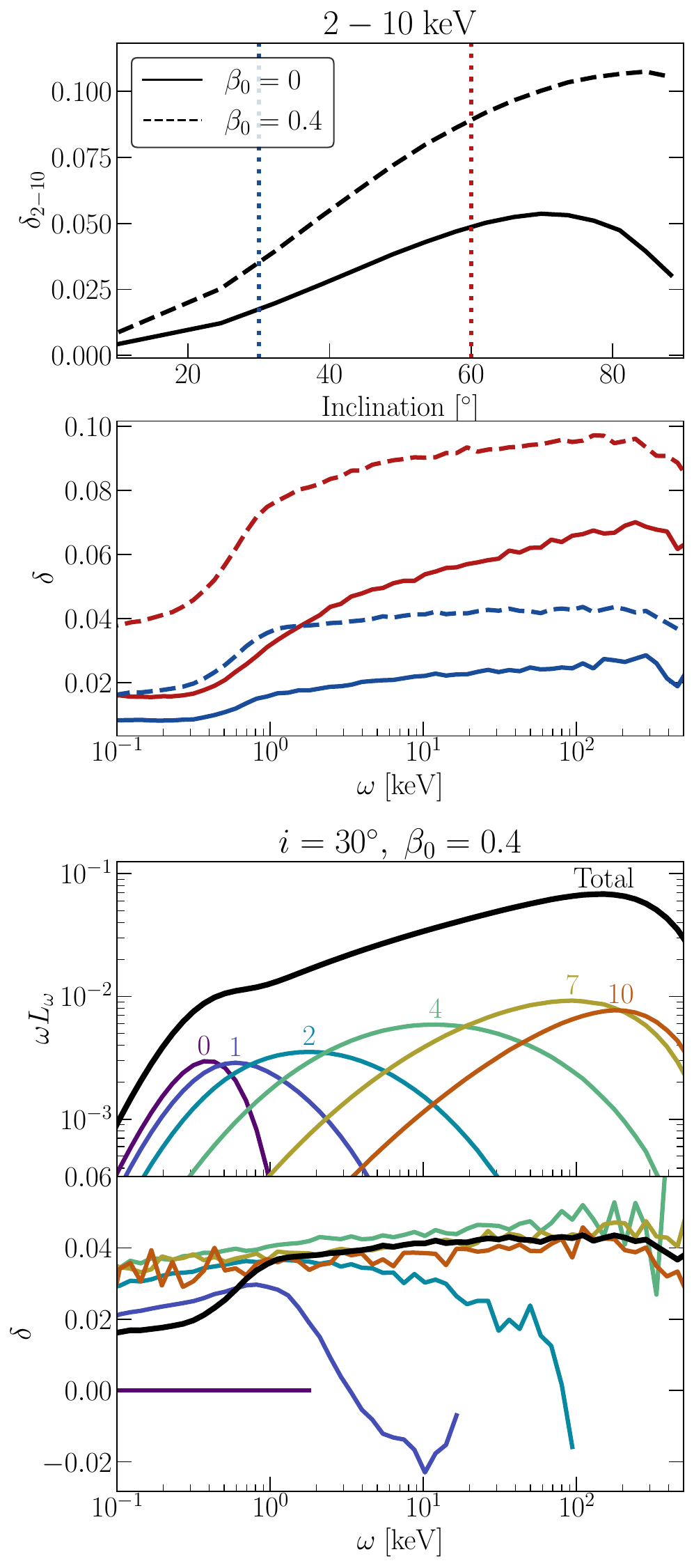}
    \caption{Polarization and spectral data for photon scattering in a Cartesian box, which can be compared
    with the independent transport model of \cite{Poutanen2023}.  First row:  polarization degree 
    for $2-8$ keV photons as a function of observer-disk inclination, for peak outflow velocity $\beta_0 = 0$ (solid) and 0.4 (dashed).  Second row:  polarization degree as a function of photon energy, for observers
    with $30^\circ$ (blue) and $60^\circ$ (red) inclination.
    Third row:  spectrum of photons escaping to an inclination of $30^\circ$, for peak outflow velocity $\beta_0 = 0.4$.  The total spectrum (black) is split into scattering orders (labeled by number and colour).
    Bottom row: corresponding polarization degree, again split
    by scattering order.
    Here, positive $\delta$ corresponds to polarization aligned with $\hat \theta$, while negative 
    $\delta$ corresponds to alignment with $\hat \phi$.}
    \label{fig:cartesian_box}
\end{figure}

To validate our routines for photon scattering and polarization evolution,
we ran a MC simulation of photons emitted from a Cartesian box with height $H$.
These results can be compared directly against geometry A in \cite{Poutanen2023}.

The vertical profile of the Thomson optical depth and electron density is (for $|z| \leq H)$:
\begin{align}
    \tau(z) &= \tau_0 (|z|/H)^{1/2}; \nn
    n_{e}(z) &= \frac{\tau_0}{2\sigma_{\rm T} H}(|z|/H)^{-1/2}.
\end{align}
This corona flows away from the mid-plane at speed (in units of $c$)
\begin{equation}
     \beta_{\cal F}(z) = \beta_0 (|z|/H)^{1/2} \quad\; (|z| < H).
\end{equation}
Photons escape the box when $|z| \geq H$ and are absorbed if they cross the mid-plane at $z=0$. Photons are isotropically emitted from $z = 0$ with energies drawn from a blackbody distribution with $k_{\rm B}T_{\rm seed} = 0.1$ keV; they are propagated in flat space, straight line geodesics between scatterings and from escape to infinity.
Upon escape to infinity, the photons are binned by escape inclination, energy, and additionally by the number of scatterings. 

In this test, the stochastic velocity of a scattering electron is not fixed in magnitude, but instead is drawn from a
Maxwell-J\"{u}ttner distribution with temperature $k_{\rm B}T = 100$ keV.
For a peak outflow velocity $\beta_0 = 0.4$, a photon index $\Gamma = 1.6$ in the spectrum
of escaping X-rays corresponds to optical depth $\tau_0 = 1.5$. 
Figure \ref{fig:cartesian_box} shows the polarization degree $\delta$ in two cases: 
(i) no outflow and (ii) an outflow with $\beta_0 = 0.4$.  These results show good agreement with 
Figure 2 of \cite{Poutanen2023}.  Also shown is the polarization degree as a function of photon
energy, as measured at $30^\circ$ and $60^\circ$ inclinations.  The agreement with Figure 3 of 
\cite{Poutanen2023} remains good up to $\omega \sim 100$ keV, where the KN cross section begins 
to diverge from the Thomson approximation used by \cite{Poutanen2023}.  A final test is provided 
by splitting the results for spectrum and polarization degree by scattering order, which may 
be compared with Figure 4 of \cite{Poutanen2023}.

\begin{figure*}
    \centering
    \includegraphics[width=0.8\linewidth]{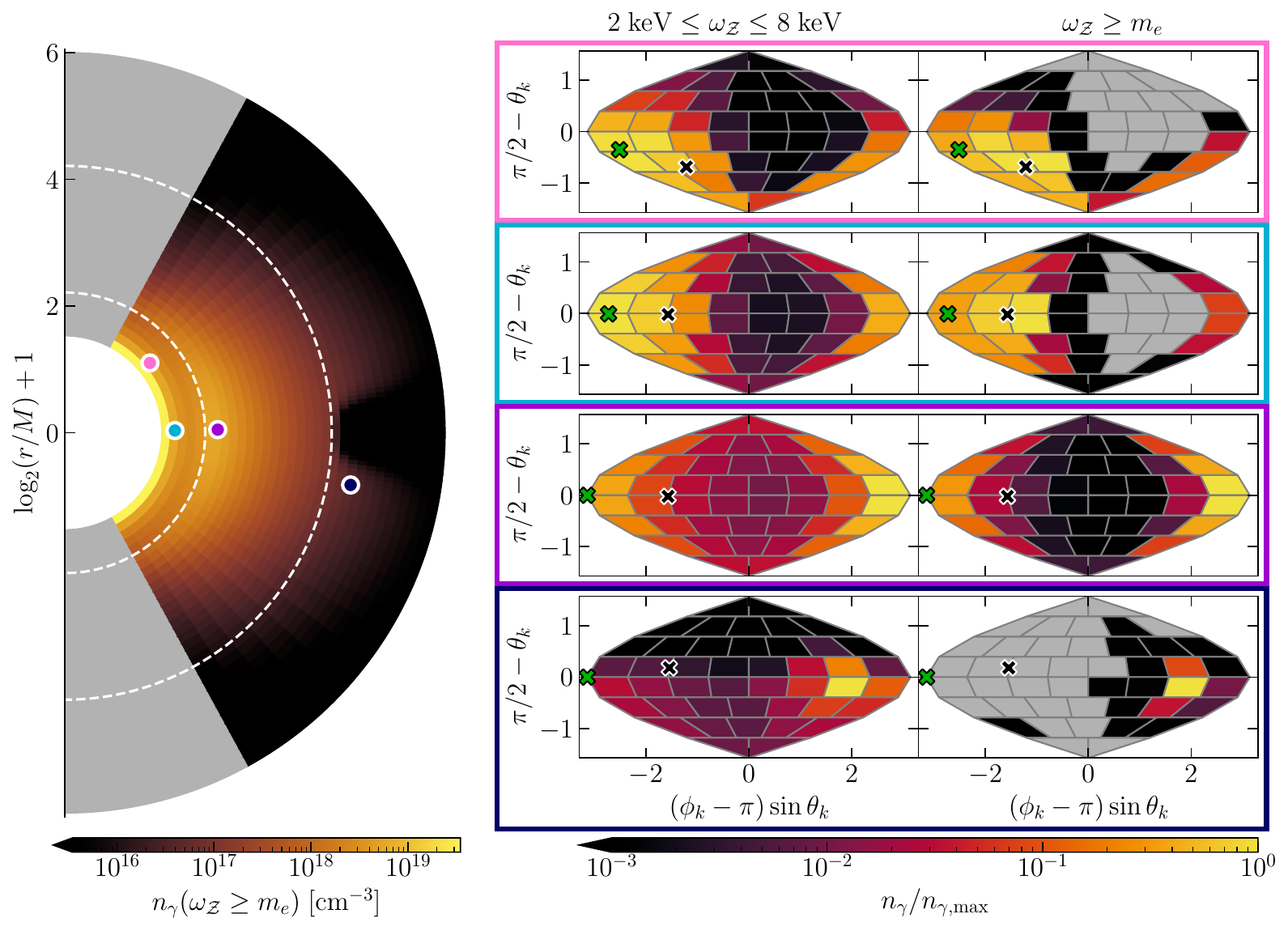}
    \caption{Left: distribution of high energy photons from model $\simbase$, in the $\phi-$averaged computational grid. 
    White dashed
    curves:  ISCO and outer boundary of the coronal disk ($\rCorona \equiv 4 \rISCO$).
    High-energy photons reach maximum density in the innermost layer of grid cells, with a secondary peak outside $\rISCO$ where the input electron-ion density peaks.  High-energy photons are rarest
    in the cold, outer disk, in which trapped photons rapidly downscatter in energy.  
    Right:  the photon angular distribution in model $\simbase$ is sampled at four color-coded 
    points representing
    different parts of the accretion flow, both in the ZAMO-frame band $2\;{\rm keV} < \omega_\ZAMO
    < 8\;{\rm keV}$ and in a high-energy band $\omega_\ZAMO > m_e$.
    Direction of the bulk flow (green cross) and the BH (black cross) are marked out within each
    angular grid.}
    \label{fig:photon_distribution}
\end{figure*}
\subsection{Recording the Photon Distribution}\label{sec:record}

The photon distribution function and electron/positron densities in each spatial cell
are recorded in the ZAMO frame (Appendix \ref{app:grid}).  The distribution function 
is proportional to the cumulative occupancy time of rays passing through the cell.
Each trajectory is recorded with respect to affine distance from the last position of
emission or scattering, and splined over a set of points separated by
\begin{equation}
\label{eq:deltalambda}
    \Delta \lambda = 0.01 \times {\rm min}\left(\frac{r}{dl/d\lambda, }, \frac{1}{d\tau/d\lambda} \right),
\end{equation}
where $dl^2 = \gr dr^2 + \gtheta d\theta^2 + \gphi(d\phi - \omega dt)^2$.  
The density of all photons propagating with
wavevector ${\bm k}_\ZAMO$ is obtained from the summed (and weighted) cell residency time $\Delta t$,
\begin{equation}\label{eq:tres}
    n_\gamma({\bm k}_\ZAMO) = \frac{\Delta t ({\bm k}_\ZAMO)}{\delta V}
    \frac{L_{\rm seed}}{E_{\rm seed}};
\end{equation}
here, $\delta V$ is the cell volume.
This quantity is binned in angular direction and energy.  The ZAMO-frame photon momentum is
$k^\mu_\ZAMO = (\tetBLZ)_{\;\nu}^\mu (dx^\nu/d\lambda)$ following Equation (\ref{eq:Lambda}).

Note that, as long as the plasma state is axisymmetric, the {\it summed} occupancy time of rays passing through cells with fixed poloidal coordinates $(r,\theta)$ is independent of whether the coordinate system is static or differentially rotating.  We only consider axisymmetric configurations and so adopt a static coordinate system (Appendix \ref{app:grid}), averaging the photon distribution in azimuth.

Photons that escape to infinity are recorded on a separate angular grid (Appendix \ref{sec:angular_grid}), 
from which spectropolarimetric data may be reconstructed for various observers (Section \ref{sec:outline}). 
We also record the last scattering location of escaping photons, binned by $i$ and $\omega$.
Finally, in order to analyze the reflected component of the spectropolarimetric data, 
photons that have scattered at least once off the cold disk at $r > \rCorona$ are tagged as 
``reflected'' and separately recorded.

Figure \ref{fig:photon_distribution} compares the distributions of pair-creating photons and
$2-8$ keV X-rays in model $\simbase$ (BH spin $a/M = 0.9$).
The angular distributions are sampled at a few locations around the BH:  close to the horizon,
just outside the ISCO, and above the outer, cold, pair-depleted disk.  
In zones of higher optical depth, the $2-8$ keV photon distribution tends to peak around the direction
of the bulk flow (marked by the green cross).  High-energy photons close to the BH instead peak in the direction of infall (marked by the black cross), due to the combined effects of gravity and a reduced scattering cross-section.
In the $\simbase$ model, plasma is only heated outside the ISCO;
hence, the appearance of a secondary maximum in photon density in the inner part of the coronal disk 
($\rISCO < r < \rCorona$). 
At the outermost sampled point above the cold disk, the distribution is dominated by outflowing photons, especially at high energy.

\section{MC Iteration of Pair Density}
\label{sec:iteration}
\begin{figure}[ht]
    \centering
    \includegraphics[width=0.9\linewidth]{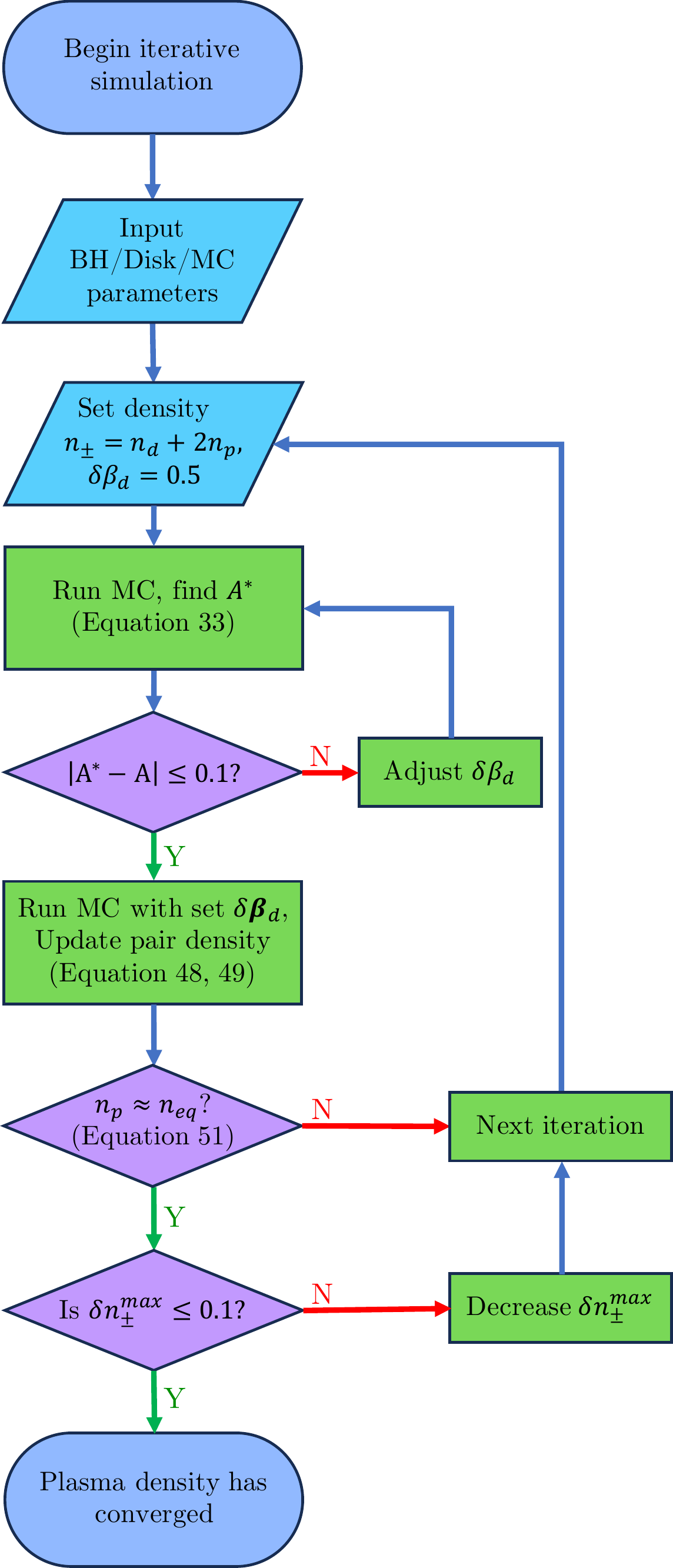}
    \caption{Flowchart of the iterative MC process used to determine a self-consistent corona. After 
    the BH, disk, and MC parameters are fixed, the random $e^\pm$ velocity magnitude $|\rmsBeta_d|$ is
    iterated upon in order to match the amplification $A$; then a larger MC simulation records
    the photon distribution and pair density field.  This process repeats until convergence in
    the $e^\pm$ density is achieved (Section \ref{sec:convergence}).}
    \label{fig:sim_flowchart}
\end{figure}

The MC procedure involves two main steps that are iteratively repeated: the first step of photon emission and scattering involves a calibration of the stochastic
velocity amplitude $|\rmsBeta_\disk|$, so as to obtain a fixed hard-photon luminosity amplification
$A = E_\infty/E_{\rm seed}$ (Equation (\ref{eq:amplif}));
the second step in the MC procedure involves updating the density of scattering charges.
We compute the pair creation rate in each spatial cell from the accumulated distribution
of hard photons, and then balance creation with annihilation as well as with advection inside the ISCO 
(Figure \ref{fig:sim_flowchart}). 
The two steps are repeated until 
energy and annihilation equilibrium are simultaneously achieved to $\sim 10\%$ accuracy 
(Section \ref{sec:convergence}).
The MC procedure is completed with a final MC run with $2 \times 10^9$ photons that sharpens
the polarization statistics.  We now describe these steps in more detail.

\subsection{Photon Collision Rate}
\label{sec:photon_collision}
The distribution of $e^\pm$ pairs around an accreting BH tracks the distribution
of high-energy photons.  In the model presented here, the heated pairs self-consistently upscatter
a small fraction of the injected soft X-rays above energy $\sim m_e$ in the ZAMO frame.

To determine the equilibrium pair density $n_p \equiv n_{e^+} (= n_{e^-}$ in the absence of ions), we first determine the pair creation rate. (To simplify notation, the index $\ZAMO$
is dropped on the particle densities $n_{p,d,e^\pm}$ and photon wavevectors ${\bm k}_i$ 
in this section.)  This is calculated in each cell as a sum over pairs of photon directions and energies, using the distribution function (Equation (\ref{eq:tres})).
For each photon direction and energy bin, a representative wavevector ${\bm k}_i = k^t_i(1, \hat{k}_i)$ is constructed, along with polarization density matrix $\rho({\bm k}_{i}) \equiv \rho_i$
(Equation (\ref{eq:denmatrix})):
\begin{equation}
    \frac{dn_p}{dt_\ZAMO} =\sum_{{\bm k}_i, {\bm k}_j}  |1 - \hat{k}_i \cdot \hat{k}_j|\:\sigma_{\gamma \gamma}(\theta, \rho_i, \rho_j) n_\gamma({\bm k}_i) n_\gamma({\bm k}_j).
\end{equation}
The polarization-dependent cross section for pair creation is given by \citep{Breit1934}:
\begin{align} 
\label{eq:pair_creation_sigma}
\sigma_{\gamma\gamma}(\theta, \rho_i, \rho_j) &= \sigma_{\gamma \gamma, \perp} + {\rm Tr}(\rho_i \rho_j)[\sigma_{\gamma \gamma, \parallel} - \sigma_{\gamma \gamma, \perp}] \nn
\sigma_{\gamma\gamma,\parallel}(\theta) &= \pi r_e^2 \biggl[-SC^{-3} - \frac{3}{2} SC^{-5} - 
\frac{3}{2} \theta C^{-6} \nn 
&+ 2\theta C^{-4} + 2\theta C^{-2}\biggr]; \nn 
\sigma_{\gamma\gamma,\perp} (\theta)&= \pi r_e^2 \biggl[-SC^{-3} - \frac{1}{2} SC^{-5} - 
\frac{1}{2} \theta C^{-6} \nn 
&+ 2\theta C^{-4} + 2\theta C^{-2}\biggr].
\end{align}
Here, $\parallel$ and $\perp$ denote collisons with the two polarization vectors parallel or perpendicular to each
other. The quantities $\{S, C\} = \{ \sinh\theta, \cosh\theta\}$ are functions of the COM energy, which is expressed as
\begin{equation}
    \label{eq:theta_COM}
    E_{\rm COM} = [(1-\hat{k}_i \cdot \hat{k}_j)/2]^{1/2}(k^t_i k^t_j)^{1/2} \equiv m_e\cosh\theta.
\end{equation}

The created pairs are quickly entrained in the disk flow.  The rate of annihilation
is most readily computed in the disk frame, where the $e^\pm$ densities  and the 
time lapsed are reduced by a factor $\gamma_{d,\ZAMO}^{-1}$ compared with the ZAMO
frame.  The random motion of the pairs in the disk frame
is found self-consistently to be sub-relativistic, and so we adopt the 
annihilation cross section for cold pairs, 
$\langle \sigma_{\rm ann} \Delta v\rangle = c \pi r_e^2$ \citep{Berestetskii1971}.
The rate of pair annihilation, as measured by a ZAMO, is then
\be
    \frac{dn_{\gamma \gamma}}{dt_\ZAMO} = 
    n_{e^+} n_{e^-} \langle \sigma_{\rm ann} \Delta v\rangle \gamma_{d,\ZAMO}^{-1}.
\ee
The pair density is iteratively updated between runs by choosing a small
time step $\delta t_\ZAMO$.  The update from run $i-1$ to run $i$ is given as
\begin{align}\label{eq:np_update}
    {n_{p, i} - n_{p, i-1}\over \delta t_\ZAMO} 
    & = \frac{dn_p}{dt_\ZAMO} - n_{p, i}(n_{p, i} + n_d) 
    \langle \sigma_{\rm ann} \Delta v\rangle \gamma_{d,\ZAMO}^{-1}; \nn
    \delta t_\ZAMO &= {\rm max}\left(1, \frac{r}{\rISCO}\right)\epsilon_t \int_{r_1}^{r_2} \sqrt{\gr} r dr \nn
    &\equiv {\rm max}\left(1, \frac{r}{\rISCO}\right) \epsilon_t t_c,
\end{align}
where $\delta t_\ZAMO$ is taken to be a fraction of the radial light crossing time across the smallest cell in the ZAMO frame. Cells outside the ISCO are given a larger timestep to allow for faster convergence towards an equilibrium solution; inside the ISCO, cells must have the same timestep to allow for continuity in advection.  The updated density $n_{p,i}$ is
obtained in each spatial cell by solving the quadratic equation (\ref{eq:np_update}).
The coefficient $\epsilon_t$ is adjusted to limit the total change in 
$n_\pm$ in each radial shell.  

Interior to the ISCO, we also include the effect of radial advection, in which case
\begin{align}\label{eq:np_update2}
    {n_{p, i}^j - n_{p, i-1}^j\over \delta t_\ZAMO} 
    &= \frac{dn_p}{dt_\ZAMO} - n_{p, i}(n_{p, i} + n_d) 
    \langle \sigma_{\rm ann} \Delta v\rangle \gamma_{d,\ZAMO}^{-1} \nn
    & + \frac{\beta_{d, \ZAMO}^r (r_1) n^j_{p, i} A^j_{r_1} - \beta_{d, \ZAMO}^r (r_2) n^{j+1}_{p, i} A^j_{r_2}}{V^j}.
\end{align}
Here $j$ labels the cell and $A_{r_1}^j (A_{r_2}^j)$ is the area of its inner (outer) radial boundary.
The quadratic equation for $n_{p,i}^j$ now depends on the solution 
for $n_{p,i}^{j+1}$ one cell out in radius.  We therefore solve Equation
(\ref{eq:np_update2}) first in the layer of cells just interior to $\rISCO$
(where $\beta_{d,\ZAMO}^r \rightarrow 0$ and the advective flow becomes small).
The remaining solution in the plunging zone is obtained by iteration,
moving inward to the horizon at constant $\theta$.  (The background flow is assumed
to have vanishing $u_d^\theta$; Equation (\ref{eq:inner_velocity}).)

\subsection{Determining Convergence}\label{sec:convergence}

To obtain convergence in the $e^\pm$ density distribution, we start by injecting a modest 
number $N_{\delta\beta} = 2\times 10^5$ of photons when adjusting $|\rmsBeta_\disk|$ to obtain energy
equilibrium, and $N_{\rm density} = 10^7$ photons to determine the photon distribution function
and $e^\pm$ density.  The parameter $\epsilon_t$ in Equation (\ref{eq:np_update}) is adjusted
iteratively so as to avoid a large overshoot past the converged pair density. Starting at value unity, $\epsilon_t$ is successively decreased by factors of 2 until the 
fractional change of $n_\pm$ in each radial shell is less than a limiting value $\maxdenschange$. 
Once the updated pair density field $n_{p, i}$ is set, we compare with it the equilibrium density $\nequil$, equivalent to taking $\Delta t_\ZAMO \rightarrow \infty$:
\begin{equation}
    \label{eq:equil}
    \frac{dn_p}{dt_\ZAMO} - \nequil(\nequil + n_d) \langle \sigma_{\rm ann} \Delta v\rangle \gamma_{d,\ZAMO}^{-1} = 0.
\end{equation}
Within the ISCO, the equilibrium density is again altered by advection;  this corresponds
to taking the right-hand side of Equation (\ref{eq:np_update2}) to vanish.

The maximum fractional variation $\maxdenschange$ is reduced by a factor 2 between iterations, as long as the variance in each radial shell (subscript $j$) meets the convergence criterion:
\begin{equation}
    \label{eq:criteria}
    \left\langle \sqrt{\left(1 - \frac{n_{\pm, i}}{n_{\pm, {\rm eq}}}\right)^2}\right\rangle_j \leq \frac{\maxdenschange}{2}.
\end{equation} 
Both $N_{\delta \beta}$ and $N_{\rm density}$ 
are correspondingly increased by factor 2, in order to improve statistics.  This iterative process
continues until $\maxdenschange$ drops to $\lesssim 10\%$, at which point convergence has been
achieved.  

The pair creation rate is highly sensitive to $|\rmsBeta_\disk|$, as photons on the high-energy
tail of the spectrum ($\omega \gtrsim m_e$) dominate pair creation. In zones where the pair density exceeds the equilibrium density, relaxation via pair annihilation can be relatively slow; to accelerate this process,
the updated pair density is replaced with its average with the equilibrium density.

\begin{figure}[t]
    \centering
    \includegraphics[width=\linewidth]{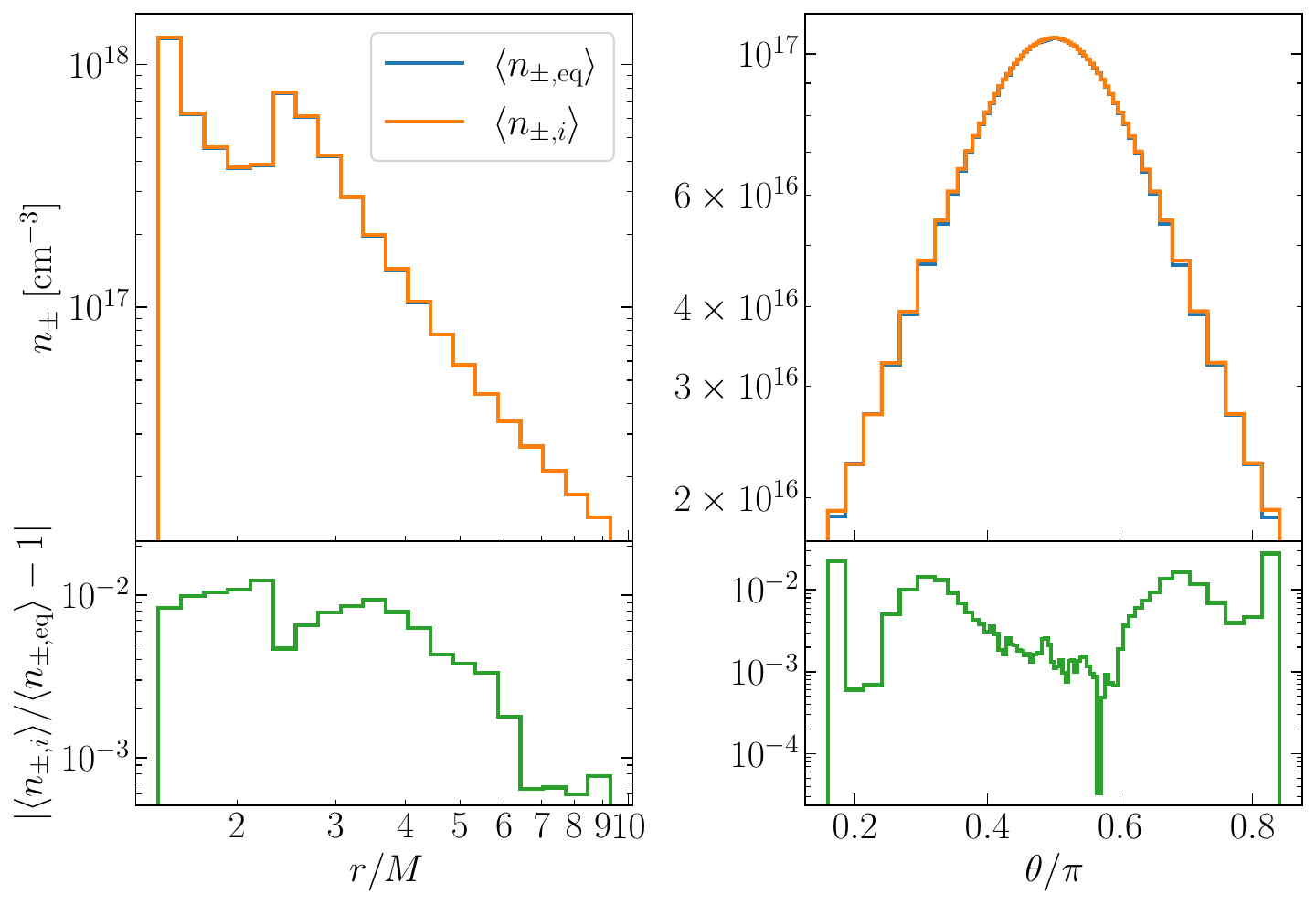}
    \caption{Top: comparison of converged $e^\pm$ density (orange) with projected equilibrium value (blue),
    both averaged over spherical shells (left) and over latitudinal bands out to a radius $\rCorona$ (right).
    Bottom:  percent difference between the two curves in each panel. Comparisons are presented for model $\simbase$.}
    \label{fig:diff}
\end{figure}

To test the convergence of the MC calculations, the pair density can be compared to the equilibrium density (\ref{eq:equil}).  Averages over radial and poloidal slices are compared for model $\simbase$ in Figure \ref{fig:diff}.
The highest-density zones which dominate photon emission and scattering are converged to within a few percent; the spectra and the polarimetric data are then stable. MC runs reach a converged density profile within $\sim 30-50$ iterations, taking a total of $\sim 50 000$ CPU-hours; an additional $50 000$ CPU-hours are used for a final production run.

\begin{figure}
    \centering
    \includegraphics[width=\linewidth]{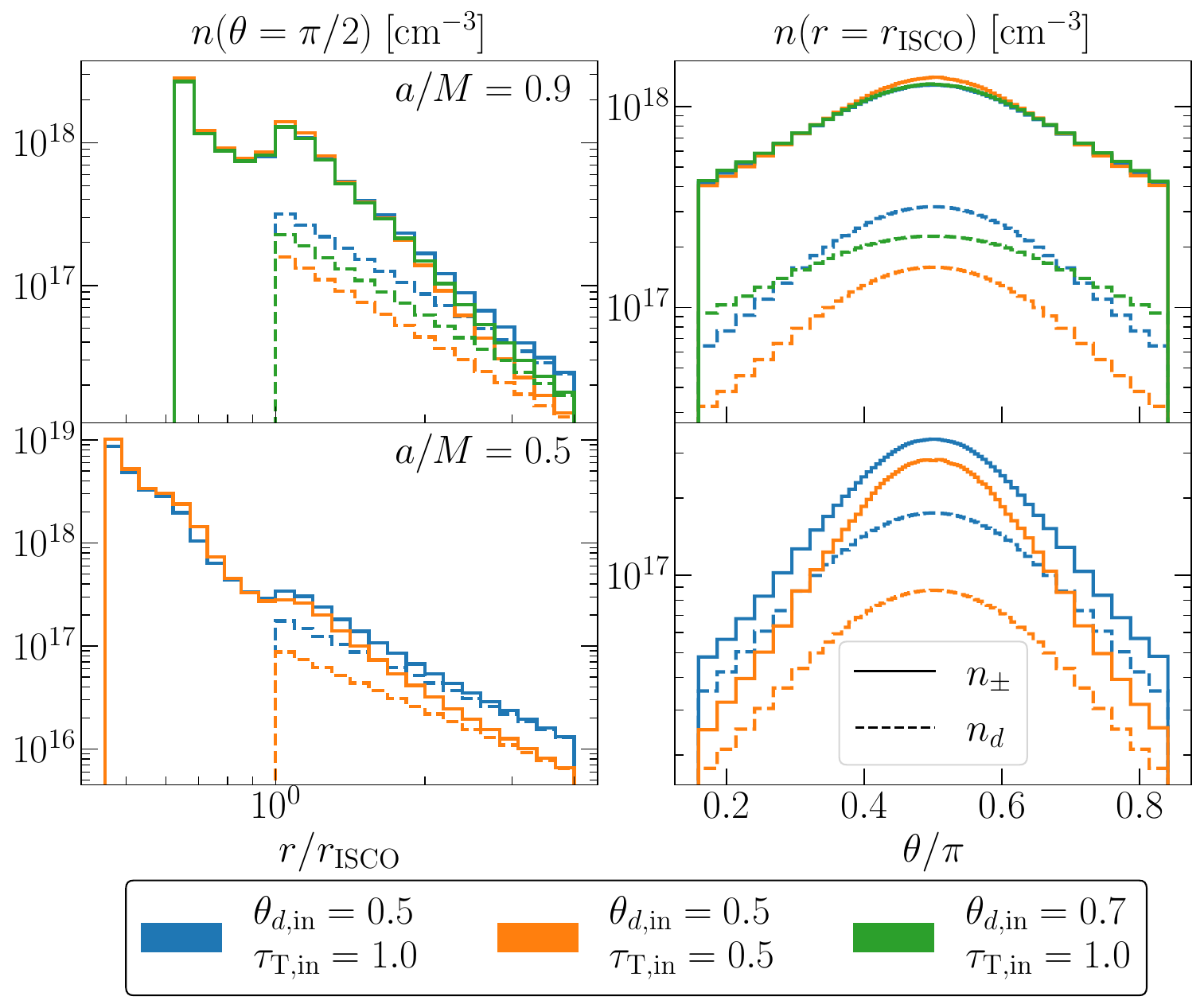}
    \caption{Effect of adjusting the input electron-ion density on the converged $e^\pm$ 
    distribution. Orange and green curves vary one parameter
    of the electron-ion profile away from the default choice (blue curves) in models $\simbase$ (top row) and $\simlowspin$ (bottom row).  
    Orange: larger disk scale height.  Green: lower input optical depth. Left column: radial distribution at the mid-plane. Right column: latitude distribution at the ISCO. The converged distribution for $a/M = 0.9$ models shows only weak dependence on the input electron-ion profile; $a/M = 0.5$ models are more sensitive as the bulk of pair production is deeper inside the ISCO.}
    \label{fig:disk_tests}
\end{figure}

\subsection{Dependence of $e^\pm$ Distribution on Ion Profile}
\label{sec:disk_test}

The $e^\pm$ profile around the BH relaxes to a characteristic shape that is found to depend weakly on the
distribution of seed electron-ion gas.   Figure \ref{fig:disk_tests} shows the effect of varying
the scale height and column density of the ion disk while fixing (i) the soft X-ray luminosity $L_{\rm seed}
= 10^{37}$ erg s$^{-1}$ and (ii) the amplification factor $A = L_\infty/L_{\rm seed} = 10$.

In model $\simbase$, where stochastic $e^\pm$ motions are only excited in the coronal disk
zone ($\rISCO \leq r \leq \rCorona$), the relaxed $e^\pm$ distribution hardly changes (less than a few
percent) when the input column or scale height
is varied by $30-50\%$. As a consequence, the observer-dependent spectra and polarization are similarly invariant. 
In the lower-spin model $\simlowspin$, soft photon emission and $e^\pm$ velocity dispersion 
are both excited inside the ISCO.
The $e^\pm$ density then peaks more strongly in the plunging zone, and the pair profile around the ISCO is somewhat more sensitive to the input electron-ion profile.

In this way, Compton heating follows the evolving pair profile.  This effect follows from
our simplified choice of uniform $|\rmsBeta_\disk|$ in the bulk of the corona.
The physical motivation here is that $e^\pm$ pair creation enhances the coupling of the photon
field to MHD wave motions and bulk orbital flows in zones where direct 
kinetic heating of $e^\pm$ is inefficient \citep{Thompson1994,Socrates2004,Beloborodov2017}.

\section{Observational Diagnostics}\label{sec:obs_results}

We now describe several ways in which the X-ray continuum spectrum and polarization may be influenced by 
(i) the BH spin;  (ii) the presence or absence of heating inside the ISCO;  (iii) the spatial distribution of 
scattering charges;  (iv) the scattering of X-rays off an outer, cold disk;
and (v) the radiative process seeding soft X-rays in the corona 
(electron-ion bremsstrahlung versus cyclo-synchrotron emission).

\subsection{Dependence of X-ray Spectrum on BH Spin} 
\label{sec:Xspec}

The X-ray spectra shown in Figure \ref{fig:spec_data} were obtained for models with very
different spin ($a/M = 0.5$ vs. $0.9$) but an identical luminosity ($L = 0.01\,L_{\rm edd}$)
and amplification relative to the seed keV luminosity ($A = 10$).
These spectra have similar shapes and dependence on observer
orientation:  each produces a power-law spectrum with high-energy cutoff that is
typical of the BHXRB hard state.  This is a manifestation of the strong self-regulation of
a Comptonizing $e^\pm$ corona.
If plasma microphysics demand a certain ratio of soft X-ray emission to global
dissipation, and if the coronal plasma is dominated by pairs, then the output spectrum depends weakly
on the depth of the gravitational potential in which the corona sits.  

\begin{figure}[t]
    \centering
    \includegraphics[width=0.9\linewidth]{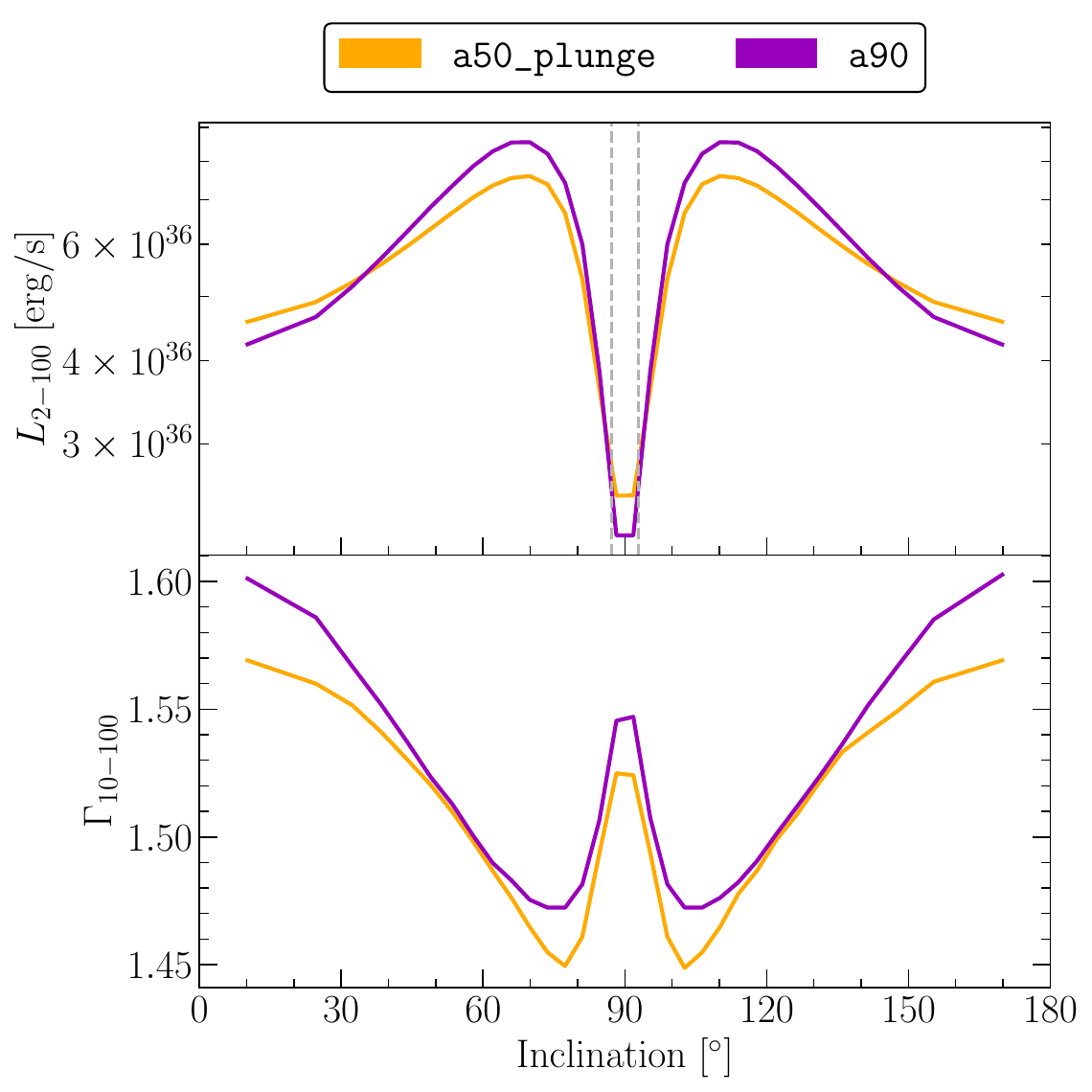}
    \caption{$2-100$ keV luminosity (top) and fitted $10-100$ keV photon index (bottom)
    as a function of observer-disk inclination.  Purple curves: high BH spin
    (model $\simbase$).  Orange curves: moderate BH spin (model $\simlowspin$). 
    Scattering by the cold, outer disk (aspect ratio marked by dashed grey lines) suppresses the luminosity for equatorial observers, and softens the spectrum.}
    \label{fig:spec_fits}
\end{figure}

We now quantify in greater detail the changes in output spectrum that do emerge from variations
in the BH spacetime (Figure \ref{fig:spec_fits}).  These are generally smaller than those
arising from a change in luminosity (Section \ref{sec:coronalum} and Figure \ref{fig:accretion_rates}).  
The lower-spin model $\simlowspin$ shows slightly less variation in apparent luminosity and 
peak energy between polar and equatorial inclinations,
a result of weaker beaming of the emission from the more slowly rotating disk.

\begin{figure}
    \centering
    \includegraphics[width=\linewidth]{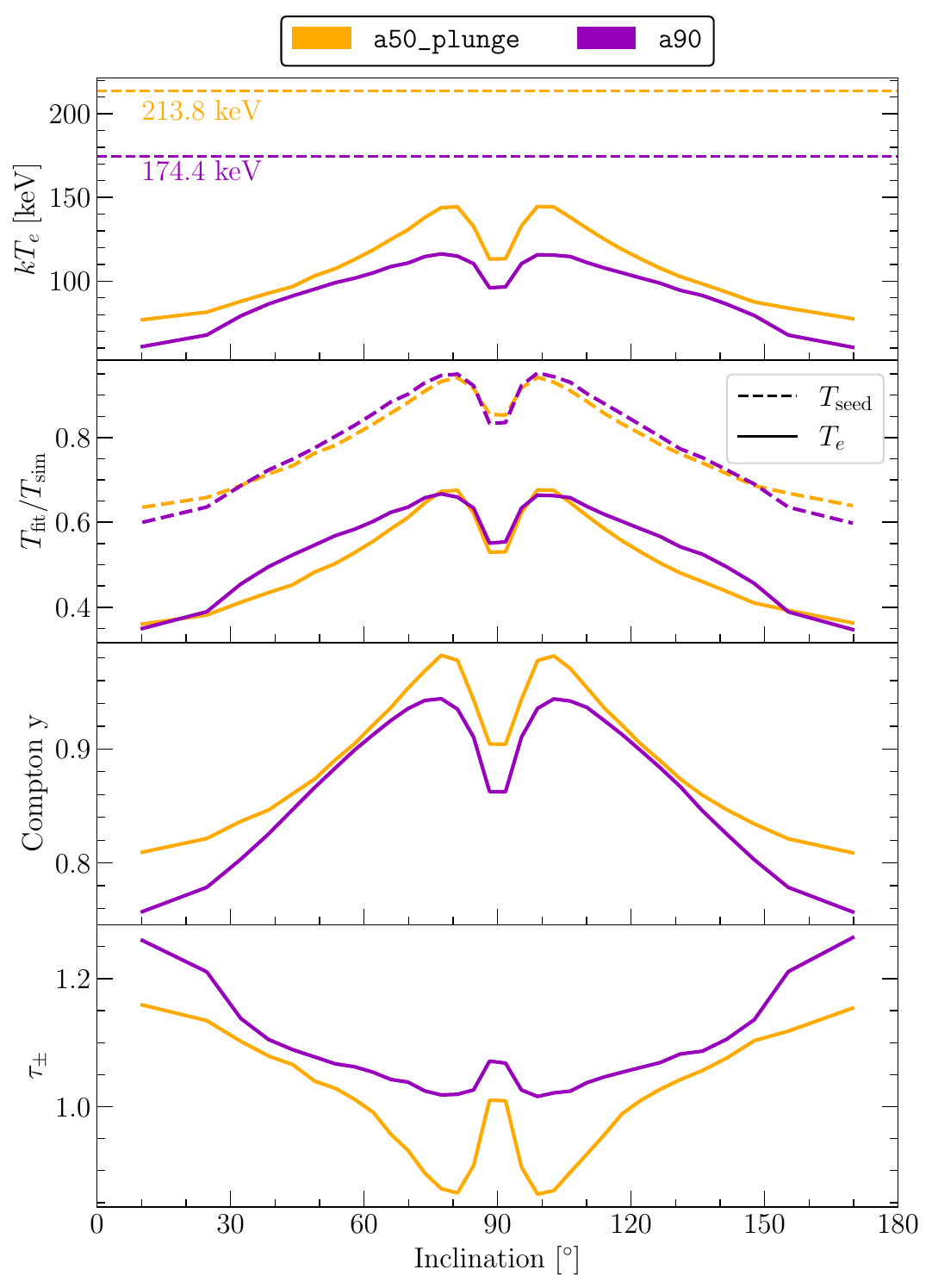}
    \caption{Parameters of the Comptonization model fitted to MC spectral data using the
    $\tt COMPPS$ package and assuming a slab geometry.  Purple lines:  high BH spin
    (model $\simbase$).  Orange lines: moderate BH spin (model $\simlowspin$). 
    Top row:  fitted $e^\pm$ temperature is compared with the local temperature derived
    from the converged random $e^\pm$ velocity $|\rmsBeta_\disk|$ (dashed lines;  
    Equation \ref{eq:sim_temp}). Second row:  ratio of {\tt COMPSS} temperature parameters to disk-frame
    seed photon temperature (dashed) and local coronal temperature (solid). 
    Third row:  fitted Compton $y$-parameter.  Bottom row:  optical depth obtained
    by combining fitted values of $y$ and $T_e$ (Equation (\ref{eq:optical_fit})).}
    \label{fig:Xspec}
\end{figure}

To compare more directly to observations,
the spectra are fitted in each inclination bin using the {\tt Xspec} {\tt COMPSS} package 
\citep{Poutanen1996}.  In these fits, we adopt a slab geometry for the corona and nominal values
for the following {\tt COMPSS} parameters:
inclination cosine $\mu = 0.5$ and reflection parameter $R = 0$.  (The fits were found to be insensitive to $\mu$ and $R$.)
The coronal temperature $T_e$, seed photon temperature $T_{\rm seed}$, and Compton $y$ parameter are 
individually fitted as a function of inclination (Figure \ref{fig:Xspec}). 
The effective slab optical depth $\tau_\pm$ is derived from the fitted parameters using
\begin{equation}
    \label{eq:optical_fit}
    {\rm max}(\tau_\pm, \tau_\pm^2) = \frac{y m_e}{4 k_{\rm B}T_e}.
\end{equation}
The fitted value of $T_e$ is also compared  with the
temperature corresponding to r.m.s. thermal speed $\rmsBeta_\disk$, 
\begin{equation}
    \label{eq:sim_temp}
    \frac{3}{2}k_{\rm B}T_{\rm sim} = (\delta\gamma_\disk - 1) m_e; \quad 
    \delta\gamma_\disk \equiv (1 - |\rmsBeta_\disk|^2)^{-1/2}.
\end{equation}

The spectral cutoff depends on a combination of factors:  most prominently Eddington ratio
and observer orientation, and secondarily BH spin.
The fitted coronal temperature is systematically higher by $\sim 20\%$ in the low-spin model $\simlowspin$
than in model $\simbase$, with a compensating reduction in $\tau_\pm$.
But this difference is smaller than the 
one arising from a change in Eddington ratio:  Figure \ref{fig:accretion_rates} shows a factor 
$\sim 40\%$ reduction in $T_e$ for a factor 3 increase in $L/L_{\rm edd}$.
Furthermore, $T_e$ and apparent luminosity correlate positively as the observer's latitude is varied,
showing a factor $\sim 0.6$ reduction from equator to pole.  This is the expected effect of Doppler 
blueshifting (redshifting) for equatorial (polar) observers, but opposite in sign to the anti-correlation 
observed between $T_e$ and angle-averaged luminosity.  The implication is that $T_e$ is a measure of
BH spin only to the extent that the orientation of the observer and the bolometric accretion rate can
be independently constrained.

\subsection{Last Scattering Distribution and Observables}

\begin{figure}[t]
    \centering
    \includegraphics[width=\linewidth]{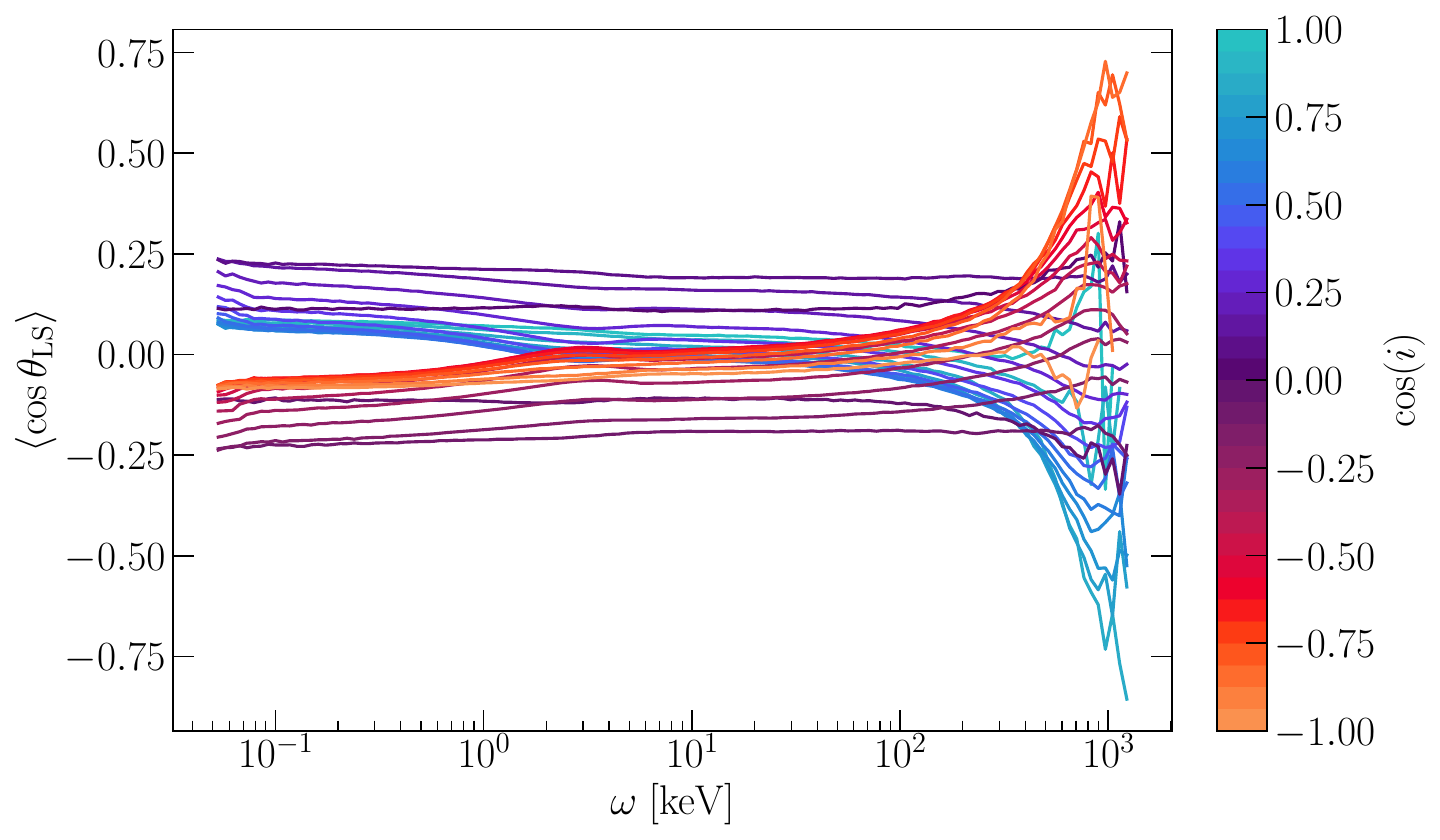}
    \caption{Average height  of last scattering of escaping photons in model $\simbase$, measured in
    terms of the polar angle $\theta_{\rm LS}$ and as a function of detected frequency.
    Colors label the observer-disk inclination.  Observers near the BH equator preferentially 
    see photons scattered at greater heights, due to occultation by the outer, cold disk.}
    \label{fig:scatter_height}
\end{figure}

Tracking the distribution of last scatterings of escaping photons
gives some insight into the dependence of spectrum and polarization on observer orientation.
In the presence of a cold, outer disk, photons reaching an equatorial observer 
are last scattered at higher latitudes $\theta_{LS}$
(see Figure \ref{fig:scatter_height} for model $\simbase$).  This effect is slightly weakened for high energy photons, due to a reduced optical depth.

\begin{figure}
    \centering
    \includegraphics[width=\linewidth]{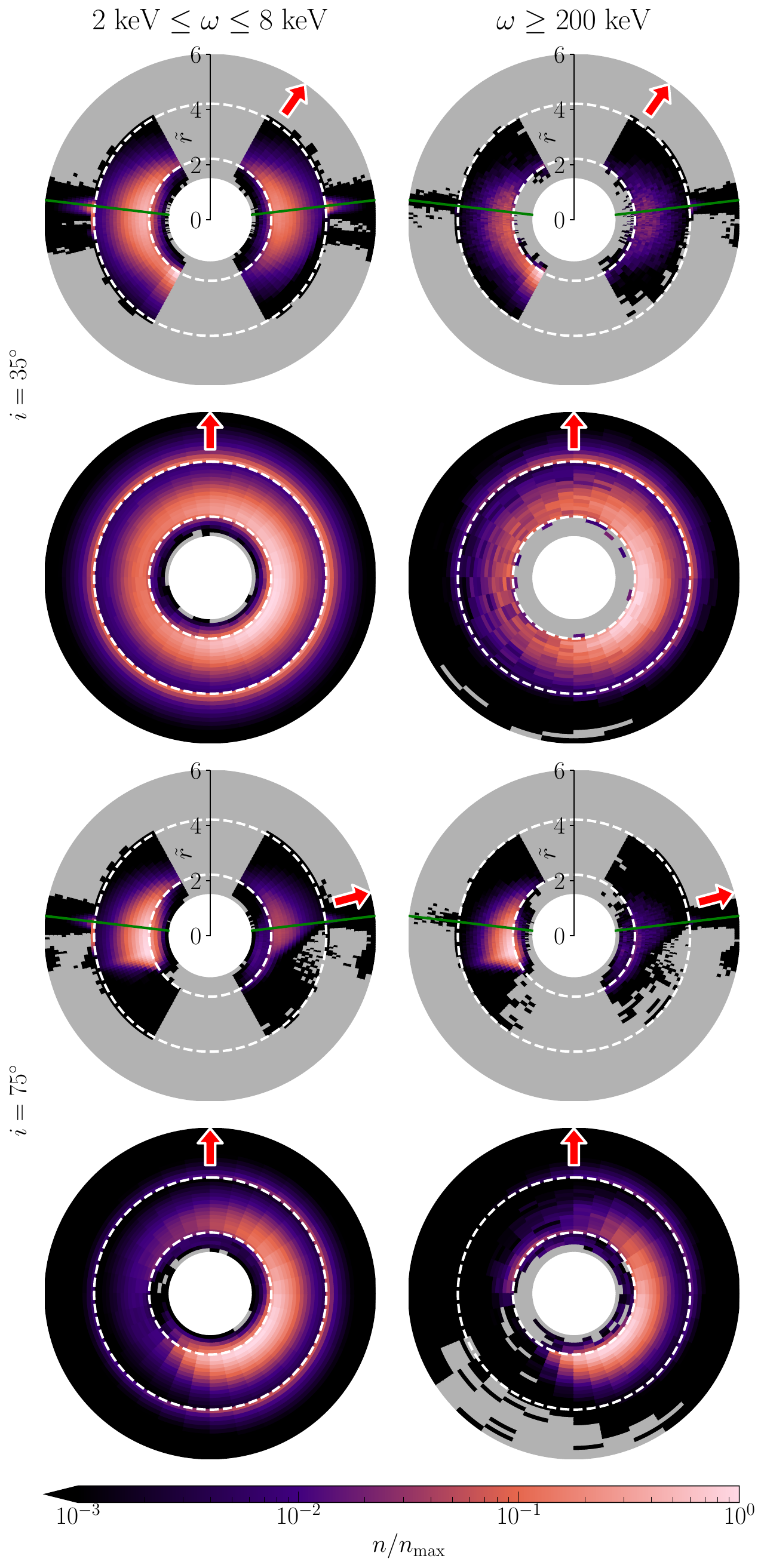}
    \caption{Density of last scattering events in model $\simbase$, for photons detected with energy $2 \; {\rm keV} \leq \omega \leq 8 \; {\rm keV}$ (left column) and $\omega \geq 200 \; {\rm keV}$ (right column).  The value in each cell is normalized by the maximum cell value.  First two rows:  observer positioned near the pole (marked by the red arrow), corresponding to an observer-disk inclination $i = 35^\circ$.   Last two rows:  observer position near the BH equator (inclination $i = 75^\circ$).  First and third rows show poloidal slices of the distribution of last scattering;  second and last rows show a projection onto the equatorial plane of the BH, with data drawn from surfaces of polar angle $\theta = 83^\circ$ (marked by the green lines in the poloidal profiles). Slices are shown with log radial scaling $\tilde{r} = \log_2(r/M) + 1$.}
    \label{fig:scattering}
\end{figure}

Polar observers are not 
so obstructed.  In the absence of disk outflow, 
the average height of last scattering shifts closer to the disk mid-plane, where
the $e^\pm$ density peaks; this is seen more clearly in the last scattering distribution shown in Figure \ref{fig:scattering}. In contrast, the scattering distribution is significantly skewed for the mid-plane observer.
Another notable feature seen in both Figures \ref{fig:scatter_height} and \ref{fig:scattering} 
is that the highest energy photons collected by
polar observers are sourced from the opposing hemisphere.   This effect carries the imprint of strong gravitational
scattering;  a high-energy photon (which tends to be emitted along the bulk disk flow)
will preferentially reach the poles by gravitational deflection.
(Note that our model excludes the possibility of direct emission from an axial jet, and
scattering is cut off at polar angles smaller than $\theta_{\rm jet}$;  Table \ref{tab:parameters}.) General relativistic effects are also clearly shown in the equatorial scattering distribution, which is asymmetric due to beaming from the bulk flow and BH spin.

\subsection{Influence of the Cold Disk}
\label{sec:disk_effects}

\begin{figure}[t]
    \centering
    \includegraphics[width=0.9\linewidth]{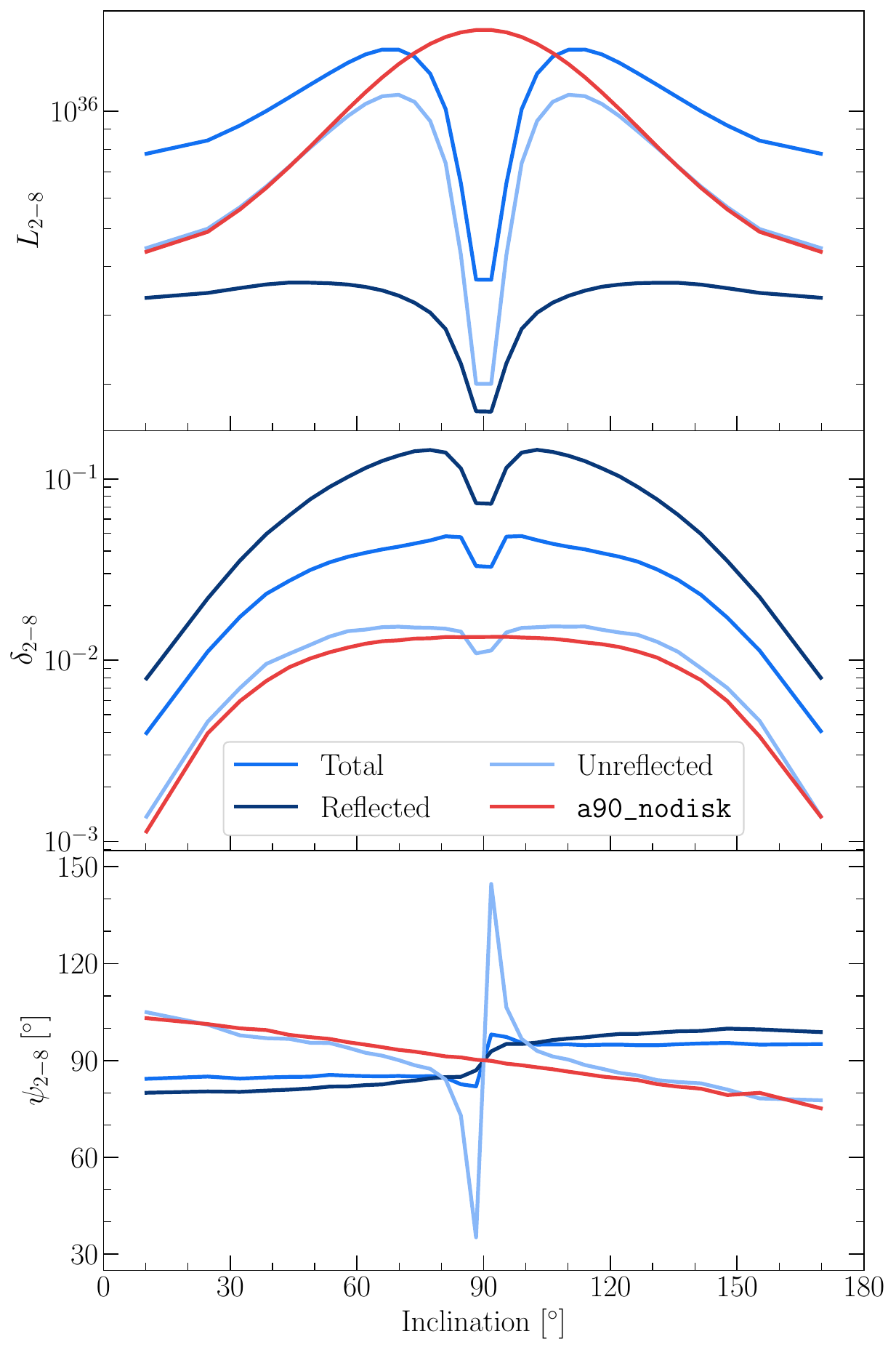}
    \caption{Relative contribution to $2-8$ keV band quantities of photons that have (have not) reflected off
    the outer, cold disk:  luminosity (top panel), polarization degree (middle panel)
    and polarization angle (bottom panel).  All quantities are plotted versus the relative 
    observer-disk inclination.  Blue curves show model $\simbase$ with outer, cold disk; 
    red curves show the effect of removing the outer disk (model $\simnodisk$).}
    \label{fig:reflection}
\end{figure}

\begin{figure}[t]
    \centering
    \includegraphics[width=\linewidth]{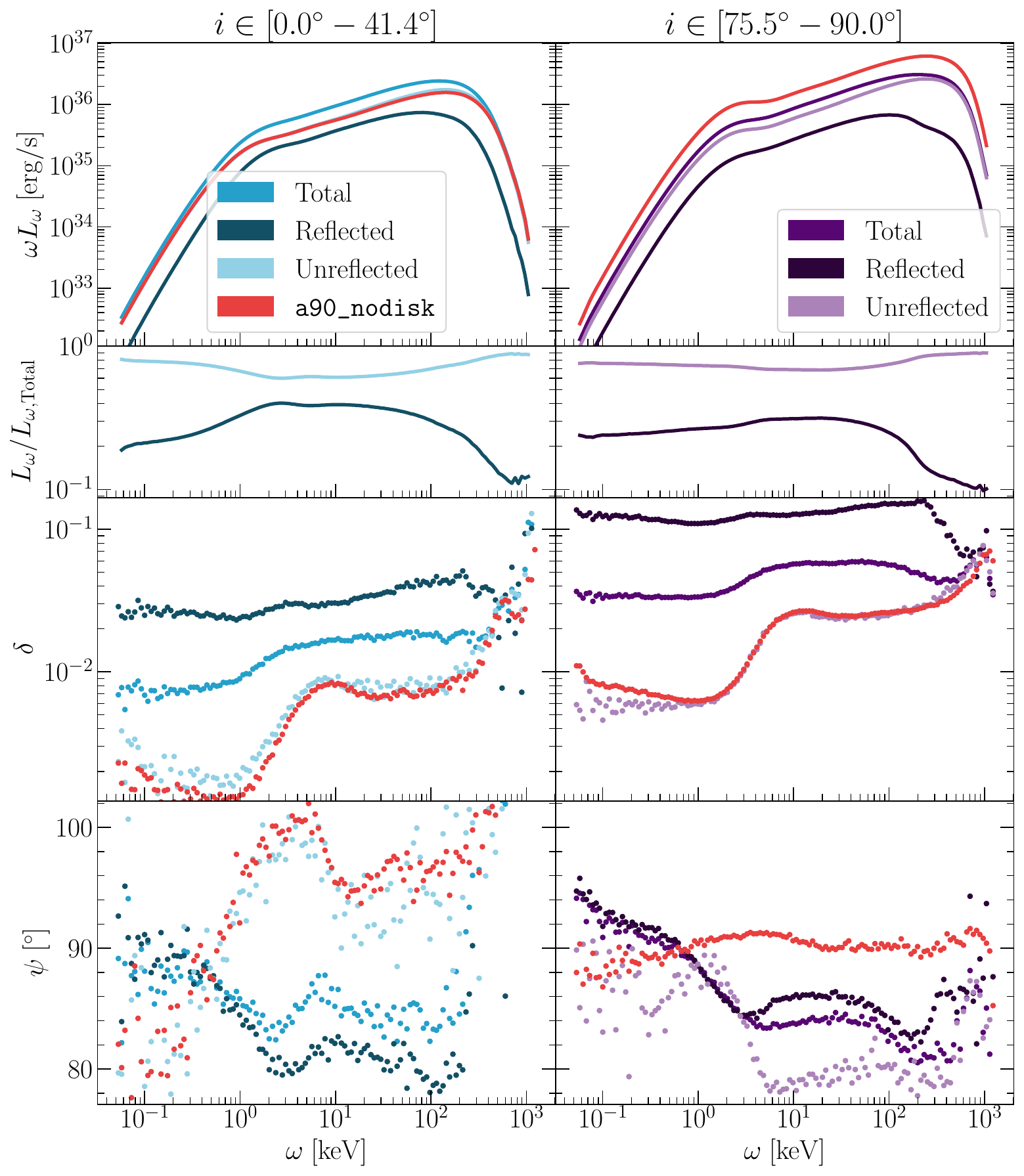}
    \caption{Spectropolarimetric data from model $\simbase$, for observers near the pole (left) and BH equator (right) as a function of detected photon energy.  Results are further split 
    (dark/light) into photons that have/have not reflected off the outer, cold disk.
    Red curves/points show the effect of removing the cold disk (model $\simnodisk$).}
    \label{fig:reflect_omega}
\end{figure}

The influence of the outer, cold disk on the measured X-ray spectrum and polarization may be explored
by removing the outer disk as a source of scattering particles.
Figures \ref{fig:reflection} and \ref{fig:reflect_omega} show the resulting model $\simnodisk$
for spin $a/M = 0.9$.
There is a negligible effect on the pair distribution in the coronal disk ($r < \rCorona$), 
but the luminosity received by equatorial observers is no longer strongly suppressed (compare
Figure \ref{fig:spec_fits}).
We separate out unreflected photons in the baseline model $\simbase$, and there remains
a clear difference with model $\simnodisk$. 

There are significant implications for polarization.  Photons emitted from the inner corona and
then reflected off an outer, cold disk (with negligible thermal and turbulent motion $\rmsBeta_\disk$) 
arrive with an order-of-magnitude higher $\delta$.
Although these reflected photons contribute $\sim 10-40\%$ of the X-ray flux in model $\simbase$
(depending on frequency and observer orientation)
they increase the net polarization by more than a factor $\sim 2$.  This effect is absent in model
$\simnodisk$:  see the middle panel of Figure \ref{fig:reflection}.  

The polarization angle also has a flatter distribution near the equator in the absence of the cold disk.  Near the equator, the $\simbase$ model shows an extreme swing in $2-8$ keV polarization angle as 
detectable non-thermal photons were forced to scatter sideways at altitude toward the observer. 

These trends extend beyond the $2-8$ keV energy range, as is seen in Figure \ref{fig:reflect_omega}.
In the absence of a cold disk, photons above 200 keV energy show a strong increase in polarization.  This is caused
by a shrinkage in the scattering region (Figure \ref{fig:scattering}), 
resulting in a reduced spread in polarization angle.
In the presence of a cold disk, equatorial observers see a slight decrease in net polarization with photon energy in the range
$100-200$ keV, even though reflected and unreflected components both have increasing polarization; this is caused by a reduction in the highly polarized reflected component. 

\subsection{Seeding Mechanism for Soft X-rays}\label{sec:seeding}

We have considered two distinct prescriptions for seeding soft X-rays around the BH (Section \ref{sec:emission_space}).
The probability of injection either tracks the $e^\pm$ density field (the default choice), 
or alternatively the seed electron-ion disk.  
These prescriptions may be connected with distinct physical processes:  cyclo-synchrotron emission
by a subset of relativistic pairs, or electron-ion bremsstrahlung by a cool subcomponent of
the accretion flow.   Although this correspondence must necessarily be rough, the spatial distribution
of the injected keV photons is very different in the two cases, with interesting implications
for the $e^\pm$ distribution and the X-ray spectrum.

In model $\simdisk$, soft photon emission is restricted to the zone outside the ISCO and follows
the electron-ion density as $n_{d,\disk}^2$.  The emission is less concentrated
in radius than in model $\simlowspin$, where it follows the inward peak in the 
$e^\pm$ cloud.
The emission also has a broader profile outside the ISCO than in model 
$\simbase$, reflecting the flatter radial distribution of the ions in the 
coronal disk.

\begin{figure}
    \centering
    \includegraphics[width=\linewidth]{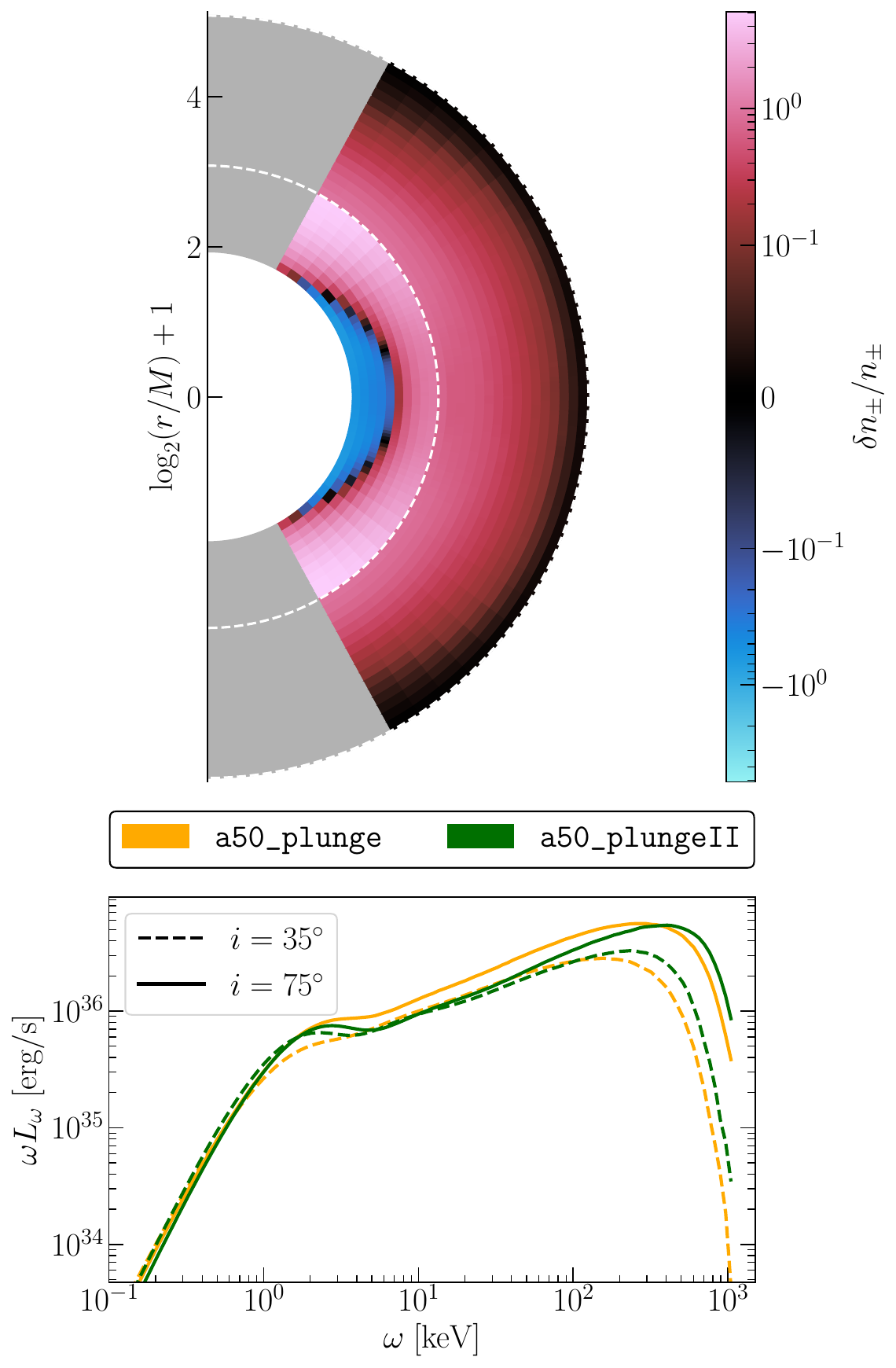}
    \caption{Changes in pair density profile and photon spectrum resulting from different seeding mechanisms
    for soft X-ray photons.  Top panel:  relative change $\delta n_\pm/n_\pm$ 
    in the converged $e^\pm$ density (Equation (\ref{eq:disk_dens_change})) when the soft photon emissivity
    tracks the input electron-ion distribution (model $\simdisk$) as opposed to the $e^\pm$ distribution 
    (model $\simlowspin$; our default choice).  Bottom panel shows the corresponding X-ray spectra as measured
    at two observer-disk inclinations.}
    \label{fig:disk_changes}
\end{figure}

Figure \ref{fig:disk_changes} shows the effect on the X-ray spectrum and the fractional change in $e^\pm$ density,
\begin{equation}
    \label{eq:disk_dens_change}
    \frac{\delta n_{\pm}}{n_\pm} \equiv \frac{n_{\pm, \simdisk} - n_{\pm, \simlowspin}}
    {n_{\pm, \simlowspin}}.
\end{equation}
The centroid of the pair cloud shifts outward as the seed photon emission
is pushed beyond the ISCO.
The seed thermal peak survives more prominently in model $\simdisk$ for observers of all orientations;
this is because the non-thermal spectral tail forms farther from the BH, where
the scattering depth is lower.
To maintain the fixed amplification $A$ in model $\simdisk$, the soft photons that do scatter must experience a higher Compton parameter so as to absorb the same net energy, producing a harder spectrum between 1 and 100 keV. Interestingly, 
the {\tt Xspec} fit is now statistically poorer, because the coronal emission
is spread over a wider range of radius and density.
The polarization data show little change in comparison with model $\simlowspin$, and are not shown.

\section{Implications for BHXRBs}
\label{sec:conclusion}

This paper presents a fully general-relativistic MC model of a pair-dominated corona near a 
slowly accreting Kerr BH.  It gives a qualitative sense of how a pair-dominated corona
may develop self-consistently around the ISCO, and a quantitative measure of how
a specific coronal geometry connects with spectropolarimetric data.

Pair creation is tied self-consistently to the formation of a non-thermal X-ray continuum;
the observer-dependent photon flux, spectra, and polarization are computed.  We
obtain a self-consistent description of energy and pair creation/annihilation equilibrium.
The $e^\pm$ density field is strongly inhomogeneous and is seen to decouple significantly 
from the ion distribution.  This process can partially be motivated by an increase in 
Compton drag in zones where otherwise there is limited direct kinetic heating of plasma particles.
Explicit magnetohydrodynamic effects are absent from the model.

Some key outcomes of the model are as follows.

\begin{enumerate}[wide, labelwidth=!, labelindent=0pt]
    \item The density of $e^\pm$ creating photons is modulated by gravitational focusing and redshift;
    the treatment of electron-photon scattering is fully relativistic.  The $e^\pm$ density field 
    extends deeper into the gravitational potential of the BH than the seed coronal ion disk.  
    As a result, even in an accretion state of low luminosity ($L \sim 0.01\,L_{\rm edd})$, the 
    hard-photon compactness is high enough to support a significant scattering depth in $e^\pm$.
    \item  The principal conjecture underlying the model -- one with broad support from observation --
    is that a modest $\sim 10\%$ of the accretion energy is released in the form of
    quasi-thermal $0.1-1$ keV X-rays.  Our baseline model ties this soft emission to the distribution of $e^\pm$,
    and so by implication identifies it with cyclo-synchrotron radiation by a fraction
    of the $e^\pm$ that reach
    relativistic energies. The remaining $90\%$ of the energy is released more gradually (e.g. as
    hydromagnetic motions and kinetic heating).  In most of the coronal volume, the amplitude of
    stochastic $e^\pm$ motions is continuously
    regulated by Compton energy loss.  The MC method self-consistently accounts for heating by orbital shear.
    
    \item The X-ray spectrum of a BHXRB in the low-hard state emerges naturally from 
    multiple Compton scattering, consistent with much previous modelling.
    The spectrum depends weakly
    on the BH spin, but shows somewhat stronger variation with observer orientation and accretion rate.
    The $e^\pm$ scattering depth self-regulates to produce an effective temperature $T_e \sim 50-150$ keV.  
    While the coronal plasma could be composed mainly
    of electrons and ions, more fine-tuning of the disk column might then be needed to match the 
    observed low-hard state spectra over 2 or more decades in luminosity.  

    \item  Relativistic gravity has measureable effects. 
    The X-ray spectrum varies with observer inclination due to beaming by
    bulk orbital motions and also gravitational deflection, with higher-spin models showing 
    a greater variation with inclination.
    The spectral cutoff is redshifted compared with the local $e^\pm$ kinetic temperature and
    the net polarization angle and polarization degree can shift dramatically due to parallel transport.  

    \item The dependence of the measured X-ray polarization on photon frequency is 
    broadly consistent with observation. 
    Outflow of the $e^\pm$, which we implement in a simplified manner, can increase the
    measured polarization to the level observed in hard-state BHXRBs.  As shown in Section \ref{sec:outflow}, 
    this can also enhance the $e^\pm$ density away from the disk mid-plane.

    \item The outer, cold disk leaves a noticeable imprint on the observer-dependent X-ray spectrum and
    polarization.  Occlusion by the outer disk causes a significant softening of both luminosity and
    spectral index for equatorial observers;  reflection by free electron scattering
    is a significant source of highly polarized photons.
\end{enumerate}

\subsection{Comparison to Observation and Future Extensions}

The X-ray spectra, \texttt{Xspec} fits, and polarization produced by the MC model are
all broadly agree with observations of BHXRBs in the low-hard state. A photon index 
$\Gamma \sim 1.6$ is consistent with the chosen luminosity of the BH; the negative correlation 
between coronal temperature and luminosity is also observationally supported \citep{Yan2020, You2023}. The choice of bolometric amplification factor $A = 10$ matches well with the values fitted to
more limited bandwiths; compare Figure \ref{fig:accretion_rates} with 
Figure 7 of \cite{Burke2017}.
The polarization angle is mostly aligned with the spin axis, in line with IXPE measurements \citep{Krawczynski2022}. Increases in the polarization degree with energy have also been observed \citep{Podgorny2024, Ewing2025}; high polarization degree measurements may be accommodated by a vertical outflow.

The MC model only follows the development of the X-ray continuum and ignores all effects of
photoelectric absorption and line emission.  Although the input electron-ion profile is schematic, the properties of the
corona are relatively insensitive to the choice of profile, as the positron density dominates the ion density
where the hard X-ray continuum is generated.  The degree of $e^\pm$ domination is expected to
diminish as the accretion luminosity rises above the default value $L = 0.01\,L_{\rm edd}$, so
that the background ion profile will begin to have a stronger effect on the $e^\pm$ distribution.
The effects of emission or scattering within a jet are ignored, and the jet zone is excised from
the computational domain.  One or more of these physical effects could realistically be included
in this MC procedure in future work.

\section*{Acknowledgments}

The authors thank Michael Grehan, Gibwa Musoke, Bart Ripperda, and Aryanna Schiebelbein-Zwack for feedback and insightful comments.
We acknowledge the support of the Natural Sciences and
Engineering Research Council of Canada, funding reference number RGPIN-2023-04612, 
and of the Simons Foundation (MP-SCMPS-00001470). 
The computational resources and services used in this work were partially provided by facilities supported by the VSC (Flemish Supercomputer Center), funded by the Research Foundation Flanders (FWO) and the Flemish Government – department EWI, and by Compute Ontario and the Digital Research Alliance of Canada (alliancecan.ca), compute allocation rrg-thompsoc.

\newpage

\appendix
\section{Monte Carlo Grid} \label{app:grid}
The grid is constructed in spherical coordinates ($r,\theta,\phi)$.
Cells are spaced log-uniformly in $r$ while allowing for separate spacings in the plunging zone
($r_H < r < \rISCO$), coronal disk ($\rISCO < r < \rCorona$) and the outer thin disk ($\rCorona < r < \rmax$).
The disk density profile sharply changes across the boundary radii $\{\rISCO, \rCorona \}$. 
The number of cells in each zone is
\begin{align}
    N_{r, {\rm inner}} &= \left \lceil N_r \frac{\log(\rISCO) - \log(r_H)}{\log(\rmax) - \log(r_H)} \right \rceil \nn 
    N_{r, {\rm corona}} &= \left \lceil N_r \frac{\log(\rCorona) - \log(\rISCO)}{\log(\rmax) - \log(r_H)} \right \rceil \nn 
    N_{r, {\rm outer}} &= N_r - N_{r, {\rm corona}} - N_{r, {\rm inner}}
\end{align}
for a total number $N_r$ of radial cells.  This choice maximizes the resolution in the inner two zones.

\begin{figure}[ht]
    \centering
    \includegraphics[width=0.4\linewidth]{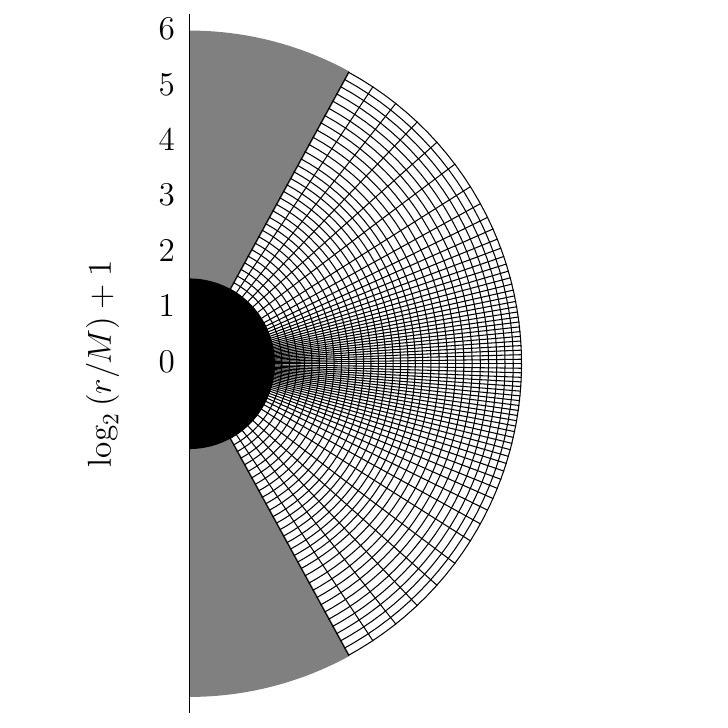}
    \caption{Poloidal slice of the computational MC grid for grid disk thickness $\theta_{d, {\rm grid}} = 0.2$, $N_r = 32$, and $N_\theta = 64$.}
    \label{fig:grid}
\end{figure}

The grid is uniformly divided in azimuth ($\Delta \phi = 2\pi / N_\phi$).  The division of cells in the $\theta$-direction
is calibrated to the density profile in Equation (\ref{eq:rho}), now choosing a scale height $\theta_{d, {\rm grid}} = 0.2$
to accommodate the peaking of the $e^\pm$ profile at the mid-plane.  The cell width is adjusted to give a uniform
column density $n_\pm \sqrt{g_{\theta\theta}}\Delta\theta$ near the equator, while limiting the width to $\Delta\theta \leq \Delta\theta_{\rm max}$ 
off the equator.  The cell distribution so defined is adjusted iteratively to give a fixed number $N_\theta$ of cells.
In addition, a zone $(\theta,\;\pi-\theta) < \theta_{\rm jet}$ is excised around the poles, given the possible presence of 
a dynamically distinct magnetorotational flow connected to the BH.

\subsection{Photon Phase Space}

The photon phase space is fully discretized in the ZAMO frame.  
Within each spatial cell $(r,\,\theta,\,\phi)$, the photon energy $k^t_\ZAMO$ has a log-uniform binning
and the photon direction $\hat k_Z$ is tabulated in an angular coordinate system
$(\theta_k,\,\phi_k)$ that is aligned with the axis $\hat z$ of the BH spin. The conversion between angular coordinates and either cylindrical vectors $\{\hat{z}, \hat{\phi}, \hat{z} \times \hat{\phi} = -\hat{\rho} \}$ or global spherical vectors $\{\hat{r}, \hat{\theta}, \hat{\phi} \}$ is defined as:
\begin{align}\label{eq:kspace_Z}
    \hat{k}_\ZAMO &= \cos\theta_k \hat{z} + \sin\theta_k \cos\phi_k \hat{\phi} + \sin\theta_k \sin\phi_k (\hat{z} \times \hat{\phi}) \nn 
    &= \cos\theta_k \hat{z} + \sin\theta_k \cos\phi_k \hat{\phi} - \sin\theta_k \sin\phi_k \hat{\rho} \nn
    &= [\cos\theta \cos\theta_k - \sin\theta \sin\theta_k \sin\phi_k] \: \hat{r} \nn 
    &+ [-\sin\theta \cos\theta_k - \cos\theta \sin\theta_k\sin\phi_k] \: \hat{\theta} \nn
    &+ [\sin\theta_k \cos\phi_k] \: \hat{\phi}.
\end{align}
Each angular bin carries an approximately uniform solid angle, with uniform division in $\theta_k$, and the number of $\phi_k$ cells scaling as
$\lceil N_{\phi_k, \rm max} \sin\theta_k\rceil$. 
The phase-space bin $(\theta_k,\,\phi_k,\,k^t_\ZAMO)_i$ of the $i^{\rm th}$ photon then has a 
representative wavevector 
${\bm k}_{\ZAMO,i} = k^t_{\ZAMO,i}\{1,\,\hat{k}_{\ZAMO,i}(\theta_k,\,\phi_k)\}$.

The polarization density matrix is also recorded in each spatial cell and for each
propagation direction $\hat k_\ZAMO$.  It can be written in terms of the orthonormal basis
\begin{align}
    \label{eq:pol_basis}
    \epsone &= \frac{\hat{k}_\ZAMO \times \hat{\theta}}{|\hat{k}_\ZAMO \times \hat{\theta}|}; \;\quad \epstwo = \hat{k}_\ZAMO \times \epsone; \nn 
    \hat\varepsilon_{\times1,\times2} &= (\pm\epsone + \epstwo)/\sqrt{2}. 
\end{align}
This basis reduces to $\{ \epsone, \epstwo \} = \{\hat{\phi}, -\hat{\theta}\}$ for photons propagating
to null infinity.  Our focus here is on photons that are partly linearly polarized but with vanishing
circular polarization.  Then the density matrix can be written in terms of the mean Stokes 
parameters $\{\bar I,\bar Q,\bar U\}$ of all photons resident in the cell,
\begin{align}\label{eq:denmatrix}
    \rho_{\mu\nu,\ZAMO} = \frac{1}{2\bar I} \bigg[&\bar I(\hat\varepsilon_{1\mu} \hat\varepsilon_{1\nu} + 
    \hat\varepsilon_{2\mu} \hat\varepsilon_{2\nu}) 
    + \bar U(\hat\varepsilon_{1\mu} \hat\varepsilon_{2\nu} + \hat\varepsilon_{2\mu} \hat\varepsilon_{1\nu})\nn 
    &\quad + \bar Q(\hat\varepsilon_{1\mu} \hat\varepsilon_{1\nu} - \hat\varepsilon_{2\mu} \hat\varepsilon_{2\nu})\bigg]. 
\end{align}
The probability that a photon propagating in direction $\hat k_\ZAMO$ has polarization $\hat\varepsilon$
is given by $\hat\varepsilon^\mu \rho_{\mu\nu,\ZAMO}\hat\varepsilon^\nu$.  This quantity depends only on the
components of $\hat\varepsilon$ perpendicular to $\hat k_\ZAMO$.  

\subsection{Photon Assignment}\label{sec:assign}

For a photon propagating through the MC grid, an array containing the coordinates, polarization, and integrated optical depth is recorded at each increment in $\Delta \lambda$ (Equation (\ref{eq:deltalambda})). The array is used to build a spline as a function of $\lambda$; the intersection of the spline in each grid cell is determined through bisection on $\lambda$, and the difference in $t(\lambda)$ at the start and end of the path is taken as the residency time. The residency time in each cell is recorded in the ZAMO frame, and multiplied by the photon weight (Equation (\ref{eq:weight})) to give the effective time $\Delta t_\ZAMO$. In a spatial cell, the photon direction and energy at the midpoint of the trajectory is used to determine the appropriate direction and energy bin.  The Stokes parameters are summed over successive photons,
weighted by $\Delta t_\ZAMO$, the emission weight $w$, and
the emission rate of soft photons (Equation (\ref{eq:tres})), to give the polarization density matrix (\ref{eq:denmatrix}).

\subsection{Polarization Effects}

Each trial photon carries polarization data $\{\delta, \psi \}$ and basis states
$\{\varepsilon_x^\mu, \varepsilon_y^\mu\}$ which evolve under parallel transport in the 
approximation of vanishing Faraday rotation.  
The basis states may be converted to the ZAMO frame and a representative polarization constructed as
\begin{equation}
    \hat \varepsilon = \cos \psi \; \hat \varepsilon_{x, \ZAMO} + \sin \psi \; \hat \varepsilon_{y, \ZAMO}.
\end{equation}
The Stokes parameters of the single photon may be expressed in terms of the projection of 
$\hat\varepsilon$ on the basis (\ref{eq:pol_basis}) and
the polarization degree $\delta = \sqrt{Q^2 + U^2}$, 
\be
Q = \delta\cdot\bigl[(\hat\varepsilon\cdot\hat\varepsilon_1)^2 - 
      (\hat\varepsilon\cdot\hat\varepsilon_2)^2\bigr];
\quad\quad
U = 2\delta\cdot(\hat\varepsilon\cdot\hat\varepsilon_1)(\hat\varepsilon\cdot\hat\varepsilon_2).
\ee
Equivalently, we may define the parameters 
$c_i^2 \equiv 0.5(1 - \delta) + \delta(\hat\varepsilon \cdot \hat{\varepsilon}_i)^2$
($i = 1,2,\times1,\times2$), in terms of which
\begin{align}\label{eq:stokes}
    I &= c_1^2 + c_2^2 = c_{\times1}^2 + c_{\times2}^2 = 1; \nn 
    Q &= c_1^2 - c_2^2; \;\quad U = c_{\times 1}^2 - c_{\times 2}^2.
\end{align}
Summing the parameters (\ref{eq:stokes}) over successive photons that pass through a given cell,
with the weighting described in Section \ref{sec:assign}, gives total Stokes parameters $\{\bar I, \bar Q, \bar U\}$ for each angular bin.

\subsection{The External Grid}\label{sec:angular_grid}

We also construct an external angular grid to record spectral and polarization data at infinity. This grid is divided uniformly in $\cos \theta_{\rm ext}$ and $\phi_{\rm ext}$. In each angular cell,
photons are further binned in BL frame energy $\mathcal{E}$. The polarization amplitudes $c_i$ of each photon are stored, to later reconstruct the aggregate Stokes parameters. 
We also define the total energy of photons reaching infinity, $\mathcal{E}_\infty$, from which the amplification factor
$A = L_\infty/L_{\rm seed}$ is determined.  A separate array stores (in each angular cell)
the spatial cell containing the last scattering (or the emission point, if no scattering occurred). 
Taking advantage of azimuthal symmetry, we only bin in $\phi$ by the offset between last scattering and the ray trajectory at infinity.  
A second copy of these arrays is created for photons that have experienced reflection off
the cold disk at $r > \rCorona$, to allow for separation of the data into ``reflected'' and 
``unreflected'' components.

\section{Klein-Nishina Scattering}\label{sec:KN}

\label{app:scattering}

Once it has been determined that a photon scatters, the energy, direction and polarization of the outgoing photon
are readily evaluated. To determine the incoming polarization direction, we first convert the Walker-Penrose constant to 
a BL-frame polarization basis using the constraints $g_{\mu \nu} \varepsilon^\mu dx^\nu/dt = 0$ and the gauge choice $\varepsilon^t = 0$ \citep{Penrose1970, Connors1977}.
For a photon with polarization degree $\delta$ and angle $\psi$, a random number $X$ is drawn from the uniform distribution $U(0, 1)$. If $X \leq \delta$, the representative polarization is taken to be at angle $\psi$; otherwise the polarization is taken from a uniform distribution \citep{Matt1996}. 
The polarization and photon wavevector are then transformed to the ZAMO frame.

As we now describe, 
the scattering procedure is carried out in the initial rest frame of the scattering charge; 
the photon variables are then readily boosted back to the ZAMO frame. For a generic velocity in the ZAMO frame ${\bm \beta} = (\beta_r, \beta_\theta, \beta_\phi)$, the corresponding matrix to Lorentz boost four-vectors is
\begin{equation}
\begin{bmatrix}
    \gamma & -\gamma\beta_r & -\gamma \beta_\theta & -\gamma \beta_\phi \\
    -\gamma\beta_r & 1 + \frac{\gamma^2}{1+\gamma}\beta_r^2 & \frac{\gamma^2}{1+\gamma}\beta_r \beta_\theta & \frac{\gamma^2}{1+\gamma}\beta_r \beta_\phi  \\ 
    -\gamma\beta_\theta & \frac{\gamma^2}{1+\gamma}\beta_r \beta_\theta & 1 + \frac{\gamma^2}{1+\gamma}\beta_\theta^2 &\frac{\gamma^2}{1+\gamma}\beta_\theta \beta_\phi  \\ 
    -\gamma\beta_\phi & \frac{\gamma^2}{1+\gamma}\beta_r \beta_\phi & \frac{\gamma^2}{1+\gamma}\beta_\theta \beta_\phi & 1 + \frac{\gamma^2}{1+\gamma}\beta_\phi^2  \\ 
\end{bmatrix}
\end{equation}
In the particle frame, the scattering direction $\hat k_1$ may be recorded in a coordinate
system defined by the initial photon direction $\hat{k}_0$ and the orthogonal polarization $\epszero$,
\begin{equation}
    \hat{k}_1 = \cos\theta \; \hat{k}_0 + \sin\theta \cos\phi \; \epszero + \sin\theta \sin\phi \; \norm.
\end{equation}
Including polarization, the KN cross section for scattering of a photon off 
an electron at rest is \citep{Berestetskii1971}
\ba\label{eq:sigmaKN}
\frac{d\sigma}{d\Omega}(E_0, \theta) &=& \frac{r_e^2}{4}\left(\frac{E_1}{E_0}\right)^2 \left[\frac{E_1}{E_0} + \frac{E_0}{E_1} + 4(\epszero \cdot \epsone)^2 - 2\right]; \nn 
E_1(E_0, \theta) &=& \frac{E_0}{1 + (E_0/m_e)(1 - \cos \theta)}.
\ea
Here, $E_0 = k^t_0$ ($E_1 = k^t_1$) is the initial (final) photon energy and $\theta = \cos^{-1}(\hat k_0\cdot\hat k_1)$ is
the scattering angle. This cross section must be summed over the final-state polarizations.  We choose a basis
$\epsone = \cos\alpha \; \hat{\varepsilon}_{1a} + \sin\alpha \; \hat{\varepsilon}_{1b} $
for the outgoing photon polarization:
\ba 
\hat{\varepsilon}_{1a} &=& \frac{\epszero \times \kf}{|\epszero \times \kf|} \nn 
&=& \frac{ -\cos \theta \; \norm + \sin\theta \sin \phi \; \ki}{\sqrt{1 - \sin^2 \theta \cos^2 \phi}} \nn 
\hat{\varepsilon}_{1b} &=& \frac{\epszero \times \kf}{|\epszero \times \kf|} \times \kf; \nn 
&=& \frac{\sin\theta \cos\theta \cos\phi \; \ki + (-1 + \sin^2 \theta \cos^2 \phi) \; \epszero + \sin^2 \theta \sin \phi \cos \phi \; \norm} {\sqrt{1 - \sin^2 \theta \cos^2 \phi} }.
\ea 
The outgoing direction is defined by the angles $\{\theta, \phi\}$. Each angle is 
determined by randomly drawing a number $X_i \in U[0, 1]$, and matching it to the corresponding
cumulative distribution function (CDF).

We first select the angle $\theta$. Taking into account that $\hat\varepsilon_{1a}\cdot\epszero = 0$ and 
$\hat\varepsilon_{1b}\cdot\epszero = \pm\sqrt{1-\sin^2\theta\cos^2\phi}$, the integration of
the cross section (\ref{eq:sigmaKN}) over $\phi$ gives
\ba 
\left\langle\frac{d\sigma_{0\rightarrow a}}{d\cos \theta}\right\rangle_\phi
&=& \frac{\pi r_e^2}{2}\left(\frac{E_1}{E_0}\right)^2 \left[\frac{E_1}{E_0} + \frac{E_0}{E_1} - 2\right]; \nn 
\left\langle\frac{d\sigma_{0 \rightarrow b}}{d\cos \theta}\right\rangle_\phi
&=& \frac{\pi r_e^2}{2} \left(\frac{E_1}{E_0}\right)^2 \left[\frac{E_1}{E_0} + \frac{E_0}{E_1} + 2 \cos^2 \theta \right]
\ea 
for each polarization state.  Summing over final polarization states gives the spin-averaged 
KN cross section,
\be\label{eq:sigmaKN2}
\frac{d\sigma}{d\cos \theta} = \pi r_e^2 \left(\frac{E_1}{E_0}\right)^2 \left[\frac{E_1}{E_0} + \frac{E_0}{E_1} - \sin^2 \theta \right].
\ee 
The cumulative distribution function $0 < {\rm CDF}_\theta(\theta) < 1$ for the scattering angle $\theta$ is obtained by
integrating Equation (\ref{eq:sigmaKN2}) over
$u \equiv \cos \theta$.  Defining $x \equiv E_0/m_e$, 
\ba
\frac{d\sigma}{du} &=& \pi r_e^2 \left(1 + x (1- u)\right)^{-2} \left[ (1 + x(1-u))^{-1} + x(1-u) + u^2\right]; \nn 
\int \frac{d\sigma}{du} du &=& \frac{\pi r_e^2}{2x^3}\left[ \frac{x^2}{(1+x(1-u))^2} - 2(x^2 - 2x - 2)\ln(1 + x(1-u)) + 2x u + \frac{4 x + 2}{1 + x(1-u)}\right] \equiv \frac{\pi r_e^2}{2x^3} F(u),\nn
\ea
we have
\be
{\rm CDF}_\theta(\cos\theta) = \frac{F(\cos\theta) - F(-1)}{F(1) - F(-1)}. 
\ee
Drawing a random variable $X_\theta$ then gives $\cos \theta = {\rm CDF}_\theta^{-1}(X_\theta)$.
This also determines the energy $E_1$ of the outgoing photon and the ratio $R = E_1 / E_0$ 
via Equation (\ref{eq:sigmaKN}).
The distribution of $\phi$ is obtained in a similar manner by integrating $d\sigma_{0\rightarrow 1a}/d\Omega +
d\sigma_{0\rightarrow 1b}/d\Omega$ over $\phi$, 
\ba 
{\rm CDF}_\phi(\phi) &=& \frac{2\phi(R + R^{-1} - \sin^2 \theta) - \sin^2\theta \sin(2\phi)}{4\pi(R + R^{-1} - \sin^2\theta)}.
\ea
Drawing a second random variable $X_\phi$ sets $\phi = {\rm CDF}_\phi^{-1}(X_\phi)$. 

After $\theta$ and $\phi$ are set, the outgoing Stokes parameters and basis states can be determined, using the method derived in \cite{Berestetskii1971} for scattering of photons off unpolarized electrons. For incoming and outgoing photon directions $\hat k_0$ and $\hat k_1$, two sets of polarization bases $\{\hat \varepsilon_{0, \perp}, \hat\varepsilon_{0, \parallel} \}$ and $\{\hat \varepsilon_{1, \perp}, \hat\varepsilon_{1, \parallel} \}$ are constructed, with

\begin{align}
    \hat \varepsilon_{0,\perp} &= \hat \varepsilon_{1, \perp} = \frac{\hat k_0 \times \hat k_1}{|\hat k_0 \times \hat k_1|} = -\sin \phi \; \hat \varepsilon_0 + \cos \phi \; \norm; \nn
    \hat \varepsilon_{0, \parallel} &= \hat k_0 \times \hat \varepsilon_{0, \perp} = -\cos\phi \; \hat \varepsilon_0 - \sin \phi \; \norm; \nn 
    \hat \varepsilon_{1, \parallel} &= \hat k_1 \times \hat \varepsilon_{1, \perp} = \sin \theta \; \hat k_0 - \cos \theta \cos\phi \; \hat \varepsilon_0 - \cos \theta \sin \phi \; \norm.
\end{align}
The Stokes parameters of the incoming photon must be defined with respect to the new basis $\{\hat \varepsilon_{0, \perp}, \hat\varepsilon_{0, \parallel} \}$. From the initial basis states $\{\hat \varepsilon_{i, 1}, \hat \varepsilon_{i, 2}\}$ and polarization data $\{\delta_i, \psi_i\}$, the characteristic polarization vector is given as:
\begin{equation}
    \hat \varepsilon_i = \cos \psi_i \; \hat \varepsilon_{i, 1} + \sin \psi_i \; \hat \varepsilon_{i, 2}
\end{equation}
Note that $\hat \varepsilon_0$ is only guaranteed to be $\hat \varepsilon_i$ if $\delta_i = 1$. The transformed incoming Stokes parameters are then
\begin{align}
    Q_0 &= \delta_i \left((\hat\varepsilon_i \cdot \hat \varepsilon_{0,\perp})^2 - (\hat\varepsilon_i \cdot \hat \varepsilon_{0,\parallel})^2 \right) = \delta_i\left(\sin^2\phi - \cos^2\phi\right); \nn
    U_0 &= 2 \delta_i (\hat\varepsilon_i \cdot \hat \varepsilon_{0,\perp}) (\hat\varepsilon_i \cdot \hat \varepsilon_{0,\parallel}) = 2 \delta_i \sin\phi \cos\phi.
\end{align}
The Stokes parameters of the outgoing polarization are then determined to be:
\begin{align}
    Q_1 &= \frac{\sin^2 \theta + (1 + \cos^2\theta) Q_0}{R + R^{-1} - (1 - Q_0)\sin^2\theta},\nn 
    U_1 &= \frac{2 \cos \theta U_0}{R + R^{-1} - (1 - Q_0)\sin^2\theta},
\end{align}
with the new basis states $\{\hat \varepsilon_{1, \perp}, \hat\varepsilon_{1, \parallel} \}$. The new basis states are transformed back to BL frame polarizations, and re-encoded into Walker-Penrose constant.

\section{High-Energy Photon Loss by Photon Collisions}
\label{app:photon_consumption}

\begin{figure}[h]
    \centering
    \includegraphics[width=\linewidth]{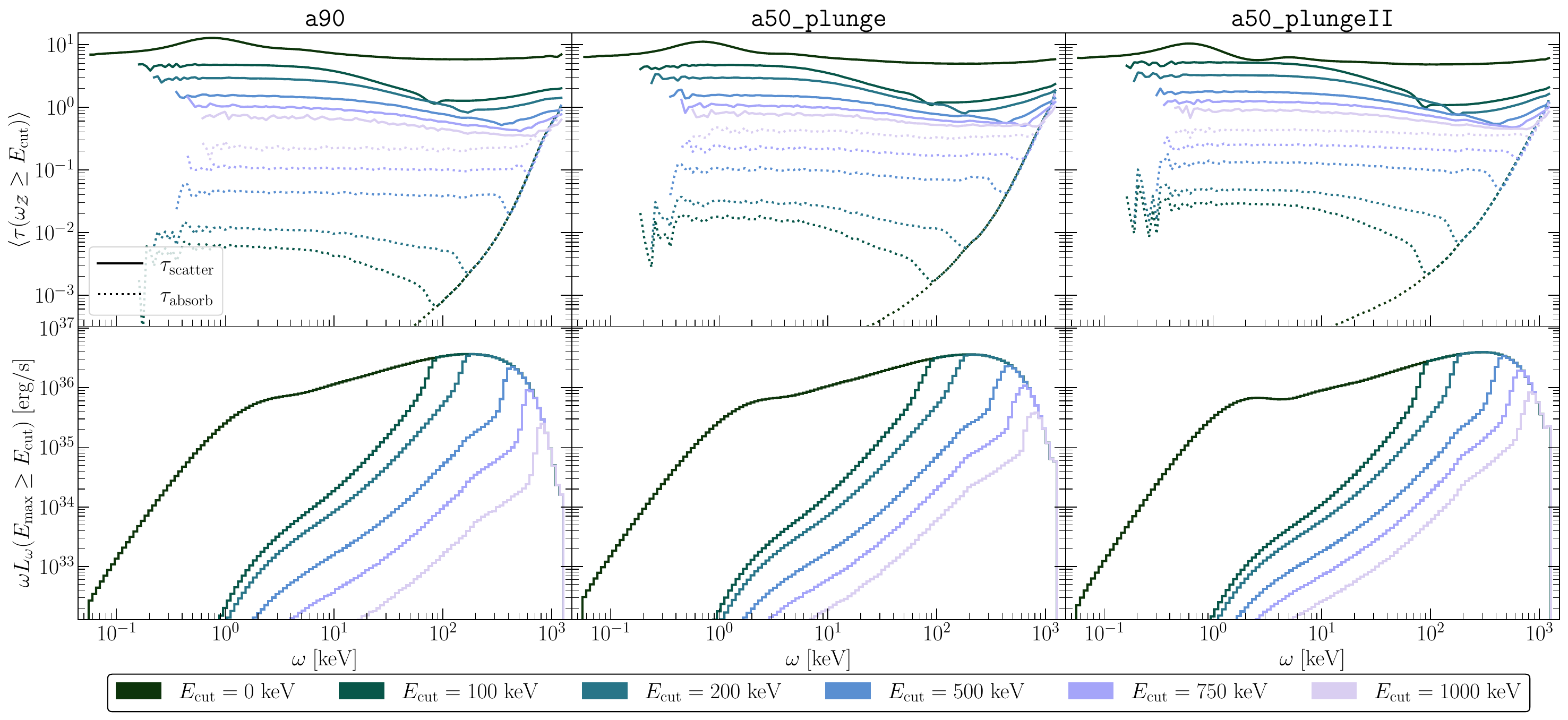}
    \caption{Comparison of the scattering optical depth to the absorption optical depth for pair-creating photons. Top: Average optical depths to scattering (solid, Equation \ref{eq:scatter_depth}) and absorption (dashed, Equation \ref{eq:absorb_depth}). The optical depths are only integrated along parts of the trajectory where the photon energy exceeds $E_{\rm cut}$, in order to highlight the conditions
    encountered by pair-creating photons. Different choices of $E_{\rm cut}$ are shown as different colours. Bottom: Inclination-summed spectra for all escaping photons. Photons with peak energy exceeding $E_{\rm cut}$ along the trajectory are also shown as separate spectra.  Low-frequency tails of the $\tau_{\rm abs}$ curves represent
    a small subset of photons that are trapped and downscatter in frequency in the outer, cold disk.}
    \label{fig:consumption}
\end{figure}

During the Monte Carlo evolution of ray trajectories, absorption via $\gamma + \gamma
\rightarrow e^+ + e^-$ is not considered.  It is relatively expensive to evaluate 
the optical depth to pair absorption during ray propagation, as this
involves integration over discrete sums of angular and energy bins in each cell.
The pair absorption optical depth in the ZAMO frame is
\begin{equation}
    \label{eq:absorb_depth}
    \frac{d\tau_{\rm absorb}}{dt_\ZAMO} ({\bm k}_\ZAMO, \varepsilon^\mu_\ZAMO) =\sum_{{\bm k}_i} |1 - \hat{k}_i \cdot \hat{k}_\ZAMO| n_\gamma({\bm k}_i) \sigma_{\gamma \gamma}(\theta({\bm k}_i, {\bm k}_\ZAMO), \rho_i, \rho(\varepsilon^\mu_\ZAMO)),
\end{equation}
where the cross section is given in Equation (\ref{eq:pair_creation_sigma}), COM quantity $\theta$ is defined in Equation (\ref{eq:theta_COM}), and the sum is taken over all representative wavevectors ${\bm k}_i$ in the angular and energy bins for a given cell. Re-emission through pair annihilation can be included as a separate emission source; the process of absorption and re-emission then effectively acts as a separate scattering channel for high energy photons. We estimate the overall importance of absorption by comparing the integrated absorption optical depth to the integrated scattering optical depth for escaping photons along the full trajectory. Both the absorption and scattering optical depths are functions of photon energy, which vary along the trajectory. To neglect non-pair creating photons, the optical depths are only integrated in parts of the trajectory where the photon energy $\omega_\ZAMO$ exceeds an energy cutoff $E_{\rm cut}$. Figure \ref{fig:consumption} shows the integrated optical depths to scattering and absorption by pair creation, as well as the total, inclination-summed spectra for escaping photons. The spectra are further divided by trajectories which peak at energies above $E_{\rm cut}$, to highlight the relative populations of pair-creating photons. The integrated optical depths along parts of trajectories with $\omega_\ZAMO \geq E_{\rm cut}$ highlights the relative effect of scattering to absorption; absorption is only a comparable effect at the strictest cut $E_{\rm cut} = 1000$ keV, which is a relatively small population of the pair-creating photons. As seen in the optical depths and spectra, photons with energy above $E_{\rm cut}$ have extended tails of low energy photons, representing photons that were trapped in the cold, outer disk.

\newpage

\bibliography{main}{}
\bibliographystyle{aasjournal}

\end{document}